\documentclass[a4paper,11pt,amsmath,mathrsfs,placeins]{article}
\usepackage{jcappub} 
\usepackage{cancel}
\usepackage{subcaption}
\usepackage{mathrsfs}
\usepackage{amsmath}   % for \overset
\usepackage{amssymb}   % for some math symbols
\usepackage{feynmp-auto,expdlist}

 \newcommand{\beq}[1]{\begin{equation}\label{#1}}
 \newcommand{\eeq}{\end{equation}}
 \newcommand{\bea}[1]{\begin{eqnarray}\label{#1}}
 \newcommand{\eea}{\end{eqnarray}}

\title{Enhanced Stochastic Gravitational Waves signals from Wess-Zumino chiral superfield}
 \author{AlexKen Lee $^{a}$, Keyun Wu $^{a}$}
 \affiliation[a]{Departament de F\'{i}sica Qu\`{a}ntica i Astrof\'{i}sica, Institut de Ci\`{e}ncies del Cosmos (ICCUB), Universitat de Barcelona, Mart\'{i} i Franqu\`{e}s 1, E-08028 Barcelona, Spain}

\emailAdd{alexkenlee@163.com}
\emailAdd{keyunwu@fqa.ub.edu}

\abstract{In this work, we investigate the possibility that supersymmetric structures may leave observable imprints in the stochastic gravitational-wave (GW) background generated during the reheating era. To this end, we construct a phenomenological interaction vertex describing the coupling between a single inflaton and the D-term sectors of a pair of chiral and anti-chiral superfields. In contrast to the conventional Yukawa coupling between the inflaton and structureless matter fields, we find that the supersymmetry-preserving chiral multiplet structure leads to a substantial enhancement, by at least one order of magnitude, in the amplitude of the resulting GWs spectrum. Our results therefore suggest that the interplay between reheating-era stochastic GWs and supersymmetric phenomenology merits further exploration and development.}

\begin{document} 
\maketitle
\flushbottom

\section{Introduction}
The detections of gravitational waves (GWs) by the LIGO–Virgo collaborations \cite{LIGOScientific:2016emj} have inaugurated a new era in which GWs serve as a powerful probe of fundamental physics, offering unprecedented opportunities to uncover the underlying mechanisms hidden behind the evolution of the universe as well as the mergers of compact and supermassive astrophysical objects. Gravitational-wave signals provide a unique window into the physics of the early Universe, carrying imprints of high-energy processes that remain inaccessible to conventional probes. A broad class of cosmological scenarios predict the existence of stochastic GWs backgrounds spanning a wide range of frequencies. Within the standard inflationary paradigm, quantum fluctuations of the metric generate nearly scale-invariant tensor modes at ultra-low frequencies, which are the primary targets of cosmic microwave background (CMB) B-mode polarization searches \cite{Caprini:2018mtu,BICEP:2021xfz}. On smaller scales, large curvature perturbations responsible for primordial black hole (PBH) formation inevitably generate scalar-induced GWs, with spectral peaks that map directly onto PBH mass ranges \cite{Inomata:2017okj,Saito:2008jc}. First-order cosmological phase transitions, accompanied by bubble collisions, sound waves, and magnetohydrodynamic turbulence, can generate peaked GWs spectra at frequencies relevant for space-based observatories \cite{Caprini:2015zlo,Hindmarsh:2017gnf,Caprini:2019egz,Ellis:2018mja,Athron:2023xlk}. Likewise, networks of topological defects such as cosmic strings or monopole–string systems radiate GWs continuously, yielding broad-band signals constrained from pulsar timing arrays to ground-based detectors \cite{Damour:2004kw,Blanco-Pillado:2017rnf,Ringeval:2017eww}. Altogether, these mechanisms demonstrate that the early Universe is a natural laboratory for GWs production, motivating a broad, synergistic program to search for stochastic backgrounds across all available frequency bands.

Most intriguingly, the dynamics of reheating, along with various extensions incorporating new physics beyond the Standard Model (SM), can act as a powerful source of high-frequency gravitational waves (GWs), with peak frequencies potentially detectable by terrestrial interferometers. In particular, a specific class of GWs signals generated through inflaton decay, often referred to as graviton bremsstrahlung, has recently attracted considerable attention \cite{Nakayama:2018ptw,Huang:2019lgd,Barman:2023ymn}. Such a stochastic GWs background could provide profound insights into the microscopic structure of physics beyond the SM, as well as crucial clues about the early evolution of the Universe. As pointed out in \cite{Barman:2023ymn}, the emission of massless gravitons from inflaton decay closely resembles the bremsstrahlung process originally described by Weinberg \cite{Weinberg:1965nx}. In particular, several works, including \cite{Nakayama:2018ptw,Huang:2019lgd,Barman:2023ymn}, have investigated graviton bremsstrahlung arising from the three-body decay of the inflaton, where the inflaton decays into a pair of massive particles, accompanied by the emission of a massless graviton. By evaluating the differential three-body decay width and solving the corresponding Boltzmann equations, one can determine the evolution of the GWs energy density and, subsequently, compute the resulting gravitational-wave spectrum. Since these pioneering studies, a broad array of follow-up works \cite{Barman:2023rpg,Kanemura:2023pnv,Tokareva:2023mrt,Choi:2024ilx,Hu:2024awd,Kaneta:2024yyn,Xu:2024fjl,Mantziris:2024uzz,Inui:2024wgj,Datta:2024tne,Jiang:2024akb,Bernal:2024jim,Xu:2025wjq,Banik:2025olw,Ai:2025fqw,Wang:2025lmf,Lee:2025lyk} has significantly extended the framework. Compared with the early investigations of stochastic GWs signals from inflaton decay, these subsequent developments have dramatically enriched the phenomenological predictions of the theory—affecting not only the amplitude of the GWs spectrum, but also its characteristic frequency range and even the detailed shape of the spectral profile.

%Complementary to the inflaton decay scenario, graviton emission in the context of inflaton scattering has been explored in \cite{Bernal:2023wus}. Meanwhile, \cite{Barman:2023rpg} analyzed how the shape of the inflaton potential during its oscillation phase affects the resulting stochastic GW spectrum. Moreover, \cite{Kanemura:2023pnv} showed that graviton bremsstrahlung from the decay of the $B$–$L$ Higgs boson can generate ultra-high-frequency gravitational waves with a notably amplified spectral amplitude. Further studies have extended the investigation of graviton bremsstrahlung to diverse physical settings, including leptogenesis scenarios \cite{Ghoshal:2022kqp}, Planck-scale physics \cite{Hu:2024awd}, the kination epoch \cite{Inui:2024wgj}, and classical field-theoretic treatments of polynomial inflaton potentials \cite{Jiang:2024akb}. //In addition, as shown by \cite{Lee:2025lyk}, it can be also extended into the studies about the induced by the trosion structure...

In \cite{Barman:2023ymn}, it was shown that within a Yukawa-type coupling framework between the inflaton and a pair of massive particles with different spins, the dimensionless strain associated with the stochastic gravitational waves generated from the three-body decay exhibits only a very slight dependence on the spin of the final-state particles. More precisely, the higher the spin of the particle produced in the inflaton decay, the larger the corresponding amplitude of the stochastic GWs dimensionless strain tends to be; however, the differences in amplitude remain well below one order of magnitude. This observation naturally leads to an intriguing question: if these particles of different spins can be organized into multiplets transforming under some internal symmetry structure, for example the chiral multiplets in the context of supersymmetry, would their associated gravitational-wave amplitudes exhibit similarly universal behavior, or could there be deviations beyond this expectation? Motivated by these considerations, we aim to investigate the stochastic GWs sourced by inflaton decay induced by a generalized Yukawa coupling between the inflaton and the D-term sectors of a pair of chiral and anti-chiral superfields. For simplicity, we extract only the quadratic contributions from the D-term sectors of the chiral and anti-chiral superfields, and we focus primarily on the three-body decay of the inflaton, which produces a pair of complex scalars or Majorana fermions together with the radiative emission of a single graviton. Since the complex scalars and Majorana fermions contained in a chiral superfield form a chiral supermultiplet and naturally share the same mass, it is reasonable to expect that the stochastic GWs generated in the inflaton decay channels with these particles as final states should exhibit nearly identical amplitudes. In addition, we aim to investigate whether the inclusion of supersymmetry could potentially generate a prominent enhancement of the gravitational-wave signal, in comparison with the typical Yukawa-coupling scenario studied in \cite{Barman:2023ymn}. If any significant improvement indeed emerges, it would imply that the stochastic GWs produced in the early universe, particularly during the reheating stage, may provide a new observational window for probing possible remnants of supersymmetry (SUSY) \cite{Wess:1974tw, Gates:1983nr, Sohnius:1985qm, Martin:1997ns,Quevedo:2010ui,Salam:1974jj,Lykken:1996xt}.

Building on the above analysis and motivations, the structure of this work is as follows. In Section~\ref{SUSYCubicInteraction}, we introduce an extended Yukawa-like interaction between the inflaton and the (D-term sector of) a pair of chiral and anti-chiral superfields, with particular emphasis on all types of cubic interactions that arise in this setup. The detailed derivations and component expansions of the chiral superfield in superspace are provided in Appendix~\ref{ReviewWZChiral}. In Appendix~\ref{PathIntegralFeynmanRule}, we further derive all Feynman rules required for these effective cubic interactions and compute the corresponding two-body decay processes of the inflaton. The corresponding three-body decays accompanied by graviton emission are analyzed in Section~\ref{ThreeBodyDecaySec}, while the explicit expressions for the squared amplitudes and the resulting three-body decay widths are provided in Appendix~\ref{CubicInflaComplexScalar} and Appendix~\ref{EvaluThreeBodyDecay}. In these calculations, we also make use of the Feynman rules for the interactions between gravitons and Majorana fermions, which are derived in Appendix~\ref{FeyRuleGravitonMajorana}. Moreover, the evaluation of the squared amplitudes for the three-body decay in the Majorana case requires a range of techniques involving two-component Majorana spinors and $\sigma$-matrix algebra, for which we follow the standard formalism summarized in \cite{Dreiner:2008tw}. Subsequently, in Section~\ref{ThreeBodyDecaySec}, we analyze the contributions arising from the cubic interactions induced by the generalized Yukawa coupling between the inflaton and the D-term sectors of the chiral and anti-chiral superfields, and we compare these results with those obtained from the standard three-body decay channels. Finally, our conclusions and discussions are presented in Section~\ref{ConcluAndDiscuss}.

\section{Effective Cubic Interactions from D-Term Construction of Chiral Superfields and the Associated Inflaton Decay Processes \label{SUSYCubicInteraction}}

In the model construction of this work, we focus on the coupling between the inflaton and a pair of chiral superfields. In particular, we formulate an effective phenomenological Lagrangian through the so-called D-term construction. In the present study, we restrict our attention to the cubic interaction terms in this effective Lagrangian and investigate their roles in the inflaton decay processes, specifically in the emission of gravitons and the generation of a stochastic gravitational-wave background. Higher-order interaction terms will be addressed in future work. Given a chiral superfield expressed in the form \cite{Martin:1997ns}
\begin{align}
\nonumber
\Phi(x,\theta,\theta^{\dagger})&=\phi(x)+\sqrt{2}\theta\psi(x)+\theta\theta F(x)+\text{i}\theta^{\dagger}\bar{\sigma}^{\mu}\theta\partial_{\mu}\phi(x)\\
&-\frac{\text{i}}{\sqrt{2}}\theta\theta\theta^{\dagger}\bar{\sigma}^{\mu}\partial_{\mu}\psi(x)-\frac{1}{4}\theta\theta\theta^{\dagger}\theta^{\dagger}\partial_{\mu}\partial^{\mu}\phi(x)
\end{align}we consider an extended Yukawa-like interaction between the inflaton and the chiral superfields,
\begin{align}
\label{InflatonSuperFieldYukawa}
&S_{\text{int}}\!=\!\mathsf{y}_{\text{D}} \kappa\int d^{8}z~\varphi[\bar{\Phi}(x,\theta,\theta^{\dagger})\Phi(x,\theta,\theta^{\dagger})]_{D}
\end{align}where the associated interaction strength is characterized by the coupling constant $\mathsf{y}_{\text{D}}$, which is dimensionless. Here, $\kappa$ is defined as $\kappa\equiv\sqrt{16\pi}/M_P$, where $M_P$ denotes the reduced Planck scale, $M_P\simeq 2.44\times 10^{18}$ GeV. The $\bar{\Phi}(x,\theta,\theta^{\dagger})$ denotes the Hermitian conjugate of the superfield $\Phi(x,\theta,\theta^{\dagger})$. The detailed explanations and derivations of the superspace expansion of \eqref{InflatonSuperFieldYukawa} are presented explicitly in Appendix~\ref{ReviewWZChiral}. Here, we simply extract all types of cubic interaction terms as follows
\begin{align}
\label{YukawaInflatonMajoranaMain}
&S_{\text{int}}^{(\varphi\psi\psi)}=\frac{\text{i}}{2}\mathsf{y}_{\text{D}}\kappa\int d^{4}x\,\varphi(\psi^{\dagger}\bar{\sigma}^{\mu}\partial_{\mu}\psi-\partial_{\mu}\psi^{\dagger}\bar{\sigma}^{\mu}\psi)\\
\label{YukawaInflatonComScalarMain}
&S_{\text{int}}^{(\varphi\phi\phi)}=\frac{1}{4}\mathsf{y}_{\text{D}}\kappa\int d^{4}x\,\varphi\big(2\eta^{\mu\nu}\partial_{\mu}\phi^{\star}\partial_{\nu}\phi+4m^{2}\phi^{\star}\phi-\phi^{\star}\partial_{\mu}\partial^{\mu}\phi-\phi\partial_{\mu}\partial^{\mu}\phi^{\star}\big)
\end{align}Note that throughout this work we adopt the two-component formalism for describing Majorana fermions. Further details can be found in \cite{Dreiner:2008tw}, as well as in Appendix \ref{ReviewFreeMajorana} of this paper.

Building upon the effective interaction given in Eq.~\eqref{YukawaInflatonMajoranaMain}, the corresponding Feynman rules are derived in detail in Appendix~\ref{CubicInflaMajorana}, and the associated Feynman diagram is shown in Fig.~\ref{FigPsiI1PsiI2Inflatonpsipsi}. In this setup, the fermion appearing in the interaction is of the Majorana type. Therefore, for completeness, we also provide a concise review of the key aspects of free Majorana field theory in Appendix~\ref{ReviewFreeMajorana}. It is worth noting that the Majorana spinor $\psi$ and the complex scalar $\phi$ share the same on-shell mass, $m$. Employing the Feynman rules summarized in Fig.~\ref{FigPsiI1PsiI2Inflatonpsipsi}, we compute the corresponding two-body decay rate, 
\begin{align}
\label{TwoBodyMajorana}
&\Gamma_{\varphi\to\psi\psi}^{(0)}=\frac{\kappa^{2}\mathsf{y}_{\text{D}}^{2}M^{3}}{8\pi}y^{2}(1-4y^{2})^{3/2}
\end{align}where the dimensionless parameter $y$ is defined as $y = m/M$, with $M$ denoting the inflaton mass. The explicit derivation of Eq.~\eqref{TwoBodyMajorana} is presented in Appendix~\ref{CubicInflaMajorana}. This result is consistent with that obtained in Ref.~\cite{Gola:2021abm}, where the decay of an axion into a pair of Majorana dark matter particles was analyzed. It should be noted that the expression in Eq.~\eqref{TwoBodyMajorana} is not applicable to the case of massless final-state Majorana fermions, since both the propagators \eqref{TwoPointPsiPsidagger}-\eqref{TwoPointPsiPsi} and the Feynman rules \eqref{ThreePointCorrelationI1I2Result}–\eqref{ThreePointCorrelationI1DaggerI2Result} are derived within the perturbative framework of a massive free Majorana field theory, which becomes ill-defined in the limit $m \to 0$. Similarly, for the effective interaction presented in Eq.~\eqref{YukawaInflatonComScalarMain}, the detailed derivation of the corresponding Feynman rules is summarized in Eq.~\eqref{ThreePointScalarPairFinal}, while the computation of the two-body decay rate,
\begin{align}
\label{TwoBodyComplexScalar}
&\Gamma_{\varphi\to\phi\phi^{\star}}^{(0)}=\frac{\kappa^{2}\mathsf{y_D}^{2}M^{3}}{64\pi}(4y^{2}-\frac{1}{2})^{2}(1-4y^{2})^{1/2}
\end{align}can be found in Appendix~\ref{CubicInflaComplexScalar}.

\section{Three-Body Decay Process and the Differential Decay Rate \label{ThreeBodyDecaySec}}

\begin{figure}[ht]
	\begin{center}
		\includegraphics[scale=0.43]{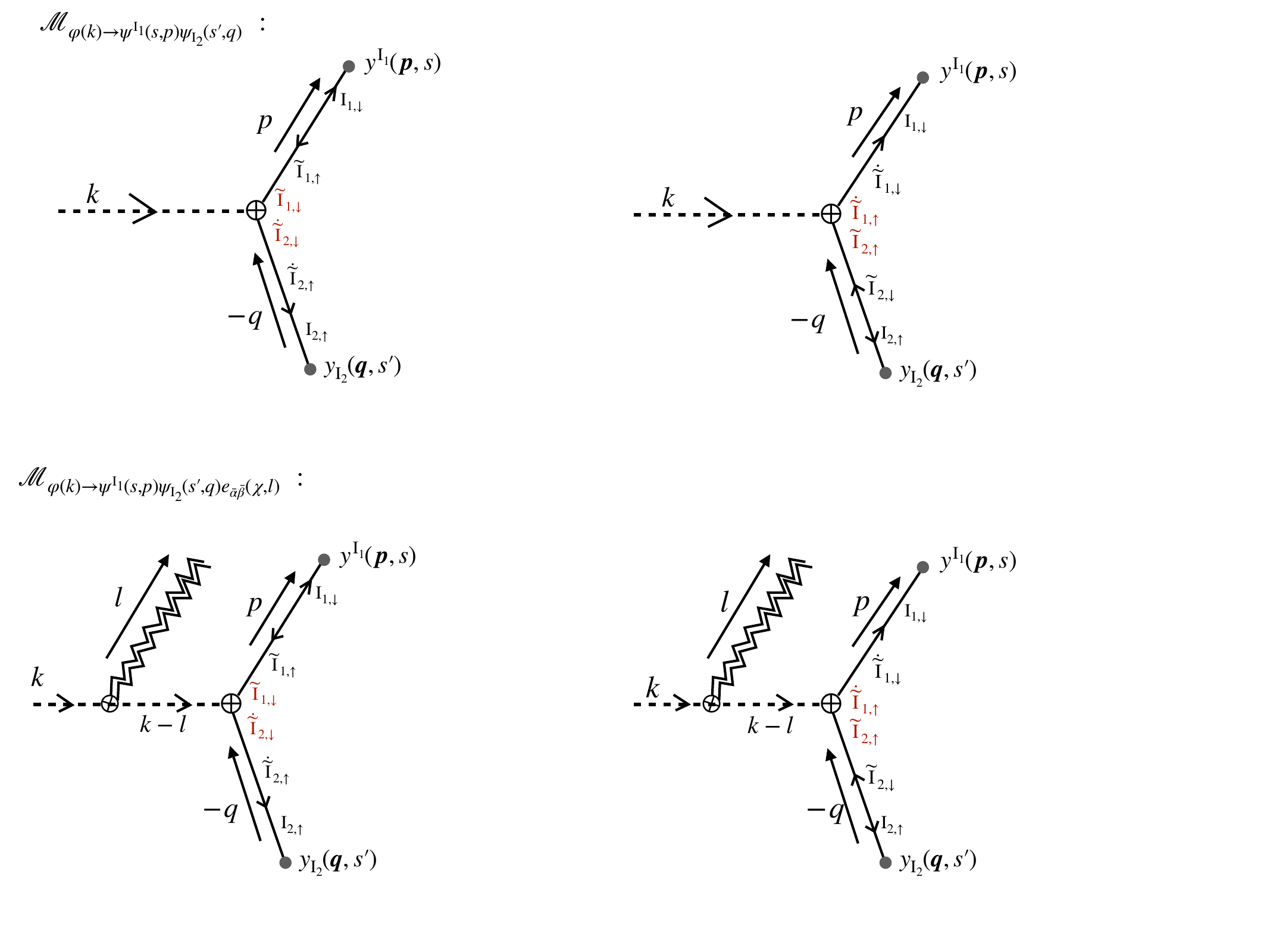}
		\caption{The upper panel illustrates the two-body decay of the inflaton into a pair of Majorana fermions, mediated by the two interaction vertices $\mathcal{V}_{(\varphi\psi\psi)}^{(q_{1},q_{2},\widetilde{\mathrm{I}}_{1,\downarrow},\dot{\widetilde{\mathrm{I}}}_{2,\downarrow})}$ and $\mathcal{V}_{(\varphi\psi\psi)}^{(q_{1},q_{2},\dot{\widetilde{\mathrm{I}}}_{1,\uparrow},\widetilde{\mathrm{I}}_{2,\uparrow})}$, presented in Appendix.  \ref{CubicInflaMajorana} and Fig. \ref{FigPsiI1PsiI2Inflatonpsipsi}, respectively. Correspondingly, the lower panel, together with Figs. \ref{ThreeBodyMajoranaCase1Case2}–\ref{ThreeBodyMajoranaCase3Case4}, depicts the full three-body decay process in which the inflaton decays into a pair of Majorana fermions accompanied by the emission of a single graviton.}
		\label{RefTwoThreeBodyMixMajorana}
	\end{center}
\end{figure}

\begin{figure}[ht]
	\begin{center}
		\includegraphics[scale=0.33]{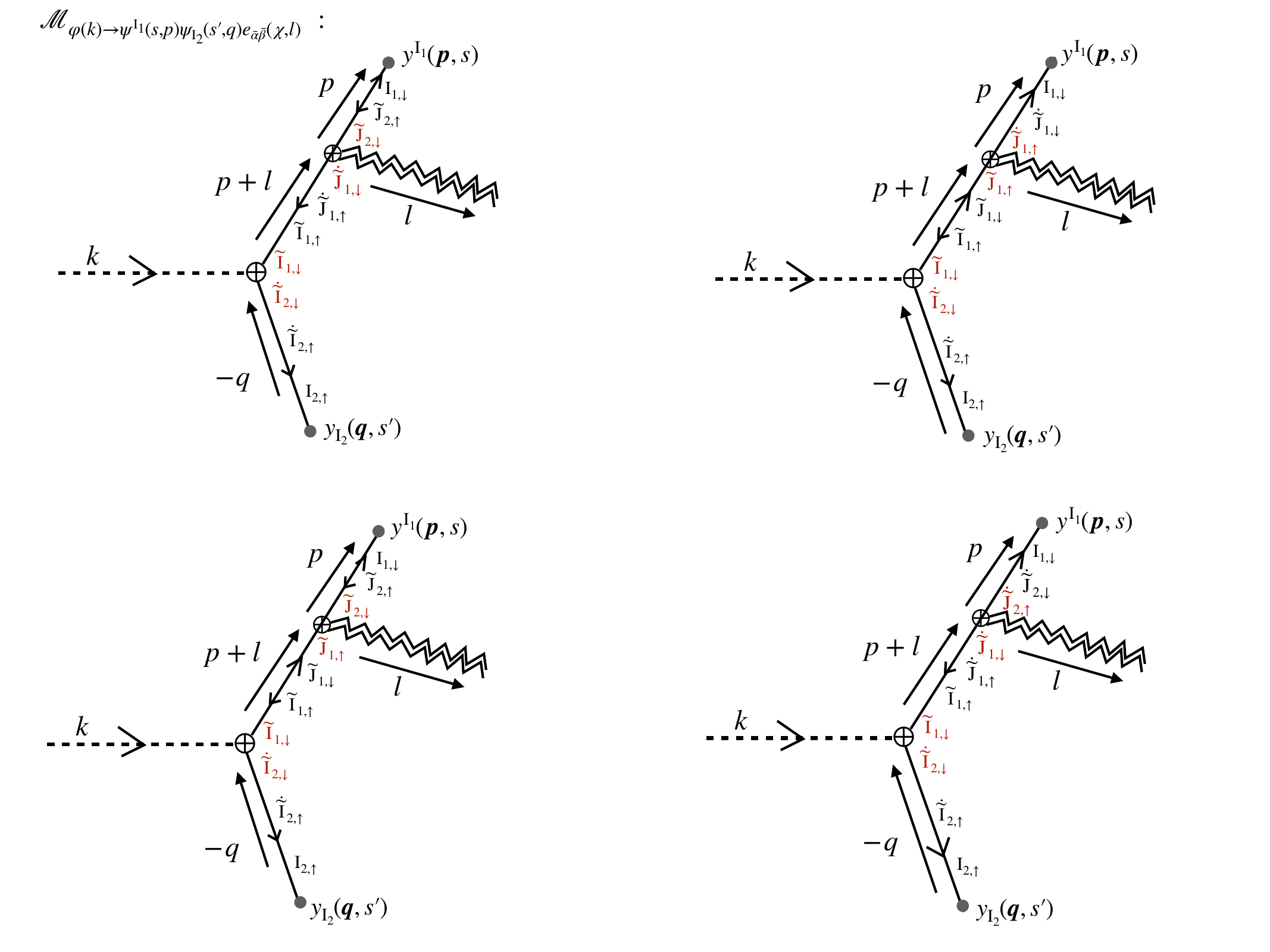}
        \includegraphics[scale=0.33]{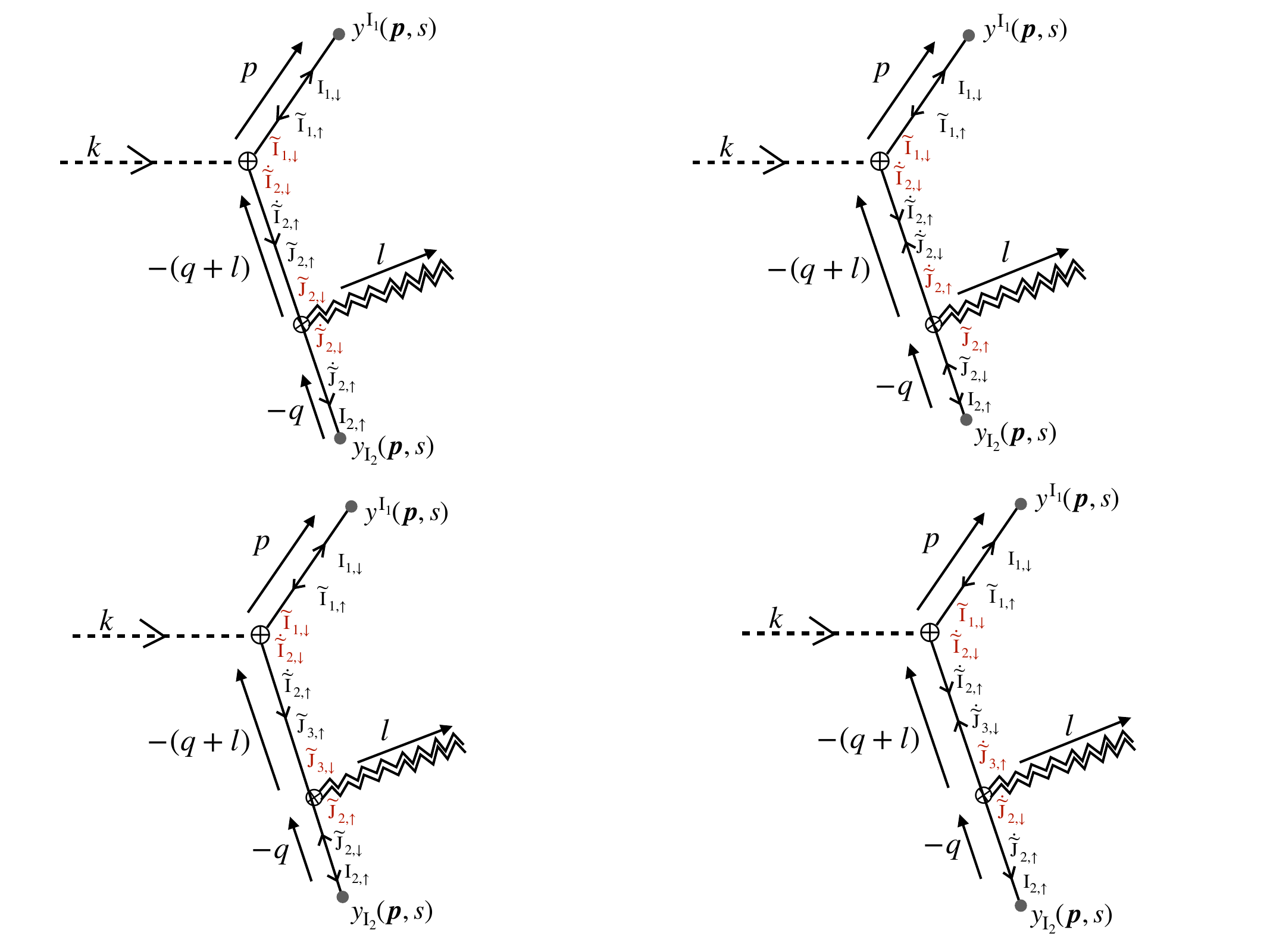}
		\caption{This figure, together with Fig. \ref{ThreeBodyMajoranaCase3Case4} and the lower panel of Fig. \ref{RefTwoThreeBodyMixMajorana}, collectively illustrates the full three-body decay process in which the inflaton decays into a pair of Majorana fermions accompanied by the emission of a single graviton.}
		\label{ThreeBodyMajoranaCase1Case2}
	\end{center}
\end{figure}

\begin{figure}[ht]
	\begin{center}
		\includegraphics[scale=0.33]{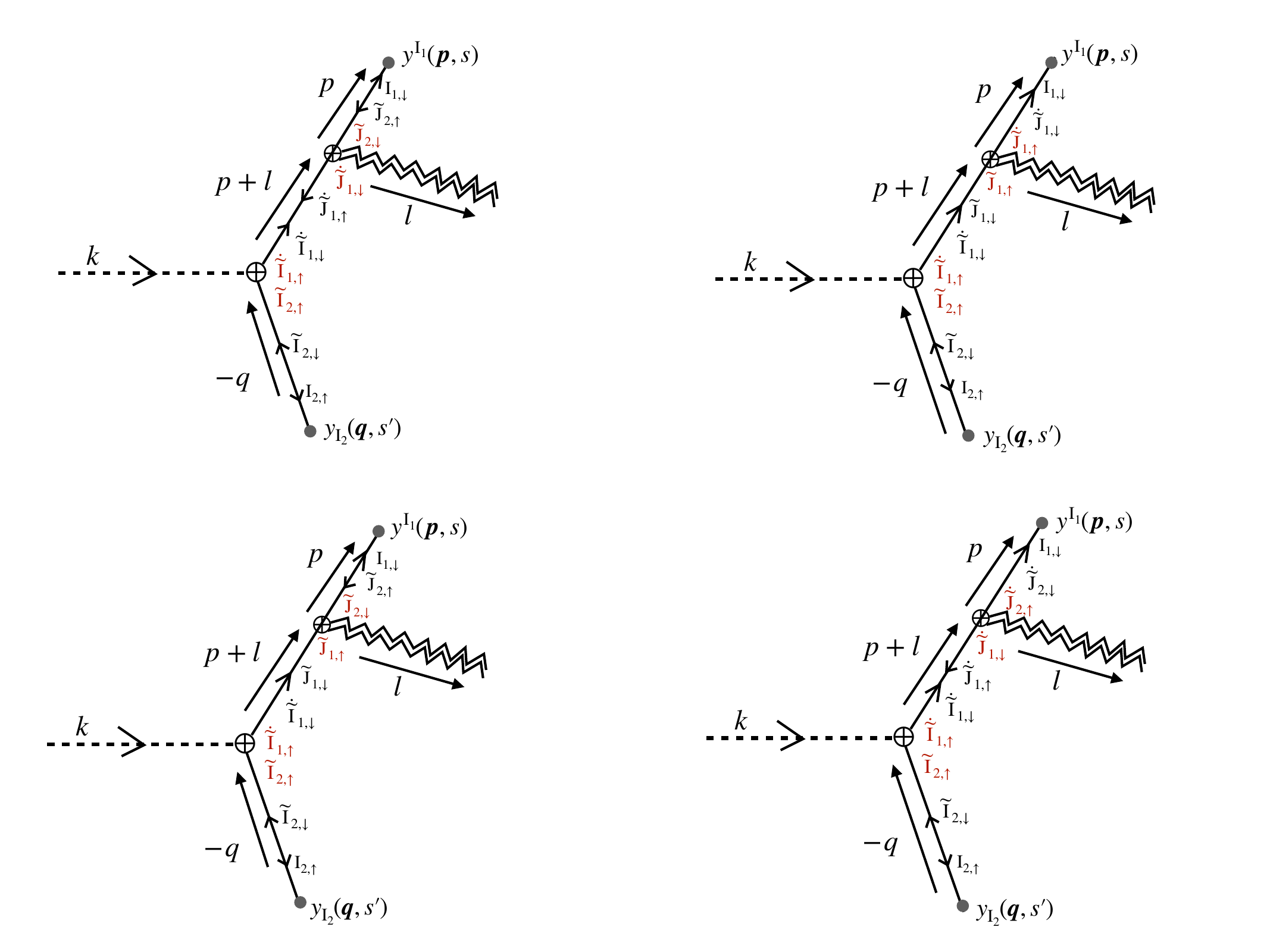}
        \includegraphics[scale=0.33]{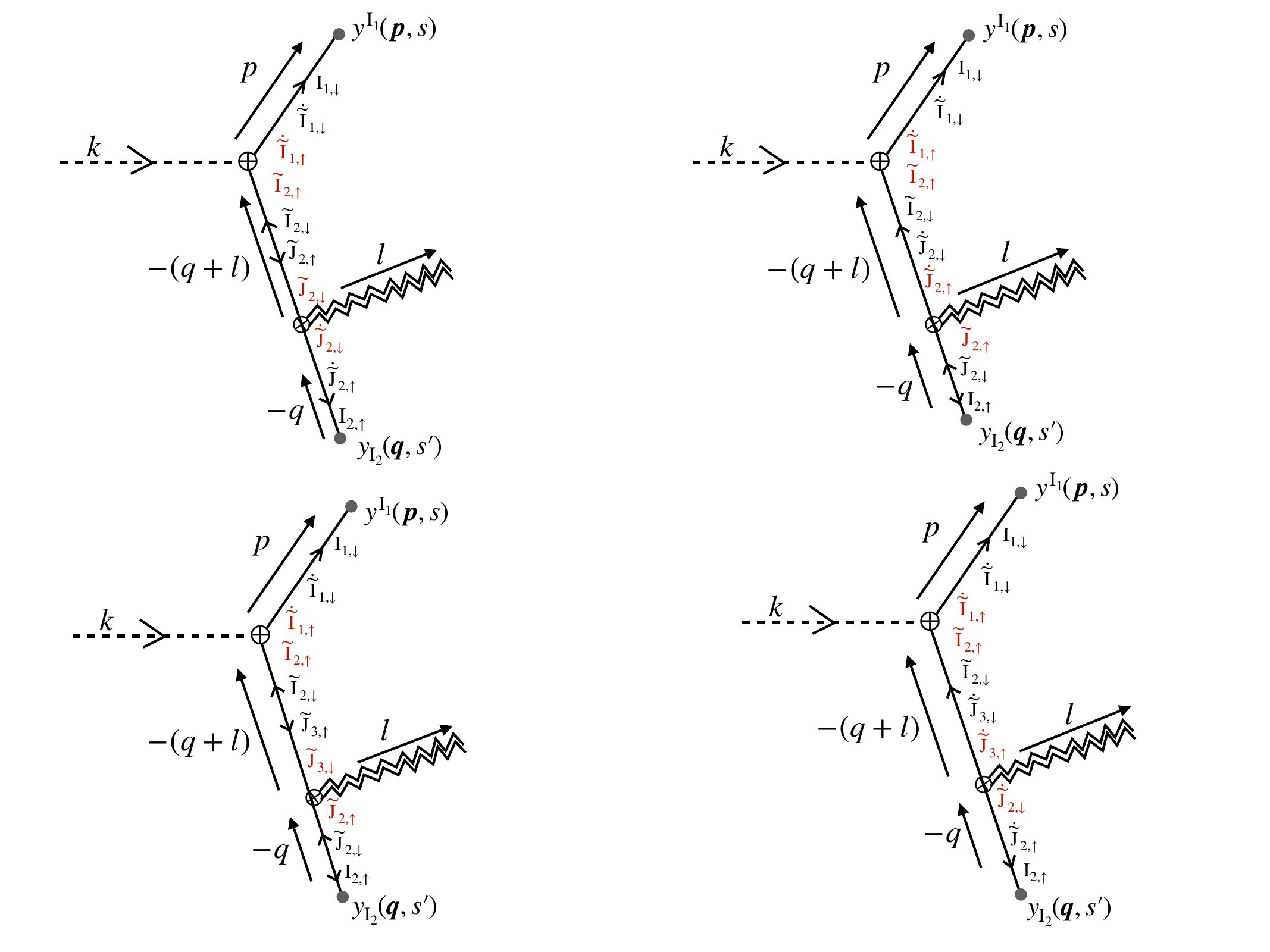}
		\caption{This figure, together with Fig. \ref{ThreeBodyMajoranaCase1Case2} and the lower panel of Fig. \ref{RefTwoThreeBodyMixMajorana}, collectively illustrates the complete three-body decay process in which the inflaton decays into a pair of Majorana fermions accompanied by the emission of a single graviton.}
		\label{ThreeBodyMajoranaCase3Case4}
	\end{center}
\end{figure}

\begin{figure}[ht]
	\begin{center}
		\includegraphics[scale=0.43]{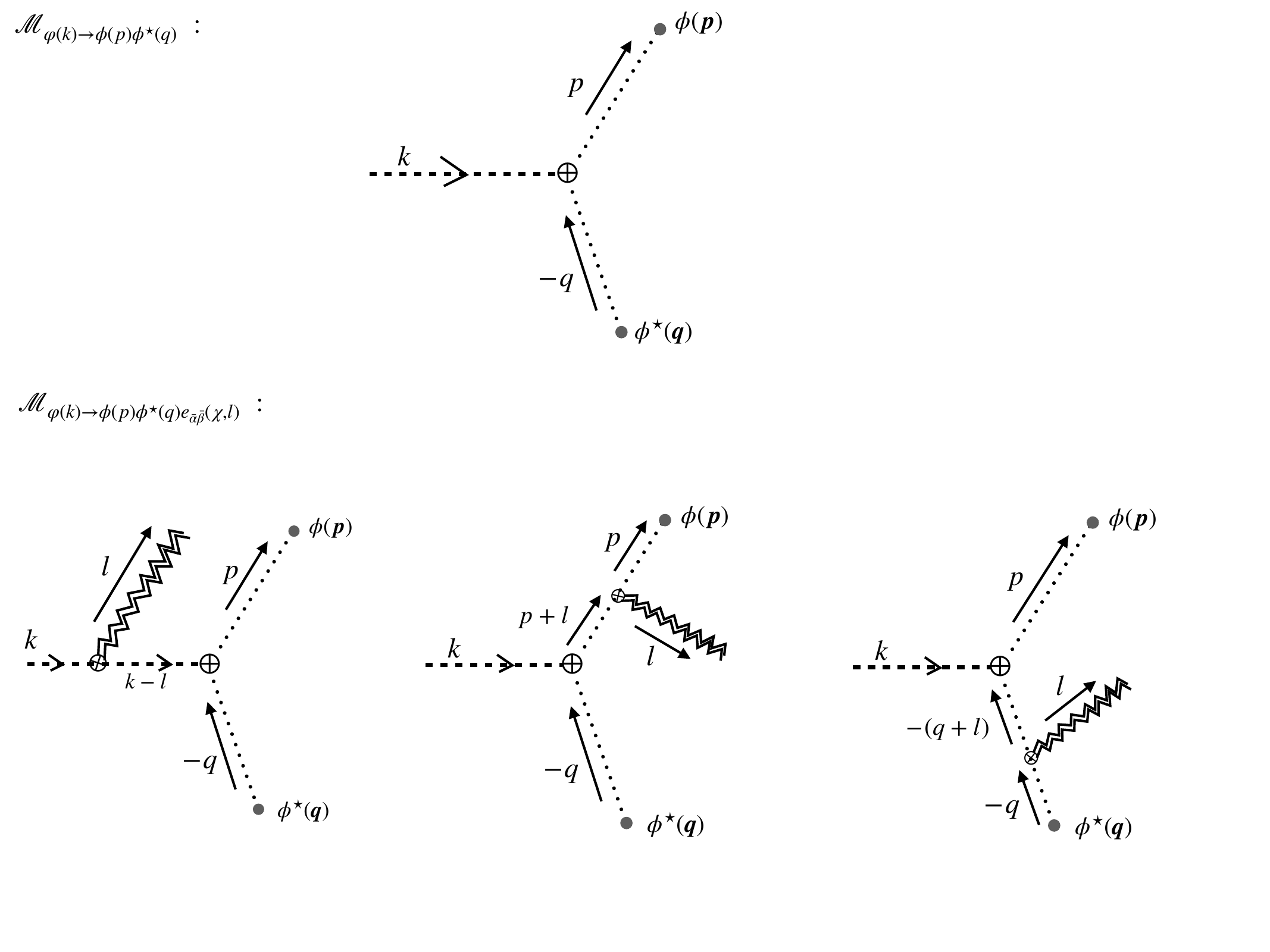}
		\caption{The upper panel illustrates the two-body decay of the inflaton into a pair of scalar particles, mediated by the interaction vertex $\mathcal{V}_{(\varphi\phi\phi)}^{(q_{1},q_{2})}$ given in \eqref{ThreePointScalarPairFinal}. Correspondingly, the lower panel depicts the complete three-body decay process in which the inflaton decays into a scalar pair accompanied by the emission of a single graviton.}
		\label{RefTwoThreeBodyMixScalar}
	\end{center}
\end{figure}

In this section, we focus primarily on the three-body decay processes illustrated in Fig. \ref{RefTwoThreeBodyMixMajorana}–\ref{ThreeBodyMajoranaCase3Case4}, which are induced by the cubic interaction given in Eq.~\eqref{YukawaInflatonMajoranaMain}, as well as those shown in Fig. \ref{RefTwoThreeBodyMixScalar}, arising from the cubic interaction specified in Eq.~\eqref{YukawaInflatonComScalarMain}. Here, $k$, $p$, $q$ and $l$ denote four-momentum of the initial and final states, respectively. The full amplitudes and the corresponding squared amplitudes associated with Figs. \ref{RefTwoThreeBodyMixMajorana}–\ref{ThreeBodyMajoranaCase3Case4} are explicitly presented in Appendix. \ref{EvaluThreeBodyDecay}. Besides, the corresponding Feynman rules for the interaction describing the coupling between a graviton and two Majorana fermions, i.e. $S^{(h\psi\psi)}_{\text{int}}$, are explicitly derived in Appendix. \ref{FeyRuleGravitonMajorana} using the path integral formalism. Meanwhile, the main results for the interaction vertex in momentum space are extracted and summarized in Fig. \ref{PsiI1PsiI2GRMajorana}. Here, we only summarize the final expressions for the squared amplitudes
\begin{align}
\nonumber
&\sum_{s}\sum_{s^{\prime}}\sum_{\chi}\vert\mathcal{M}_{\varphi(k)\to\psi^{(s),\mathrm{I}_{1}}(p)\psi_{\mathrm{I}_{2}}^{(s^{\prime})}(q)e_{\alpha\beta}^{(\chi)}(l)}^{\text{(total)}}\vert^{2}=\frac{-\kappa^{4}\mathsf{y}_{\text{D}}^{2}}{16E_{l}^{4}M^{2}(M-2E_{p})^{2}\big(M-2(E_{p}+E_{l})\big)^{2}}\\
\nonumber
&\times\big(4M^{2}(E_{p}^{2}+3E_{p}E_{l}+E_{l}^{2})-4M^{3}(E_{p}+E_{l})-8E_{l}E_{p}M(E_p+E_l)+4E_l^2m^2+M^{4}\big)^{2}\\
\nonumber
&\times\bigg(M^{2}(M-2E_{p})\big((M-2E_{p})^{3}-4E_{l}^{3}\big)-4E_{l}M(M-2E_{p})^{2}\big(M(M-2E_{p})+2m^{2}\big)\\
\label{SquAmpliThreeBodyMajorana}
&+2E_{l}^{2}\big(2m^{2}M(3M-8E_{p})+3M^{2}(M-2E_{p})^{2}+8m^{4}\big)\bigg)
\end{align}
where, $E_l$ and $E_p$ denotes energies of the massless graviton and final $\psi$ states, respectively. It is worth noting that, when evaluating the squared amplitude Eq. \eqref{SquAmpliThreeBodyMajorana}, the cross terms between the contributions from Fig. \ref{RefTwoThreeBodyMixMajorana} and those from Fig. \ref{ThreeBodyMajoranaCase1Case2}–\ref{ThreeBodyMajoranaCase3Case4} vanish. Meanwhile, it is often necessary to employ the two-component spinor techniques given in \cite{Dreiner:2008tw} to evaluate traces involving multiple products of the $\sigma$ matrices. Besides, the amplitudes and the corresponding squared amplitudes associated with the three-body decay process, illustrated in the lower panel of Fig. \ref{RefTwoThreeBodyMixScalar}, are explicitly derived in Appendix. \ref{CubicInflaComplexScalar}. The final expressions for the squared amplitudes are given by
\begin{align}\nonumber
&\sum_\chi\vert\mathcal{M}^{\text{(total)}}_{\varphi(k)\to\phi(p)\phi^{\star}(q)e_{\alpha\beta}^{(\chi)}(l)}\vert^{2}=\frac{\kappa^4\mathsf{y}_{\text{D}}^{2}}{128E_l^4M^2(M-2E_p)^2\left(M-2(E_p+E_l)\right)^2}\\ \nonumber
&\times\bigg(4M^2(E_p^2+3E_pE_l+E_l^2)-4M^3(E_p+E_l)-8E_pE_lM(E_p+E_l)+4E_l^2m^2+M^4\bigg)^2\\ 
\label{SquAmpliThreeBodyScalar}
&\times\bigg(E_lM(3M-8E_p)-2M(M-2E_p)^2+8E_lm^2\bigg)^2 
\end{align}\label{amplitudeS}where the Feynman rule for the minimal coupling interaction between a graviton and two complex scalar fields is required. This Feynman rule can be found in the literature \cite{Gleisberg:2003ue,deAquino:2011ix,Lee:2025lyk}.

\begin{figure}[ht]
	\begin{center}
		\includegraphics[scale=0.4]{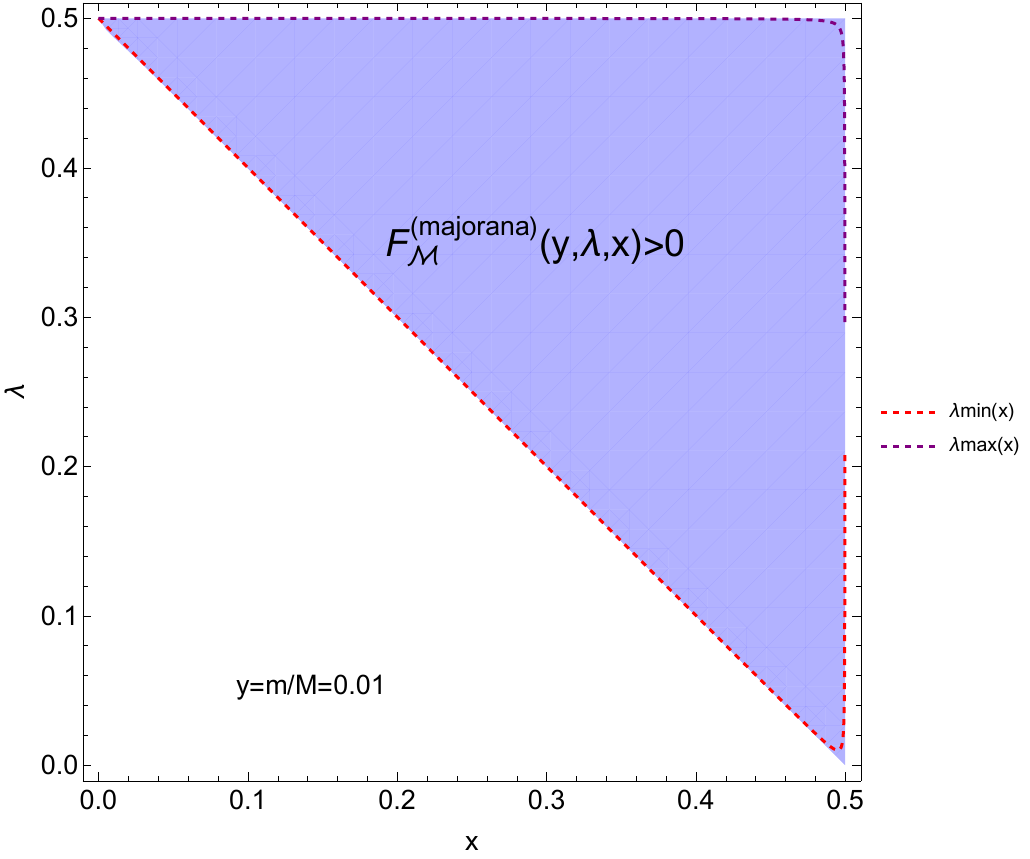}
        \includegraphics[scale=0.4]{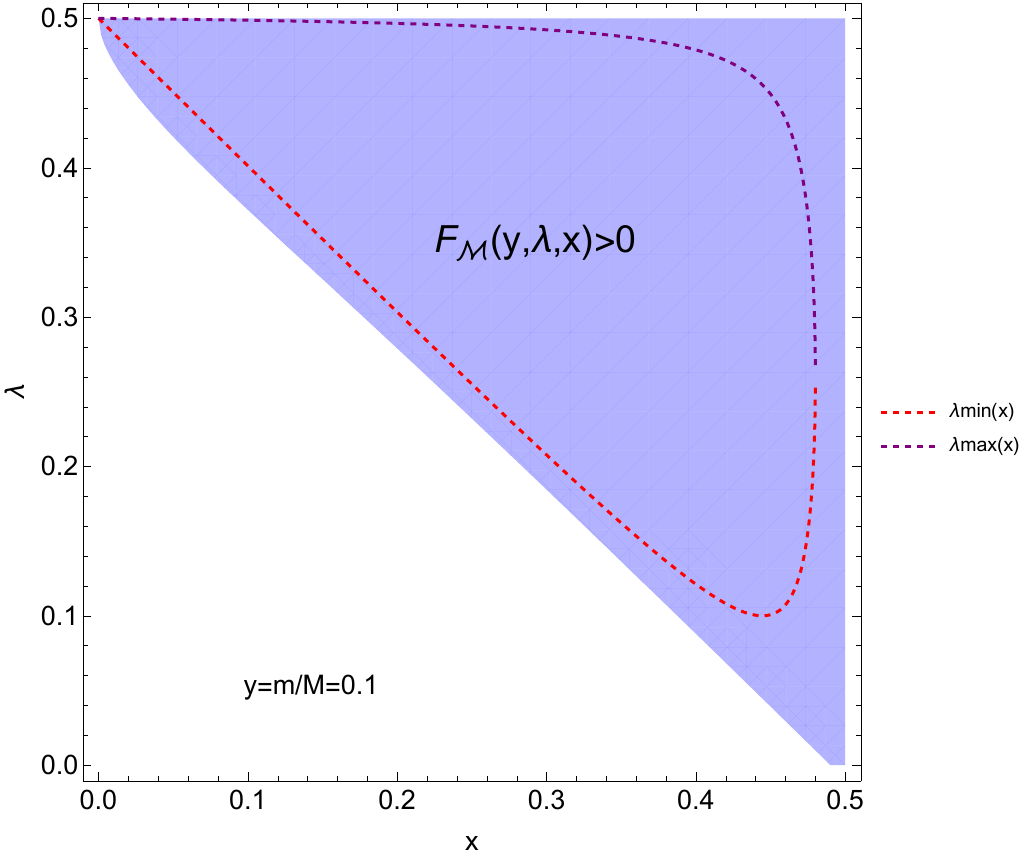}\\
        \includegraphics[scale=0.4]{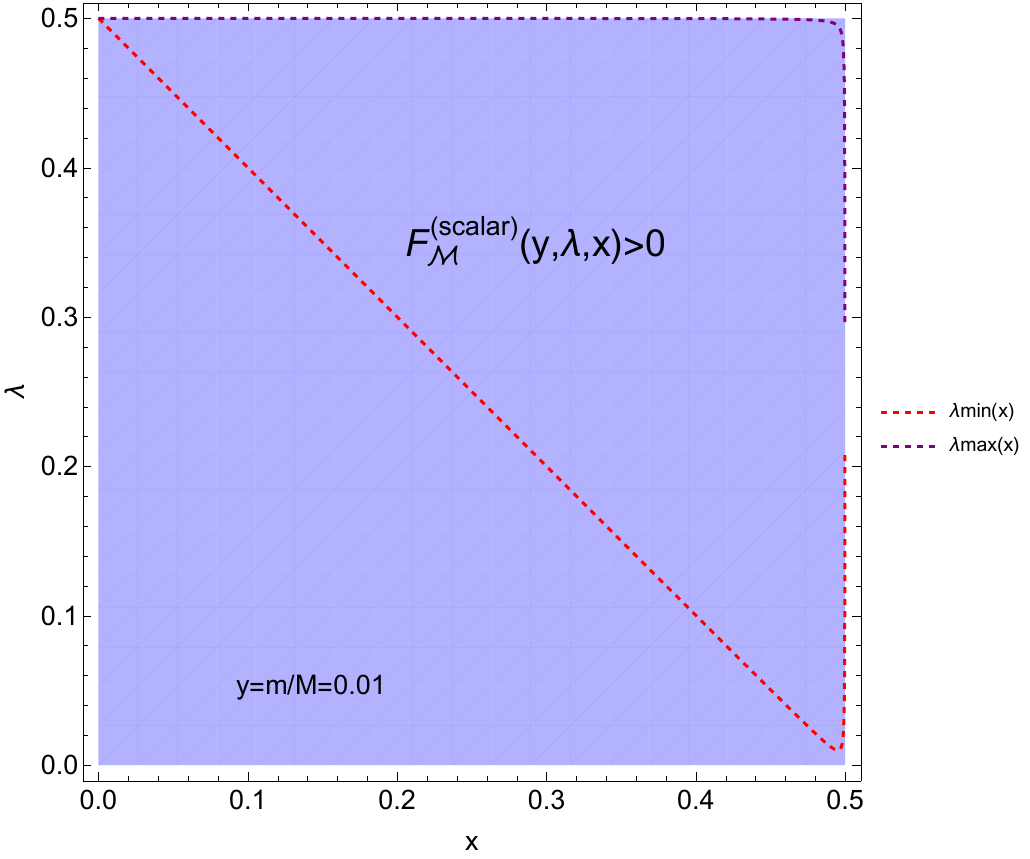}
        \includegraphics[scale=0.4]{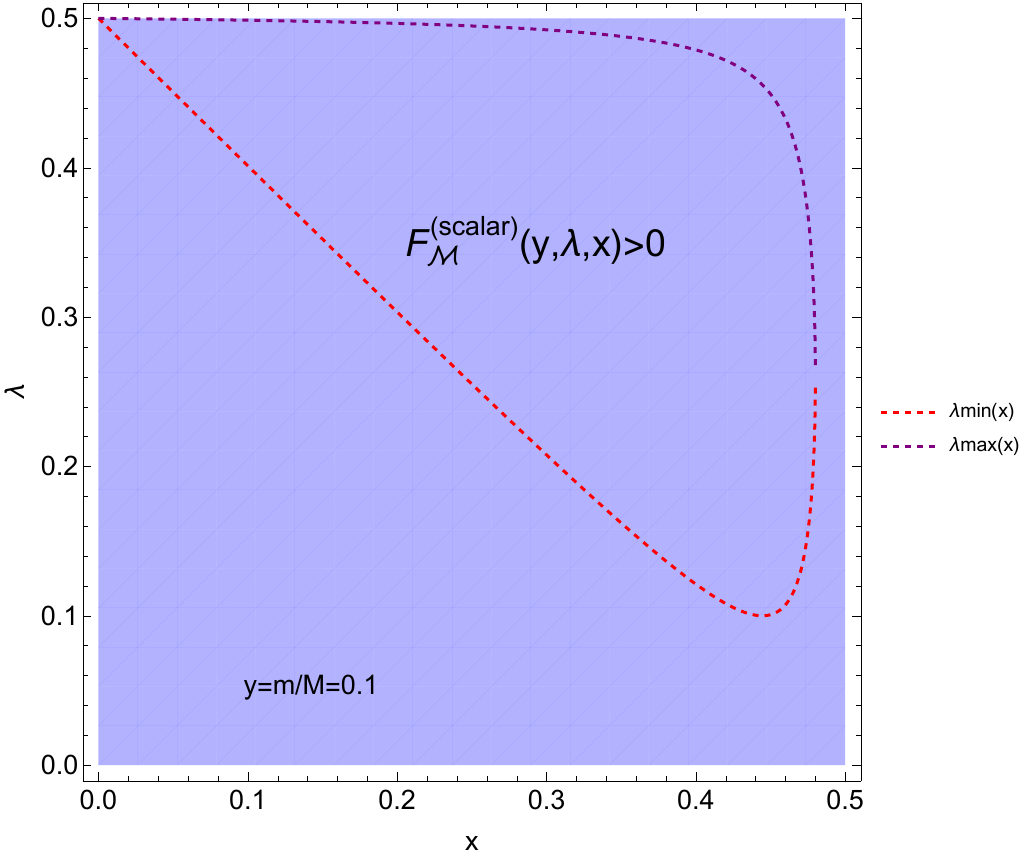}
		\caption{In the upper panel, the shaded regions indicate the parameter space in the $\lambda$–$x$ plane where $f_{\mathcal{M}}^{\text{(majorana)}}(y,\lambda,x)>0$ for fixed $y=0.01, 0.1$ situations. The red dashed and purple dashed curves denote the constraint lines $\lambda_{\min}(x)=E_{p,\min}(x)/M$ and $\lambda_{\max}(x)=E_{p,\max}(x)/M$ (given by Eq. \eqref{ThreeBodyEpmax}–\eqref{ThreeBodyEpmin}), respectively. Note that, according to the differential three-body decay formula Eq. \eqref{GenericThreeBodyDifferentialDecay}, the physically relevant parameter region is given by the intersection of the shaded area with domain bounded by $\lambda_{\min}(x)$ and $\lambda_{\max}(x)$. Similarly to the upper panel, the lower panel presents the corresponding result for the case of scalar final-state particles.}
	\label{RefThreeBodyMajoranaSquAmpli}
	\end{center}
\end{figure}

After substituting Eq. \eqref{SquAmpliThreeBodyMajorana} and \eqref{SquAmpliThreeBodyScalar} into the standard expression for the three-body differential decay width with respect to the energy of the emitted graviton, namely
\begin{align}
\label{GenericThreeBodyDifferentialDecay}
\frac{d\Gamma_{\varphi}^{(1)}}{dE_{l}}=\frac{1}{2^{6}\pi^{3}M}\int_{E_{p,min}(M,m,E_{l})}^{E_{p,max}(M,m,E_{l})}dE_{p}\big(\sum_{\text{polar}}\big\vert\mathcal{M}_{\varphi(k)\to A(p)\bar{A}(q)e_{\alpha\beta}(l)}\big\vert^{2}\big)
\end{align}
we obtain
\begin{align}\nonumber
\frac{d\Gamma^{(1)}_{\varphi,\text{scalar}}}{d E_{l}}=&\frac{\kappa^4M^4\mathsf{y}_{\text{D}}^{2}}{122880\pi^3x}\bigg[60y^2(8y^2-1)\left(3y^2(8x-3)-2x+8y^4+1\right)\log\left[\frac{1+\alpha}{1-\alpha}\right]\\ \nonumber
&+\alpha\bigg(1902y^6(1-2x)+32y^2\left(2x(168x-95)+15\right)\\
&+2y^2(2x-1)\left(8x(56x-75)+105\right)+(1-2x)^2\left(8x(4x-5)+15\right)\bigg)\bigg],
\end{align}
and
\begin{align}\nonumber
\frac{d\Gamma^{(1)}_{\varphi,\text{fermion}}}{d E_{l}}=&\frac{\kappa^4M^4\mathsf{y}_{\text{D}}^{2}}{3840\pi^3x}\bigg[60y^2\left(x(x+2)+5y^2-12xy^2-4y^4-1\right)\log\left[\frac{1+\alpha}{1-\alpha}\right]\\ \nonumber
&+\alpha\bigg(3x^2(1-2x)^2+120y^6(2x-1)-2y^4\left(2x(168x-95)+15\right)\\
&-2xy^2(x-8)(6x-5)+15y^2\bigg)\bigg]
\end{align}
in which $x$ denotes $x = E_{l}/M$, while $\alpha$ denotes $\alpha=\sqrt{1-\frac{4y^{2}}{(1-2x)}}$. Note that the derivation of Eq. \eqref{GenericThreeBodyDifferentialDecay} is reviewed in detail in Appendix. \ref{CubicInflaComplexScalar}; see also the kinematics review in \cite{ParticleDataGroup:2024cfk} for an alternative presentation. Since the expressions above are not applicable in the limit $m = 0$, we must determine suitable parameter ranges for some representative values of $y$ before incorporating them into the gravitational-wave spectrum. To proceed in a systematic way, we first introduce a dimensionless function constructed from the squared amplitudes, defined as
\begin{align}
\nonumber
&f_{\mathcal{M}}^{\text{(Majorana)}}(y,\lambda,x)\!=\!\sum_{s}\sum_{s^{\prime}}\sum_{\chi}\frac{16}{\mathsf{y}_{\text{D}}^{2}\kappa^{4}M^{4}}\vert\mathcal{M}_{\varphi(k)\to\psi^{(s),\mathrm{I}}(p)\psi_{\mathrm{J}}^{(s^{\prime})}(q)e_{\alpha\beta}^{(\chi)}(l)}^{\text{(total)}}\vert^{2}\\
\nonumber
&=\frac{\big(4(\lambda^{2}+3\lambda x+x^{2})-4(\lambda+x)-8x\lambda(\lambda+x)+4x^{2}y^{2}+1\big)^{2}}{x^{4}(1-2\lambda)^{2}(1-2\lambda-2x)^{2}}\times\big\{4x(1-2\lambda)^{2}(1-2\lambda+2y^{2})\\
&-(1-2\lambda)\big((1-2\lambda)^{3}-4x^{3}\big)-2x^{2}\big(2y^{2}(3-8\lambda)+3(1-2\lambda)^{2}+8y^{4}\big)\big\}
\end{align}where the dimensionless parameters $x$, $y$, and $\lambda$ denote the ratios of the characteristic energy scales of the final-state particles produced in the three-body decay to the inflaton mass. Explicitly, they are defined as $x = E_{l}/M$, $y = m/M$, and $\lambda = E_{p}/M$. Similarly, for the case with scalar final states, we define
\begin{align}
\nonumber
f_{\mathcal{M}}^{\text{(scalar)}}(y,\lambda,x)&=\sum_{\chi}\frac{128}{\mathsf{y}_{\text{D}}^{2}\kappa^{4}M^{4}}\vert\mathcal{M}_{\varphi(k)\to\phi(p)\phi^{\star}(q)e_{\alpha\beta}^{(\chi)}(l)}^{\text{(total)}}\vert^{2}\\
\nonumber
&=\frac{\big(x(3-8\lambda)-2(1-2\lambda)^{2}+8xy^{2}\big)^{2}}{x^{4}(1-2\lambda)^{2}(1-2\lambda-2x)^{2}}\times\big(4(\lambda^{2}+3\lambda x+x^{2})\\
&-4(\lambda+x)-8\lambda x(\lambda+x)+4x^{2}y^{2}+1\big)^{2}
\end{align}
As shown in Fig. \ref{RefThreeBodyMajoranaSquAmpli}, for several representative fixed values of $y$, we display the regions satisfying $f_{\mathcal{M}}^{\text{(Majorana)}}(y,\lambda,x)>0$ and $f_{\mathcal{M}}^{\text{(scalar)}}(y,\lambda,x)>0$, together with the constraint curves $\lambda_{\min}(x)=E_{p,\min}(x)/M$ and $\lambda_{\max}(x)=E_{p,\max}(x)/M$ (given by \eqref{ThreeBodyEpmax}–\eqref{ThreeBodyEpmin}) in the $\lambda$–$x$ plane. Note that the role of Fig. \ref{RefThreeBodyMajoranaSquAmpli} is to help determine the physically allowed range of $E_l$ for fixed values of $y = m/M$. Meanwhile, as indicated by Eq. \eqref{GravitonEnerToPresentUniverFre}, this admissible range of $E_l$ is directly related to the corresponding gravitational-wave frequency in the present Universe. Here, we clarify in advance that although the case $y=0.01$ is physically allowed from the viewpoint of the (differential) three-body decay analysis, we will not consider this parameter choice when discussing the resulting stochastic gravitational-wave signal in Sec.~\ref{GWsSpectrum}, where the reasons for this exclusion will be analyzed in detail.

\section{Spectrum of gravitational Waves from Inflaton decay \label{GWsSpectrum}}
A single graviton can be produced during the inflaton decay via bremsstrahlung emission when the inflaton decays into either fermionic or scalar final states. The corresponding Feynman diagrams for the two-body decay and the graviton-emitting three-body decay are shown in Fig. \ref{RefTwoThreeBodyMixMajorana}-\ref{ThreeBodyMajoranaCase3Case4}. The total inflaton decay width is given by the sum of two-body and three-body decay contributions,
\begin{align}
\Gamma_{\varphi}=\Gamma_{\varphi}^{(0)}+\Gamma_{\varphi}^{(1)}
\end{align}
where $\Gamma_{\varphi}^{(0)}=\Gamma^{(0)}_{\varphi\to\phi\phi(\psi\psi)}$ is the two-body decay width without graviton emission, including the inflaton decays into scalar $\phi$ and Majorana fermion $\psi$, while $\Gamma_{\varphi}^{(1)}$ denotes the three-body decay width with a graviton emitted. 

The energy densities of inflaton ($\rho_{\varphi}$), the graviton ($\rho_{\text{GW}}$), and the radiation bath  ($\rho_{\text{R}}$) evolve according to the Boltzmann equations, which incorporate the differential graviton-emission rate $d\Gamma^{(1)}_{\varphi}/dE_l$ \cite{Barman:2023ymn}
\begin{align}
&\dot{\rho}_{\varphi}+3H\rho_{\varphi}=-(\Gamma^{(0)}_{\varphi}+\Gamma^{(1)}_{\varphi})\rho_{\varphi}
\label{decayall}\\
&\dot{\rho}_{\text{GW}}+4H\rho_{\text{GW}}=\int dE_l\frac{d\Gamma^{(1)}_{\varphi}}{dE_l}\frac{E_l}{M}\rho_{\varphi}\\
&\dot{\rho}_{\text{R}}+4H\rho_{\text{R}}=\Gamma^{(0)}_{\varphi}\rho_{\varphi}+\int dE_l\frac{d\Gamma^{(1)}_{\varphi}}{E_l}\frac{M-E_l}{M}\rho_{\varphi}
\end{align}
in which, the total graviton-emission contribution $\Gamma_{\varphi}^{(1)}\rho_{\varphi}$ can be decomposed as
\begin{align}
\Gamma_{\varphi}^{(1)}\rho_{\varphi}=\int \frac{d\Gamma^{(1)}_{\varphi}}{dE_l}dE_l\rho_{\varphi}=\int\frac{d\Gamma_{\varphi}^{(1)}}{dE_l}\frac{M-E_l}{M}dE_l\rho_{\varphi}+\int\frac{d\Gamma_{\varphi}^{(1)}}{dE_l}\frac{E_l}{M}\rho_{\varphi}
\end{align}
where the factors $(M-E_l)/M$ and $E_l/M$ represent the fractions of inflaton energy transferred into SM radiation and gravitational waves, respectively. As discussed in our previous work \cite{Lee:2025lyk}, the graviton energy integral exhibits an infrared divergent in the soft limit $E_l\to0$. To regularize this divergence, we introduce an infrared (IR) cutoff and set $E_{l,\text{min}}=10^{-10}$ M. Since kinematics also requires $E_l<M/2$, the integration range is restricted to $10^{-10}M<E_l<M/2$.

With this regularization implemented, we focus on the reheating regime $a_{\text{Max}}\ll a\ll a_{\text{rh}}$ (corresponding to $T_{\text{Max}}\gg T\gg T_{\text{rh}}$), in which the Hubble dilution term $3H\rho_{\varphi}$ overwhelmingly exceeds the decay contribution $(\Gamma_{\varphi}^{(0)}+\Gamma_{\varphi}^{(1)})\rho_{\varphi}$. In this regime, the inflaton energy density and temperature follow from solving Eq. (\ref{decayall}).
\begin{align}
\frac{\rho_{\varphi}(a)}{\rho_{\varphi}(a_{\text{rh}})}\simeq\left(\frac{a_{\text{rh}}}{a}\right)^3
\end{align}
Here, $a_{\text{M}}$ and $a_{\text{rh}}$ denote the scale factors at the end of inflation and at the end of reheating, respectively. The radiation temperature scales as
\begin{align}
\frac{T(a)}{T_{\text{rh}}}=\left(\frac{a_{\text{rh}}}{a}\right)^{3/8}
\end{align}
Using this relation, the Hubble parameter during reheating can be written as
\begin{align}
&H(T)\simeq H(T_{\text{rh}})\left(\frac{a_{\text{rh}}}{a}\right)^{3/2}=H(T_{\text{rh}})\left(\frac{T}{T_{\text{rh}}}\right)^{4}
%&\rho_R(a)\simeq \rho(a_{\text{rh}})\left(\frac{a_{\text{rh}}}{a}\right)^{3/2}=\rho_R(T_{\text{rh}})\left(\frac{T}{T_{\text{rh}}}\right)^{4}
\end{align}
The evolution of the differential ratio $d(\rho_{\text{GW}}/\rho_{\text{R}})/dE_l$ follows these equations. Expressing the result in terms of the scale factor $a$, we obtain
\begin{align}
\frac{d}{da}\frac{d(\rho_{\text{GW}}/\rho_{\text{R}})}{dE_l}&=\frac{1}{aH}\frac{\rho_{\varphi}}{\rho_{\text{R}}}
\left[\frac{d\Gamma^{(1)}_{\varphi}}{dE_l}\frac{E_l}{M}-\frac{d(\rho_{GW}/\rho_{\text{R}})}{dE_l}\Gamma^{(0)}_{\varphi}\right]\\ \nonumber
&=\frac{1}{a \Gamma_{\varphi}^{(0)}}\left[\frac{d\Gamma^{(1)}_{\varphi}}{dE_l'}\frac{E_l'}{M}-\frac{d(\rho_{GW}/\rho_{\text{R}})}{dE_l}\Gamma^{(0)}_{\varphi}\right]
\end{align}
 where, at the end of reheating, the condition $\rho_{\text{R}}(T_{\text{rh}})=\rho_{\varphi}(T_{\text{rh}})$ is satisfied. We also assume that at the beginning of the reheating the Universe contains neither SM radiation nor GWs, and that $\Gamma^{(0)}_{\varphi}\simeq H(T_{\text{rh}})$ when reheating completes. The $E_{l}$ denotes the graviton energy evaluated at $a=a_{\text{rh}}$, while $E_{l}'(a)=E_l'(a)\frac{a_{\text{rh}}}{a}$ represents the redshift of graviton energy appearing only in the source term. For analytical convenience, we treat the quantity $d(\rho_{\text{GW}}/\rho_{\text{R}})/dE_l$ as an independent variable, which casts the system into a first-order ordinary differential equation. Under these assumptions, the differential gravitational wave spectrum is approximated by
\begin{align}
\frac{d\left(\rho_{\text{GW}}(T_{\text{rh}})/\rho_{\text{R}}(T_{\text{rh}})\right)}{dE_l}\simeq\frac{d\Gamma_{\varphi}^{(1)}}{dE_l}\frac{E_l}{M}\frac{1}{\Gamma_{\varphi}^{(0)}}\left[1-\left(\frac{T_{\text{rh}}}{T_{\text{Max}}}\right)^{8/3}\right]
\end{align}
 where, $T_{\text{Max}}$ denotes the maximum temperature reached during reheating. For an instantaneous decay, we take $T_{\text{Max}}\simeq T_{\text{rh}}$.
 
\begin{figure}[ht]
	\begin{center}
		\includegraphics[scale=1.0]{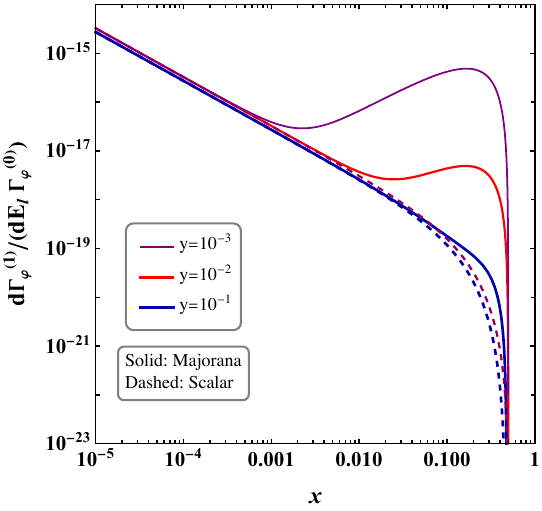}
		\caption{This figure shows how the ratio of the differential three-body decay rate and two decay rate: $d\Gamma_{\varphi}^{(1)}/(dE_l\Gamma_{\varphi}^{(0)})$, varies with $x=E_l/M$ for different values of the mass ratio $y=m/M$. The solid curves correspond to Majorana cases, while the dashed curves denote scalar cases. }
		\label{fig:yy}
	\end{center}
\end{figure}

The gravitational wave spectrum is given by
\begin{align}
\Omega_{\text{GW}}h^2&=\frac{1}{\rho_{c}h^{-2}}\frac{d\rho_{\text{GW}}}{d\ln f}=\Omega_{\gamma}^{(0)}h^2\frac{g_{\star}(T_{\text{rh}})}{g_{\star}(T_{0})}\bigg(\frac{g_{\star s}(T_0)}{g_{\star s}(T_{\text{rh}})}\bigg)^{4/3}\frac{d\left(\rho_{\text{GW}}(T_{\text{rh}})/\rho_{\text{R}}(T_{\text{rh}})\right)}{d\ln E_l}\\
&=\Omega_{\gamma}^{(0)}h^2\frac{g_{\star}(T_{\text{rh}})}{g_{\star}(T_{0})}\bigg(\frac{g_{\star s}(T_0)}{g_{\star s}(T_{\text{rh}})}\bigg)^{4/3}\frac{d\Gamma_{\varphi}^{(1)}}{dE_l}\frac{E_l}{M}\frac{1}{\Gamma_{\varphi}^{(0)}}
\end{align}
in which, 
$\rho_c$ denotes the present critical density and $\Omega_{\gamma}^{(0)}h^2\simeq 2.47\times10^{-5}$ is the observed photon abundance \cite{Planck:2018vyg}. The CMB temperature is $T_0=2.73 K=8.6\times10^{-5}$ eV, and $g_{\star}(T)$ and $g_{\star s}(T)$ represent the effective numbers of relativistic degrees of freedom in energy and entropy, respectively \cite{Drees:2015exa}.
Using the above relation, the GWs spectrum produced by the inflaton differential decay rate becomes 
\begin{align}
\Omega_{\text{GW}}h^2
%\simeq8.66\times10^{-6}\times\frac{\kappa^2ME_l}{480\pi^2}(1-2\frac{E_l}{M})^2\left(8\frac{E_l}{M}(4\frac{E_l}{M}-5)+15\right)\\
%&=8.66\times10^{-6}\times\frac{16\pi M }{480\pi^2M_p^2}\times2.165\times10^{-9}\frac{f}{\text{Hz}}T_{\text{rh}}\\
%&=8.66\times10^{-8}\frac{M }{M_p}\times2.165\times10^{-9}\frac{f}{\text{Hz}}\frac{T_{\text{rh}}}{Mp}\\
&\simeq\mathcal{C}_{\text{GW}}\times\frac{T_{\text{rh}}}{5.5\times10^{15}}\frac{M}{M_p}\frac{f}{10^{12}\text{Hz}}
\end{align}
where, we have used the relation
\begin{align}
\label{GravitonEnerToPresentUniverFre}
&E_l=2\pi f\left(\frac{a_0}{a_{\text{rh}}}\right)=2\pi f\left(\frac{T_{\text{rh}}}{T_0}\right)\times\left(\frac{g_{\star s}(T_{\text{rh}})}{g_{\star s}(T_0)}\right)^{1/3}
\end{align}
to convert the GWs energy $E_l$ at reheating to the present frequency $f$. The numerical coefficient is $\mathcal{C}_{\text{GW}}\simeq 43.3\times 10^{-8}$ for scalar channel, while typical channels studied in \cite{Barman:2023ymn}, which include both scalar and fermionic finial states, yield $\mathcal{C}_{\text{GW}}\simeq1.4\times 10^{-8}$ and $2.8\times 10^{-8}$, respectively. In contrast, the Majorana case does not yield a fixed value of $\mathcal{C}_{\text{GW}}$, because--as shown in Fig. \ref{fig:yy}, the decay rate exhibits a non-trivial dependence on the mass rate $y=m/M$, unlike the scalar cases, which are y-independent. For illustration, choosing $y=0.1$ suppresses the y-dependent enhancement and leads to a value close to the scalar result, $\mathcal{C}_{\text{GW}}\simeq43.3\times 10^{-8}$. Moreover, the result from BICEP/Keck experiment \cite{BICEP:2021xfz} impose an updated upper bound on the reheating temperature, $T_{\text{rh}}\lesssim5.5\times10^{15}$ GeV.

\begin{figure}[ht]
\centering
\begin{minipage}[t]{0.49\textwidth}
  \centering
  \includegraphics[width=\textwidth]{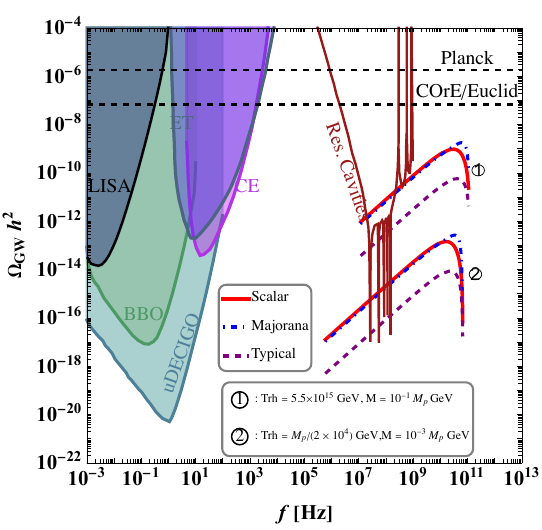}
  \label{fig:scalar-omega}
\end{minipage}%
\hfill
\begin{minipage}[t]{0.49\textwidth}
  \centering
  \includegraphics[width=\textwidth]{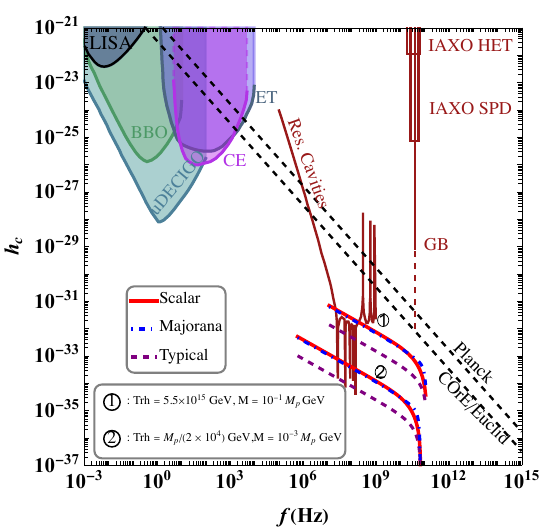}
  \label{fig:hc}
\end{minipage}
\caption{GWs spectrum $\Omega_{\text{GW}}h^2$ (left) and characteristic strain $h_c$ (right) as function of the GWs frequency $f$. The curves correspond to inflaton decay into scalar pairs (red),  Majorana fermions (blue dot-dashed), and typical scalar fields (purple dashed). The shaded regions indicate the expected sensitivity reaches of various GWs detector. Black dashed lines represent current and forthcoming CMB bounds derived from Planck and COrE/Euclid. Two benchmarks choices are illustrated: [(1) $M=M_P/10$ with $T_{\text{rh}}=5.5\times10^{15}$ GeV, and (2) $M=M_P/10^3$ with $T_{\text{rh}}=M_P/(2\times10^4)$ GeV.] }
\label{figure:Gh}
\end{figure}

Since the $10^{-10}M<E_l<0.5M$, and we used $g_{\star s(T_0)}=3.94$ and $g_{\star s(T_{\text{rh}})}=106.75$, the observed frequencies $f$ satisfy
\begin{align}
f\lesssim4.1\times10^{12}\left(\frac{M}{M_P}\right)\left(\frac{5.5\times10^{15}\text{GeV}}{T_{\text{rh}}}\right)\text{Hz}
\end{align}
Thus, both $M$ and $T_{\text{rh}}$ directly control the accessible frequency range. To characterize the GWs background in the terms of its strain amplitude $h_c(f)$ \cite{Maggiore:1999vm}, we use
\begin{align}
h_c(f)=\frac{H_0}{f}\sqrt{\frac{3\Omega_{\text{GW}}(f)}{2\pi^2}}
\end{align}
Here, $H_0\equiv1.44\times10^{-42}$ GeV is the present Hubble parameter. The Fig. \ref{figure:Gh} displays the resulting GWs spectrum (left) and dimensionless GWs strain (right) for inflaton inflaton decays into scalar (red) and Majorana fermions (blue dot-dashed). For comparison, a typical scalar benchmark is also shown (purple dashed). We set two representative parameter:  \textcircled{1} $M=M_P/10$, $T_{\text{rh}}=5.5\times 10^{15}$ GeV, and \textcircled{2} $M=M_P/10^{3}$, $T_{\text{rh}}=M_{P}/{2\times 10^4}$ GeV. We also overlay the projected sensitivities of LISA \cite{LISA:2017pwj}, BBO \cite{Crowder:2005nr,Harry:2006fi,Corbin:2005ny}, DECIGO(uDECIGO) \cite{Seto:2001qf}, CE \cite{Reitze:2019iox} and ET \cite{Punturo:2010zz,Hild:2010id,Sathyaprakash:2012jk,ET:2019dnz}, as well as high-frequency  resonant cavity proposals  \cite{Herman:2020wao,Herman:2022fau} and the IAXO sensitivity band \cite{Armengaud:2014gea,IAXO:2019mpb}. The BBN bound $\Omega_{\text{GW}}<1.3\times10^{-6}$ \cite{Yeh:2022heq} is also shown. 

Interestingly, the peak amplitude of our new results reaches $\Omega_{\text{GW}}h^2\sim10^9$  for benchmark \textcircled{1} and $\sim10^{12}$ for benchmark \textcircled{2}. The resulting signal exceeds typical inflaton-decay predictions by over an order of magnitude, placing it squarely within the discovery reach of next-generation high-frequency GWs detectors. This enhancement substantially enlarges the experimentally accessible parameter space, opening regions that would otherwise remain untestable.

%\begin{align}
%&\frac{d}{dt}(a^4\rho_{\text{GW}})=a^4(\dot{\rho}_{\text{Gw}}+4H\rho_{\text{GW}})\\
%&dt=\frac{da}{aH}
%\end{align}
%\begin{align}
%&\dot{\rho}_{\text{GW}}+4H\rho_{\text{GW}}=\Gamma^{(1)}\rho_{\varphi}\to\\
%&\frac{d}{dt}(a^4\rho_{\text{GW}})=a^4\Gamma^{(1)}\rho_{\varphi}\to \\
%&\frac{d}{da}(a^4\rho_{\text{GW}})=\frac{a^3}{H}\Gamma^{(1)}\rho_{\varphi}
%\end{align}
%for $d(\rho_{GW}/\rho_{\text{R}})/da$
%\begin{align}
%&\frac{d}{dt}(\frac{\rho_{\text{GW}}}{\rho_{\text{R}}})=\frac{1}{\rho_{\text{R}}}\frac{d\rho_{\text{GW}}}{dt}-\frac{\rho_{\text{GW}}}{\rho_{\text{R}}^2}\frac{d\rho_{\text{R}}}{dt}\\
%&=\frac{1}{\rho_{\text{R}}}[-4H\rho_{\text{GW}}+\int dE_l\frac{d\Gamma_{\varphi}^{(1)}}{dE_l}\frac{E_l}{M}\rho_{\varphi}]\\
%&-\frac{\rho_{\text{GW}}}{\rho_{\text{R}}^2}[-4H\rho_{\text{R}}+\Gamma^{(0)}\rho_{\varphi}+\int dE_l\frac{d\Gamma^{(1)}}{dE_l}\frac{M-E_l}{M}\rho_{\varphi}]\\
%&=\frac{1}{\rho_{\text{R}}}\int dE_l\frac{d\Gamma_{\varphi}^{(1)}}{dE_l}\frac{E_l}{M}\rho_{\varphi}\\
%&-\frac{\rho_{\text{GW}}}{\rho_{\text{R}}^2}[\Gamma^{(0)}\rho_{\varphi}+\int dE_l\frac{d\Gamma^{(1)}}{dE_l}\frac{M-E_l}{M}\rho_{\varphi}]\\
%&\simeq\frac{\rho_{\varphi}}{\rho_{\text{R}}}[\int dE_l\frac{d\Gamma^{(1)}}{dE_l}\frac{E_l}{M}-\frac{\rho_{\text{GW}}}{\rho_{\text{R}}}\Gamma^{(0)}]\to\\
%&\frac{d(\rho_{\text{GW}}/\rho_{\text{R}})}{da}=\frac{1}{aH}\frac{\rho_{\varphi}}{\rho_{\text{R}}}[\int dE_l\frac{d\Gamma^{(1)}}{dE_l}\frac{E_l}{M}-\frac{\rho_{\text{GW}}}{\rho_{\text{R}}}\Gamma^{(0)}]
%\end{align}

\section{Conclusion and Discussion \label{ConcluAndDiscuss}}

In this paper, we investigated the stochastic gravitational-wave background sourced by inflaton three-body decay mediated by a cubic interaction arising from a generalized Yukawa coupling between the inflaton and the D-term sectors of a pair of chiral and anti-chiral superfields. Compared with the conventional Yukawa coupling between the inflaton and structureless matter fields, the supersymmetry-preserving structure of the field content leads to a significant enhancement: we find an improvement of roughly one order of magnitude in the amplitudes of both the GWs spectrum and the GWs strain. 

In fact, our analysis focused on a toy model based on the simplest Wess–Zumino chiral superfield. Even within this minimal setup, we already observe clearly detectable enhancements in the resulting stochastic gravitational-wave signals. In addition, from a technical perspective, our analysis employed the component-expansion approach to treat the chiral superfield, together with extensive use of two-component spinor techniques to compute the squared amplitudes. Even for the single chiral-superfield setup considered here, the calculation already generates a substantial number of Feynman diagrams. If one were to incorporate more realistic supersymmetric models in the future, such as those involving a nontrivial Kähler potential, multiple chiral superfields, vector multiplets, and gauge symmetries, the component-expansion method would likely render the computation exceedingly cumbersome. Therefore, future work should consider performing these calculations using the supergraph formalism directly in superspace \cite{Grisaru:1979wc,Capper:1974ff}. 

Besides, in comparison with the standard Yukawa interaction $\mathrm{y}_{\phi}\,\varphi \phi_{(i)}\phi_{(i)}$ (where $(i)$ denotes different species of matter fields and $y_{\phi}$ denotes the corresponding coupling constant), the coupling involving chiral superfields,$\kappa \mathrm{y}_{_{\text{D}}}\varphi[\bar{\Phi}\Phi]_{\theta^{2}\theta^{\dagger2}}$, generates derivative terms after performing the component expansion, as shown in Eq. \eqref{YukawaInflatonMajoranaMain}–\eqref{YukawaInflatonComScalarMain}. These derivative contributions are likely to be largely responsible for the enhancement of the gravitational-wave amplitude observed in our analysis. This motivates further exploration of extended Yukawa couplings between the inflaton and sectors involving higher-derivative superfield structures, for example, interactions of the form $\mathrm{y}_{_{\text{D}}}\kappa^{3}\varphi\,[\mathcal{D}^{2}\Phi\bar{\mathcal{D}}^{2}\bar{\Phi}]_{\theta^{2}\theta^{\dagger2}}$.Alternatively, one may also consider a vector-type inflaton coupled to the D-term sector of a higher-derivative superfield, for instance, $\mathrm{y}_{_{\text{D}}}\kappa^{2} A_{\mu}\big[\bar{\mathcal{D}}_{\dot{\mathrm{I}}}\bar{\Phi}(\bar{\sigma}^{\mu})^{\dot{\mathrm{I}}\mathrm{J}}\mathcal{D}_{\mathrm{J}}\Phi\big]_{\theta^{2}\theta^{\dagger2}}$. This class of interactions could likewise induce derivative-enhanced contributions and may lead to further amplification of the resulting stochastic gravitational-wave signal. In summary, our work demonstrates that studying the gravitational-wave signal generated by inflaton decay during reheating can provide an indirect probe of supersymmetric physics. The supersymmetric framework indeed enhances the theoretical detectability of such signals. Therefore, the interplay between reheating-era stochastic gravitational waves and supersymmetric phenomenology merits further exploration and development.

\section{ACKNOWLEDGMENTS}
This research was supported in part by funding from the China Scholarship Council (CSC). K. W. also acknowledges support from the research Grants No. PID2022-126224NB-C21 and No. 2021-SGR-249 provided by the Generalitat de Catalunya. The authors would like to thank Federico Mescia for valuable discussions and suggestions.

\appendix

\section{A brief review about the SUSY-invariant Wess-Zumino model \label{ReviewWZChiral}}

In superspace \cite{Martin:1997ns,Gates:1983nr}, the chiral superfield is given by the expression 
\begin{align}
\nonumber
\Phi(x,\theta,\theta^{\dagger})&=\phi(x)+\sqrt{2}\theta\psi(x)+\theta\theta F(x)+\text{i}\theta^{\dagger}\bar{\sigma}^{\mu}\theta\partial_{\mu}\phi(x)\\
\label{ChiralSuperField}
&-\frac{\text{i}}{\sqrt{2}}\theta\theta\theta^{\dagger}\bar{\sigma}^{\mu}\partial_{\mu}\psi(x)-\frac{1}{4}\theta\theta\theta^{\dagger}\theta^{\dagger}\partial_{\mu}\partial^{\mu}\phi(x)		
\end{align}When chiral covariant derivatives are introduced
\begin{align}
&\mathcal{D}_{\mathrm{I}}=\frac{\partial}{\partial\theta^{\mathrm{I}}}-\text{i}(\sigma^{\mu}\theta^{\dagger})_{\mathrm{I}}\partial_{\mu}~,~\mathcal{D}^{\mathrm{I}}=-\frac{\partial}{\partial\theta_{\mathrm{I}}}+\text{i}(\theta^{\dagger}\bar{\sigma}^{\mu})^{\mathrm{I}}\partial_{\mu}\\
&\bar{\mathcal{D}}^{\dot{\mathrm{I}}}=\frac{\partial}{\partial\theta_{\dot{\mathrm{I}}}^{\dagger}}-\text{i}(\bar{\sigma}^{\mu}\theta)^{\dot{\mathrm{I}}}\partial_{\mu}~,~\bar{\mathcal{D}}_{\dot{\mathrm{I}}}=-\frac{\partial}{\partial\theta^{\dagger\dot{\mathrm{I}}}}+\text{i}(\theta\sigma^{\mu})_{\dot{\mathrm{I}}}\partial_{\mu}
\end{align}
 the chiral superfield must satisfy the condition
 \begin{align}
 &\bar{\mathcal{D}}_{\dot{\mathrm{I}}}\Phi=\bar{\mathcal{D}}^{\dot{\mathrm{I}}}\Phi=0~ ,~ \mathcal{D}_{\mathrm{I}}\bar{\Phi}=\mathcal{D}^{\mathrm{I}}\bar{\Phi}=0
 \end{align}in which $\bar{\Phi}$ denotes the Hermitian conjugate of the chiral superfield \eqref{ChiralSuperField}. As an example, let us expand $\bar{\mathcal{D}}_{\dot{\mathrm{I}}}\Phi$ explicitly. Before expanding it, we first carry out the following preparatory steps
 \begin{align}
 \label{ChiralDerivaOnCSFTerm2}
 &\sqrt{2}\bar{\mathcal{D}}_{\dot{\mathrm{I}}}(\theta\psi)\!=\!\sqrt{2}\text{i}\theta^{\mathrm{I}_{2}}(\sigma^{\mu})_{\mathrm{I}_{2}\dot{\mathrm{I}}}\theta^{\mathrm{I}_{1}}\partial_{\mu}\psi_{\mathrm{I}_{1}}\!=\!\sqrt{2}\text{i}(-\frac{1}{2}\varepsilon^{\mathrm{I}_{2}\mathrm{I}_{1}}\theta\theta)(\sigma^{\mu})_{\mathrm{I}_{2}\dot{\mathrm{I}}}\partial_{\mu}\psi_{\mathrm{I}_{1}}=-\frac{\text{i}}{\sqrt{2}}\theta\theta(\partial_{\mu}\psi\sigma^{\mu})_{\dot{\mathrm{I}}}
 \end{align}
 and
 \begin{align}
 \nonumber
 &\quad\quad\quad\quad \text{i}\bar{\mathcal{D}}_{\dot{\mathrm{I}}}(\theta^{\dagger}\bar{\sigma}^{\mu}\theta\partial_{\mu}\phi)=-\text{i}(\theta\sigma^{\mu})_{\dot{\mathrm{I}}}\partial_{\mu}\phi+\underbrace{\theta^{\mathrm{I}_{1}}\theta_{\mathrm{J}_{2}}\theta_{\dot{\mathrm{J}}_{1}}^{\dagger}(\sigma^{\nu})_{\mathrm{I}_{1}\dot{\mathrm{I}}}(\bar{\sigma}^{\mu})^{\dot{\mathrm{J}}_{1}\mathrm{J}_{2}}\partial_{\mu}\partial_{\nu}\phi}_{\theta^{\mathrm{I}_{1}}\theta_{\mathrm{J}_{2}}=\frac{1}{2}\delta_{\mathrm{J}_{2}}^{\mathrm{I}_{1}}\theta\cdot\theta}\\
  \label{ChiralDerivaOnCSFTerm4}
 &=-\text{i}(\theta\sigma^{\mu})_{\dot{\mathrm{I}}}\partial_{\mu}\phi+\frac{1}{2}\theta\cdot\theta\times(\theta^{\dagger}\bar{\sigma}^{(\mu}\sigma^{\nu)})_{\dot{\mathrm{I}}}\partial_{\mu}\partial_{\nu}\phi=-\text{i}(\theta\sigma^{\mu})_{\dot{\mathrm{I}}}\partial_{\mu}\phi+\frac{1}{2}\theta\cdot\theta\times\theta_{\dot{\mathrm{I}}}^{\dagger}\eta^{\mu\nu}\partial_{\mu}\partial_{\nu}\phi
 \end{align}With the above preparations, we now act the chiral covariant derivative $\bar{\mathcal{D}}_{\dot{\mathrm{I}}}$ on the chiral superfield \eqref{ChiralSuperField}, which yields
\begin{small}
\begin{align}
 \nonumber
 &\bar{\mathcal{D}}_{\dot{\mathrm{I}}}\Phi =\text{i}(\theta\sigma^{\mu})_{\dot{\mathrm{I}}}\partial_{\mu}\phi-\frac{\text{i}}{\sqrt{2}}\theta\theta(\partial_{\mu}\psi\sigma^{\mu})_{\dot{\mathrm{I}}}+0+\big(\frac{1}{2}\theta\theta\times\theta_{\dot{\mathrm{I}}}^{\dagger}\eta^{\mu\nu}\partial_{\mu}\partial_{\nu}\phi-\text{i}(\theta\sigma^{\mu})_{\dot{\mathrm{I}}}\partial_{\mu}\phi\big)\\
 \nonumber
 &\quad\quad\quad-\frac{\text{i}}{\sqrt{2}}\frac{\partial}{\partial\theta^{\dagger\dot{\mathrm{I}}}}(\theta\theta\partial_{\mu}\psi\sigma^{\mu}\theta^{\dagger})+\frac{1}{4}\frac{\partial}{\partial\theta^{\dagger\dot{\mathrm{I}}}}(\theta\theta\theta^{\dagger}\theta^{\dagger}\partial_{\mu}\partial^{\mu}\phi)\\
\nonumber
 &=\!\text{i}(\theta\sigma^{\mu})_{\dot{\mathrm{I}}}\partial_{\mu}\phi\!-\frac{\text{i}}{\sqrt{2}}\theta\theta(\partial_{\mu}\psi\sigma^{\mu})_{\dot{\mathrm{I}}}\!+\!0\!+\frac{1}{2}\theta\theta\!\times\!\theta_{\dot{\mathrm{I}}}^{\dagger}\!\eta^{\mu\nu}\partial_{\mu}\partial_{\nu}\phi\!-\!\text{i}(\theta\sigma^{\mu})_{\dot{\mathrm{I}}}\!\partial_{\mu}\phi\\
 \label{ChiralDeriveToChiralSuperField}
  &\quad\quad\quad+\frac{\text{i}}{\sqrt{2}}(\theta\theta\partial_{\mu}\psi\sigma^{\mu})_{\dot{\mathrm{I}}}\!-\frac{1}{2}\theta_{\dot{\mathrm{I}}}^{\dagger}\theta\theta\partial_{\mu}\partial^{\mu}\phi\!=\!0
 \end{align}
 \end{small}Note that in the first line of \eqref{ChiralDeriveToChiralSuperField}, we have made use of the relations given in \eqref{ChiralDerivaOnCSFTerm2} and \eqref{ChiralDerivaOnCSFTerm4}. Under supersymmetric transformations, each component field in the superfield follows the transformation rules 
\begin{align}
\label{SUSYTransComponentField}
&\delta_{\epsilon}\phi=\epsilon^{\mathrm{I}_{1}}\psi_{\mathrm{I}_{1}}\,,\,\delta_{\epsilon}\psi_{\mathrm{I}}=F\epsilon_{\mathrm{I}}-\text{i}\partial_{\mu}\phi(\sigma^{\mu}\epsilon^{\dagger})_{\mathrm{I}}\,,\,\delta_{\epsilon}F=-\text{i}(\epsilon^{\dagger}\bar{\sigma}^{\mu}\partial_{\mu}\psi)\\
\label{SUSYTransComponentComplexField}
&\delta_{\epsilon}\phi^{\star}=\epsilon_{\dot{\mathrm{I}}_{1}}^{\dagger}\psi^{\dagger\dot{\mathrm{I}}_{1}}\,,\,\delta_{\epsilon}\psi_{\dot{\mathrm{I}}}^{\dagger}=F^{\star}\epsilon_{\dot{\mathrm{I}}}^{\dagger}+\text{i}\partial_{\mu}\phi^{\star}(\epsilon\sigma^{\mu})_{\dot{\mathrm{I}}}\,,\,\delta_{\epsilon}F^{\star}=\text{i}(\partial_{\mu}\psi^{\dagger}\bar{\sigma}^{\mu}\epsilon)
\end{align}Subsequently, let us examine the closure of the SUSY algebra \eqref{SUSYTransComponentField}, which implies that the commutator of two supersymmetry transformations, parameterized by two spinors $\epsilon_{1}$ and $\epsilon_{2}$, yields another symmetry transformation of the theory. As for real scalar, we have
\begin{align}
\label{CommuCloseureScalar}
&[\delta_{\epsilon_{1}},\delta_{\epsilon_{2}}]\phi=\epsilon_{2}^{\mathrm{I}_{2}}\delta_{\epsilon_{1}}\psi_{\mathrm{I}_{2}}-\epsilon_{1}^{\mathrm{I}_{1}}\delta_{\epsilon_{2}}\psi_{\mathrm{I}_{1}}=(\epsilon_{1}\sigma^{\mu}\epsilon_{2}^{\dagger}+\epsilon_{1}^{\dagger}\bar{\sigma}^{\mu}\epsilon_{2})\text{i}\partial_{\mu}\phi 
\end{align}Regarding to Majorana fermion, it yields
\begin{align}
\nonumber
&\quad\quad\quad~~ [\delta_{\epsilon_{1}},\delta_{\epsilon_{2}}]\psi_{\mathrm{I}}=\big((-\text{i}\epsilon_{1}^{\dagger}\bar{\sigma}^{\mu}\partial_{\mu}\psi)\epsilon_{2,\mathrm{I}}-\text{i}(\epsilon_{1}\partial_{\mu}\psi)(\sigma^{\mu}\epsilon_{2}^{\dagger})_{\mathrm{I}}\big)-\epsilon_{1}\leftrightarrow\epsilon_{2}\text{ terms}\\
\label{CommuCloseureMajorana}
&=\!\text{i}\epsilon_{1}^{\dagger}\bar{\sigma}^{\mu}\epsilon_{2}\partial_{\mu}\psi_{\mathrm{I}}\!-\!\text{i}\big((\sigma^{\mu}\epsilon_{1}^{\dagger})_{\mathrm{I}}\epsilon_{2}\!+\!(\sigma^{\mu}\epsilon_{2}^{\dagger})_{\mathrm{I}}\epsilon_{1}\big)\partial_{\mu}\psi\!-\!\epsilon_{1}\!\leftrightarrow\!\epsilon_{2}\text{ terms}\!=\!\text{i}(\epsilon_{1}^{\dagger}\bar{\sigma}^{\mu}\epsilon_{2}\!+\!\epsilon_{1}\sigma^{\mu}\epsilon_{2}^{\dagger})\partial_{\mu}\psi_{\mathrm{I}}
\end{align}With respect to auxiliary field, we could get
\begin{align}
\label{CommuCloseureAuxiliary}
&[\delta_{\epsilon_{1}},\delta_{\epsilon_{2}}]F\!=\!-\text{i}\epsilon_{2}^{\dagger}\bar{\sigma}^{\mu}\epsilon_{1}\partial_{\mu}F\!-\!\epsilon_{2}^{\dagger}\bar{\sigma}^{(\mu}\sigma^{\nu)}\epsilon_{1}^{\dagger}\partial_{\mu}\partial_{\nu}\phi\!-\!\epsilon_{1}\!\leftrightarrow\!\epsilon_{2}\text{ terms}\!=\!\text{i}(\epsilon_{1}\sigma^{\mu}\epsilon_{2}^{\dagger}+\epsilon_{1}^{\dagger}\bar{\sigma}^{\mu}\epsilon_{2})\partial_{\mu}F
\end{align}It is worth noting that in \eqref{CommuCloseureMajorana}, we have employed the Fierz identities, specifically
\begin{align}
\nonumber
&(-\text{i}\epsilon_{1}^{\dagger}\bar{\sigma}^{\mu}\partial_{\mu}\psi)\epsilon_{2,\mathrm{I}}\!=\!\text{i}(\epsilon_{1}^{\dagger}\bar{\sigma}^{\mu})^{\mathrm{I}_{1}}\epsilon_{2}^{\mathrm{I}_{2}}\partial_{\mu}\psi^{\mathrm{J}_{1}}(\varepsilon_{\mathrm{I}_{1}\mathrm{J}_{1}}\varepsilon_{\mathrm{I}\mathrm{I}_{2}})\!=\!\text{i}(\epsilon_{1}^{\dagger}\bar{\sigma}^{\mu})^{\mathrm{I}_{1}}\epsilon_{2}^{\mathrm{I}_{2}}\partial_{\mu}\psi^{\mathrm{J}_{1}}\underbrace{(-\varepsilon_{\mathrm{I}_{1}\mathrm{I}}\varepsilon_{\mathrm{I}_{2}\mathrm{J}_{1}}\!-\!\varepsilon_{\mathrm{I}_{1}\mathrm{I}_{2}}\varepsilon_{\mathrm{J}_{1}\mathrm{I}})}\\
&\quad\quad\quad\quad\quad\quad\quad\quad \xrightarrow{\varepsilon^{\dot{\mathrm{I}}\dot{\mathrm{I}}_{1}}(\sigma^{\mu})_{\mathrm{I}\dot{\mathrm{I}}_{1}}=\varepsilon_{\mathrm{I}\mathrm{I}_{1}}(\bar{\sigma}^{\mu})^{\dot{\mathrm{I}}\mathrm{I}_{1}}}=\text{i}\epsilon_{1}^{\dagger}\bar{\sigma}^{\mu}\epsilon_{2}\partial_{\mu}\psi_{\mathrm{I}}-\text{i}(\sigma^{\mu}\epsilon_{1}^{\dagger})_{\mathrm{I}}\,\epsilon_{2}\partial_{\mu}\psi
\end{align}From \eqref{CommuCloseureScalar}-\eqref{CommuCloseureAuxiliary},  it is straightforward to summarize the following commutation relations
\begin{align}
\label{CommuSUSYvariation}
&[\delta_{\epsilon_{1}},\delta_{\epsilon_{2}}]=-(\epsilon_{1}\sigma^{\mu}\epsilon_{2}^{\dagger}+\epsilon_{1}^{\dagger}\bar{\sigma}^{\mu}\epsilon_{2})\hat{P}_{\mu}=-(\epsilon_{1}\sigma^{\mu}\epsilon_{2}^{\dagger}-\epsilon_{2}\sigma^{\mu}\epsilon_{1}^{\dagger})\hat{P}_{\mu}
\end{align}By decomposing the SUSY variation $\delta_{\epsilon}$ into the supercharge operators, namely $\delta_{\epsilon} =\epsilon^{\mathrm{J}}\hat{Q}_{\mathrm{J}}+\hat{Q}_{\dot{\mathrm{J}}}^{^{\dagger}}\epsilon^{\dagger\dot{\mathrm{J}}}$, one can recover the anticommutation relations among the supercharge operators from the commutation relation \eqref{CommuSUSYvariation}
\begin{align}
&\{\hat{Q}_{\mathrm{J}},\hat{Q}_{\dot{\mathrm{I}}}^{\dagger}\}=-(\sigma^{\mu})_{\mathrm{J}\dot{\mathrm{I}}}\hat{P}_{\mu}\,,\,\{\hat{Q}_{\mathrm{J}},\hat{Q}_{\mathrm{I}}\}=0\,,\,\{\hat{Q}_{\dot{\mathrm{J}}}^{\dagger},\hat{Q}_{\dot{\mathrm{I}}}^{\dagger}\}=0
\end{align}

\subsubsection*{SUSY-invariant kinetic term from the D-term construction}
Next, we expand the chiral superfield in superspace coordinates and construct the component form of the supersymmetric invariant action in real space. For the kinetic term, let us consider the D-term of the product $\bar{\Phi}(x,\theta,\theta^{\dagger})\Phi(x,\theta,\theta^{\dagger})$, namely
\begin{align}
\nonumber
&[\bar{\Phi}(x,\theta,\theta^{\dagger})\Phi(x,\theta,\theta^{\dagger})]_{D}=\theta^{\dagger}\theta^{\dagger}\theta\theta\big(\eta^{\mu\nu}\partial_{\mu}\phi^{\star}(x)\partial_{\nu}\phi(x)+\vert F\vert^{2}+\text{i}\psi^{\dagger}(x)\bar{\sigma}^{\mu}\partial_{\mu}\psi(x)\big)\\
\label{DtermPhibarPhi}
&\quad\quad\quad\quad\quad-\frac{1}{4}\theta^{\dagger}\theta^{\dagger}\theta\theta\underbrace{\partial_{\mu}\big(2\text{i}\psi^{\dagger}(x)\bar{\sigma}^{\mu}\psi(x)+\phi^{\star}(x)\partial^{\mu}\phi(x)+\phi(x)\partial^{\mu}\phi^{\star}(x)\big)}_{\text{total derivative}}
\end{align}in which we have primarily employed the following identities
\begin{align}
\nonumber
&\hspace{-11mm}\theta^{\dagger}\bar{\sigma}^{\mu}\theta\partial_{\mu}\phi^{\star}(x)\times\theta^{\dagger}\bar{\sigma}^{\nu}\theta\partial_{\nu}\phi(x)=-\theta_{\dot{\mathrm{I}}}^{\dagger}(\bar{\sigma}^{\mu})^{\dot{\mathrm{I}}\mathrm{I}}\theta_{\mathrm{I}}\partial_{\mu}\phi^{\star}(x)\times\theta^{\mathrm{J}}(\sigma^{\nu})_{\mathrm{J}\dot{\mathrm{J}}}\theta^{\dagger\dot{\mathrm{J}}}\partial_{\nu}\phi(x)\\
\nonumber
&=-\varepsilon^{\dot{\mathrm{J}}\dot{\mathrm{J}}_{1}}\varepsilon^{\mathrm{J}\mathrm{J}_{1}}\theta_{\dot{\mathrm{I}}}^{\dagger}\theta_{\dot{\mathrm{J}}_{1}}^{\dagger}\theta_{\mathrm{I}}\theta_{\mathrm{J}_{1}}(\bar{\sigma}^{\mu})^{\dot{\mathrm{I}}\mathrm{I}}(\sigma^{\nu})_{\mathrm{J}\dot{\mathrm{J}}}\partial_{\mu}\phi^{\star}(x)\partial_{\nu}\phi(x)\\
\nonumber
&=-\varepsilon^{\dot{\mathrm{J}}\dot{\mathrm{J}}_{1}}\varepsilon^{\mathrm{J}\mathrm{J}_{1}}(-\frac{1}{2}\varepsilon_{\dot{\mathrm{I}}\dot{\mathrm{J}}_{1}}\theta^{\dagger}\theta^{\dagger})(\frac{1}{2}\varepsilon_{\mathrm{I}\mathrm{J}_{1}}\theta\theta)(\bar{\sigma}^{\mu})^{\dot{\mathrm{I}}\mathrm{I}}(\sigma^{\nu})_{\mathrm{J}\dot{\mathrm{J}}}\partial_{\mu}\phi^{\star}(x)\partial_{\nu}\phi(x)\\
&=\frac{1}{4}\theta^{\dagger}\theta^{\dagger}\theta\theta\,\text{Tr}(\bar{\sigma}^{\mu}\sigma^{\nu})\partial_{\mu}\phi^{\star}(x)\partial_{\nu}\phi(x)\!=\!\frac{1}{2}\theta^{\dagger}\theta^{\dagger}\theta\theta\partial^{\nu}\phi^{\star}(x)\partial_{\nu}\phi(x)
\end{align}and
\begin{align}
\nonumber
&\quad\quad\quad \theta^{\dagger}\psi^{\dagger}(x)\!\times\!\theta^{\dagger}\bar{\sigma}^{\mu}\partial_{\mu}\psi(x)\!=\!\psi_{\dot{\mathrm{I}}}^{\dagger}(x)\theta^{\dagger\dot{\mathrm{I}}}\theta_{\dot{\mathrm{J}}}^{\dagger}\big(\bar{\sigma}^{\mu}\partial_{\mu}\psi(x)\big)^{\dot{\mathrm{J}}}\\
&=\!-\frac{1}{2}\varepsilon_{\dot{\mathrm{J}}\dot{\mathrm{J}}_{1}}\varepsilon^{\dot{\mathrm{J}}_{1}\dot{\mathrm{I}}}\theta^{\dagger}\theta^{\dagger}\psi_{\dot{\mathrm{I}}}^{\dagger}(x)\big(\bar{\sigma}^{\mu}\partial_{\mu}\psi(x)\big)^{\dot{\mathrm{J}}}\!=-\frac{1}{2}\theta^{\dagger}\theta^{\dagger}\!\times\!\psi^{\dagger}(x)\bar{\sigma}^{\mu}\partial_{\mu}\psi(x)
\end{align}and
\begin{align}
\nonumber
&\theta\psi(x)\!\times\!\theta\sigma^{\mu}\partial_{\mu}\psi^{\dagger}(x)\!=\!\theta^{\mathrm{J}}\theta^{\mathrm{I}}\psi_{\mathrm{I}}(x)\big(\sigma^{\mu}\partial_{\mu}\psi^{\dagger}(x)\big)_{\mathrm{J}}=-\frac{1}{2}\varepsilon^{\mathrm{J}\mathrm{I}}\theta\cdot\theta\psi_{\mathrm{I}}(x)\big(\sigma^{\mu}\partial_{\mu}\psi^{\dagger}(x)\big)_{\mathrm{J}}\\
&\quad\quad\quad=-\frac{1}{2}\theta\cdot\theta\!\times\!\psi(x)\sigma^{\mu}\partial_{\mu}\psi^{\dagger}(x)\!=\!\frac{1}{2}\theta\cdot\theta\!\times\!\partial_{\mu}\psi^{\dagger}(x)\bar{\sigma}^{\mu}\psi(x)
\end{align}After discarding the total derivative term in \eqref{DtermPhibarPhi}, the supersymmetry-invariant kinetic term for the Lagrangian of the chiral superfield is built as
\begin{align}
\nonumber
&\quad\quad\quad\quad S^{(D)}=\int d^{8}z[\bar{\Phi}(x,\theta,\theta^{\dagger})\Phi(x,\theta,\theta^{\dagger})]_{D}\\
&=\!\! \int d^{4}x\,\big(\eta^{\mu\nu}\partial_{\mu}\phi^{\star}(x)\partial_{\nu}\phi(x)\!+\!F^{\star}(x)F(x)\!+\!\text{i}\psi^{\dagger}(x)\bar{\sigma}^{\mu}\partial_{\mu}\psi(x)\big)
\end{align}After considering the variation of the D-term Lagrangian under the supersymmetric variation \eqref{SUSYTransComponentField}-\eqref{SUSYTransComponentComplexField}, we could show that it just produces the total derivative terms
\begin{align}
\nonumber
&\delta_{\epsilon}S^{(\text{D})}\!=\!\int d^{4}x\big\{\eta^{\mu\nu}\partial_{\nu}\phi\!\times\!\epsilon^{\dagger}\partial_{\mu}\psi^{\dagger}\!+\!\eta^{\mu\nu}\partial_{\mu}\phi^{\star}\!\times\!\epsilon\partial_{\nu}\psi\!+\!\text{i}F\times\partial_{\mu}\psi^{\dagger}\bar{\sigma}^{\mu}\epsilon\!-\!\text{i}F^{\star}\times\epsilon^{\dagger}\bar{\sigma}^{\mu}\partial_{\mu}\psi\\
\nonumber
&\quad\quad~~+\text{i}\big(F^{\star}\epsilon_{\dot{\mathrm{I}}_{1}}^{\dagger}\!+\!\text{i}\partial_{\nu}\phi^{\star}(\epsilon\sigma^{\nu})_{\dot{\mathrm{I}}_{1}}\big)(\bar{\sigma}^{\mu})^{\dot{\mathrm{I}}_{1}\mathrm{I}_{2}}\partial_{\mu}\psi_{\mathrm{I}_{2}}\!+\!\text{i}\psi_{\dot{\mathrm{I}}_{1}}^{\dagger}(\bar{\sigma}^{\mu})^{\dot{\mathrm{I}}_{1}\mathrm{I}_{2}}\big(\partial_{\mu}F\epsilon_{\mathrm{I}_{2}}\!-\!\text{i}\partial_{\mu}\partial_{\nu}\phi(\sigma^{\nu}\epsilon^{\dagger})_{\mathrm{I}_{2}}\big)\big\}\\
\label{DtermLagrangianSUSYvariation}
&\quad\quad~~=\!\int d^{4}x\partial_{\mu}(\eta^{\mu\nu}\partial_{\nu}\phi\epsilon^{\dagger}\psi^{\dagger}\!+\!\eta^{\mu\nu}\phi^{\star}\epsilon\partial_{\nu}\psi\!-\!\phi^{\star}\epsilon\sigma^{\mu}\bar{\sigma}^{\nu}\partial_{\nu}\psi\!+\!\text{i}F\psi^{\dagger}\bar{\sigma}^{\mu}\epsilon)
\end{align}

\subsubsection*{SUSY-invariant interaction term from the F-term construction}
 
Let us now proceed to extract the component fields from the F-term, which is associated with the superpotential. In this work, we focus exclusively on the study of renormalizable superpotentials, namely,
\begin{align}
\label{RenorSuperPotential}
&W(\Phi)=\Lambda+\lambda\Phi(x,\theta,\theta^{\dagger})+\frac{m}{2}\Phi(x,\theta,\theta^{\dagger})^{2}+\frac{g}{3!}\Phi(x,\theta,\theta^{\dagger})^{3}
\end{align}For brevity, we just set $\Lambda=\lambda=0$ and take into account
\begin{align}
\label{SimpliRenorSuperPotential}
&W(\Phi)=\frac{m}{2}\Phi^{2}+\frac{g}{3!}\Phi^{3}~,~\bar{W}(\bar{\Phi})=\frac{m}{2}\bar{\Phi}^{2}+\frac{g}{3!}\bar{\Phi}^{3}
\end{align}For the quadratic and cubic products of chiral superfields in $\eqref{SimpliRenorSuperPotential}$, the corresponding $F$-terms are expanded as follows
\begin{align}
\label{DoubleProductChiralSuperfield}
&[\Phi\Phi]_{\theta\theta}=\theta\theta(2\phi F-\psi\psi)~,~[\bar{\Phi}\bar{\Phi}]_{\theta^{\dagger}\theta^{\dagger}}=\theta^{\dagger}\theta^{\dagger}(2\phi^{\star}F^{\star}-\psi^{\dagger}\psi^{\dagger})\\
\label{CubicProductChiralSuperfield}
&[\Phi\Phi\Phi]_{\theta\theta}=\theta\theta(3\phi^{2}F-3\phi\psi\psi)~,~[\bar{\Phi}\bar{\Phi}\bar{\Phi}]_{\theta^{\dagger}\theta^{\dagger}}=\theta^{\dagger}\theta^{\dagger}(3\phi^{\star2}F^{\star}-3\phi^{\star}\psi^{\dagger}\psi^{\dagger})
\end{align}
Based on the results \eqref{DoubleProductChiralSuperfield}–\eqref{CubicProductChiralSuperfield}, the SUSY-invariant interaction terms corresponding to \eqref{SimpliRenorSuperPotential} can be constructed
\begin{align}
\nonumber
&\quad\quad\quad~ S^{(F)}\!=\!\int d^{6}z(\frac{m}{2}[\Phi\Phi]_{\theta\theta}\!+\!\frac{g}{3!}[\Phi\Phi\Phi]_{\theta\theta}\big)\!+\!\int d^{6}\bar{z}\big(\frac{m}{2}[\bar{\Phi}\bar{\Phi}]_{\bar{\theta}\bar{\theta}}\!+\!\frac{g}{3!}[\bar{\Phi}\bar{\Phi}\bar{\Phi}]_{\bar{\theta}\bar{\theta}}\big)\\
&=\!\int d^{4}x\big(m\phi F\!-\frac{m}{2}\psi\psi\!+\frac{g}{2}(\phi^{2}F\!-\!\phi\psi\psi)\!+\!m\phi^{\star}F^{\star}\!-\frac{m}{2}\psi^{\dagger}\psi^{\dagger}\!+\frac{g}{2}(\phi^{\star2}F^{\star}\!-\!\phi^{\star}\psi^{\dagger}\psi^{\dagger})\big)
\end{align}Similar to \eqref{DtermLagrangianSUSYvariation}, under the supersymmetric transformations \eqref{SUSYTransComponentField}-\eqref{SUSYTransComponentComplexField}, it is straightforward to verify that
\begin{align}
\label{FtermLagrangianSUSYvariation}
&\delta_{\epsilon}S^{(F)}=\int d^{4}x\,\big\{\partial_{\mu}(\text{i}m\phi\,\psi\sigma^{\mu}\epsilon^{\dagger}+\frac{\text{i}g}{2}\phi^{2}\,\psi\sigma^{\mu}\epsilon^{\dagger})+\text{h.c.}\big\}
\end{align}

\subsubsection*{The elimination of the auxiliary field $F(x)$ via the on-shell EOMs}

By varying the total action $S^{(D)} + S^{(F)}$ with respect to the auxiliary field $F(x)$ (and $F^{\star}$), we obtain the following equations of motion
\begin{align}
\label{OnshellEOMsF}
&F=-m\phi^{\star}-\frac{g}{2}\phi^{\star2}\,,\,F^{\star}=-m\phi-\frac{g}{2}\phi^{2}
\end{align}Subsequently, by substituting \eqref{OnshellEOMsF} back into the total action $S^{(D)} + S^{(F)}$, we obtain the following on-shell form of the total action
\begin{align}
\nonumber
&S\!=\!S^{(D)}\!+\!S^{(F)}\!=\! \int d^{4}x\big\{\eta^{\mu\nu}\partial_{\mu}\phi^{\star}(x)\partial_{\nu}\phi(x)\!-\!m^{2}\vert\phi\vert^{2}\!-\frac{gm}{2}\phi\phi^{\star2}\!-\frac{mg}{2}\phi^{\star}\phi^{2}\!-\frac{g^{2}}{4}\vert\phi\vert^{4}\\
\label{OnshellTotalWZ}
&\quad\quad\quad\quad\quad\quad\quad\quad +\text{i}\psi^{\dagger}(x)\bar{\sigma}^{\mu}\partial_{\mu}\psi(x)-\frac{m}{2}(\psi\psi+\psi^{\dagger}\psi^{\dagger})-\frac{g}{2}(\phi\psi\psi+\phi^{\star}\psi^{\dagger}\psi^{\dagger})\big\}
\end{align}Note that in this work, we focus exclusively on tree-level physical processes. Therefore, the on-shell level Lagrangian given in \eqref{OnshellTotalWZ} is sufficient for our purposes.

\section{Decay Process $\varphi \to [\bar{\Phi}\Phi]_{D}$ and the relevant  Feynman Rules from the Path Integral approach \label{PathIntegralFeynmanRule}}

In this work, we are interested in exploring the Yukawa-like interaction that governs the decay of the inflaton into a pair of chiral superfields
\begin{align}
\nonumber
&S_{\text{int}}\!=\!\mathsf{y}_{\text{D}} \kappa\int d^{8}z\,\varphi[\bar{\Phi}(x,\theta,\theta^{\dagger})\Phi(x,\theta,\theta^{\dagger})]_{D}=\!\mathsf{y}_{\text{D}} \kappa\int d^{4}x\big\{\varphi\,\vert F\vert^{2}+\frac{\text{i}}{2}\varphi\,\big(\psi^{\dagger}\bar{\sigma}^{\mu}\partial_{\mu}\psi-\partial_{\mu}\psi^{\dagger}\bar{\sigma}^{\mu}\psi\big)\\
\label{IntInflatonToChiralSuperfield}
&\quad\quad\quad\quad\quad\quad\quad\quad +\frac{1}{4}\varphi\big(-\phi^{\star}\partial_{\mu}\partial^{\mu}\phi-\phi\partial_{\mu}\partial^{\mu}\phi^{\star}+2\eta^{\mu\nu}\partial_{\mu}\phi^{\star}\partial_{\nu}\phi\big)\big\}
\end{align}For further details, one could refer to \eqref{DtermPhibarPhi}. It is important to emphasize that, in this context, the total derivative term in \eqref{DtermPhibarPhi} is retained and not discarded. After eliminating the auxiliary field $F(x)$ through the equation of motions \eqref{OnshellEOMsF}, \eqref{IntInflatonToChiralSuperfield} is rewritten as
\begin{align}
\nonumber
&\quad\quad S_{\text{int}}=\mathsf{y}_{\text{D}}\kappa\int d^{4}x\big\{\frac{1}{4}\varphi\big(-\phi^{\star}\partial_{\mu}\partial^{\mu}\phi-\phi\partial_{\mu}\partial^{\mu}\phi^{\star}+2\eta^{\mu\nu}\partial_{\mu}\phi^{\star}\partial_{\nu}\phi\big) \\
&+\varphi\,(m^{2}\vert\phi\vert^{2}\!+\frac{gm}{2}\phi\phi^{\star2}\!+\frac{gm}{2}\phi^{\star}\phi^{2}\!+\frac{g^{2}}{4}\vert\phi\vert^{4})\!+\frac{\text{i}}{2}\varphi\,(\psi^{\dagger}\bar{\sigma}^{\mu}\partial_{\mu}\psi\!-\!\partial_{\mu}\psi^{\dagger}\bar{\sigma}^{\mu}\psi)\big\}
\end{align}For simplicity, we set the dimensionless coupling $g=0$ in this work, as our analysis is restricted to the two-body decay process (which also corresponds to the three-body decay when the graviton is included). In summary, one of the key objectives of this work is to investigate the three-point interactions
\begin{align}
\label{YukawaInflatonMajorana}
&S_{\text{int}}^{(\varphi\psi\psi)}=\frac{\text{i}}{2}\mathsf{y}_{\text{D}}\kappa\int d^{4}x\,\varphi(\psi^{\dagger}\bar{\sigma}^{\mu}\partial_{\mu}\psi-\partial_{\mu}\psi^{\dagger}\bar{\sigma}^{\mu}\psi)\\
\label{YukawaInflatonComScalar}
&S_{\text{int}}^{(\varphi\phi\phi)}=\frac{1}{4}\mathsf{y}_{\text{D}}\kappa\int d^{4}x\,\varphi\big(2\eta^{\mu\nu}\partial_{\mu}\phi^{\star}\partial_{\nu}\phi+4m^{2}\phi^{\star}\phi-\phi^{\star}\partial_{\mu}\partial^{\mu}\phi-\phi\partial_{\mu}\partial^{\mu}\phi^{\star}\big)
\end{align}In parallel, as shown in later, we also aim to incorporate the interaction between a single graviton and two Majorana fermions, given by \eqref{YukawaGravitonMajorana}.

\subsection{The derivations of the two-point correlations for the free Majorana fermions \label{ReviewFreeMajorana}}

\begin{figure}[ht]
	\begin{center}
		\includegraphics[scale=0.43]{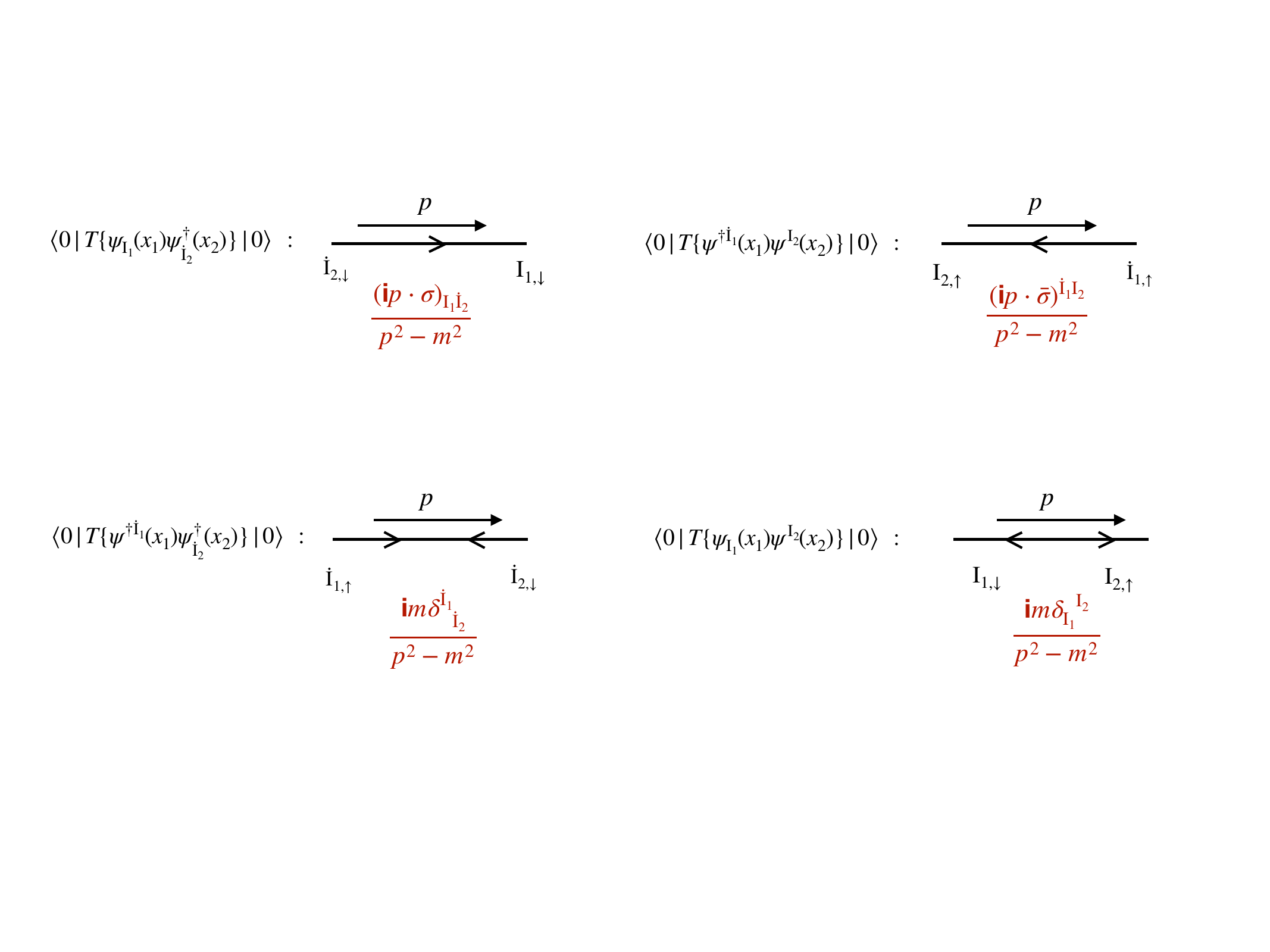}
		\caption{Propagators in momentum space extracted from the various two-point correlation functions for a neutral two-component Majorana fermion with mass $m$. The conventions for the spinor indices are chosen to be consistent with Appendix. \ref{ReviewWZChiral}.}
		\label{VariousTwoPointCorrelationLabel}
	\end{center}
\end{figure}

Before deriving the interaction vertices corresponding to \eqref{YukawaInflatonMajorana} and \eqref{YukawaInflatonComScalar}, let us begin with the derivation of the two-point correlation functions within the framework of free Majorana fermions, as a warm-up exercise. The Lagrangian of free Majorana fermion is given by
\begin{small}
\begin{align}
\nonumber
\mathcal{L}_{\text{Majorana}}^{\text{(free)}}&\!=\!\text{i}\psi^{\dagger}\bar{\sigma}^{\mu}\partial_{\mu}\psi\!-\frac{m}{2}(\psi\psi\!+\!\psi^{\dagger}\psi^{\dagger})\!=\frac{\text{i}}{2}\psi^{\dagger}\bar{\sigma}^{\mu}\partial_{\mu}\psi\!-\frac{\text{i}}{2}\partial_{\mu}\psi\sigma^{\mu}\psi^{\dagger}\!-\frac{m}{2}(\psi\psi\!+\!\psi^{\dagger}\psi^{\dagger})\\
\label{FreeMajoranaLagrangian}
&\equiv\frac{\text{i}}{2}\psi^{\dagger}(x)\bar{\sigma}^{\mu}\partial_{\mu}\psi(x)+\frac{\text{i}}{2}\psi(x)\sigma^{\mu}\partial_{\mu}\psi^{\dagger}(x)-\frac{m}{2}(\psi\psi+\psi^{\dagger}\psi^{\dagger})+\text{total derivative}
\end{align}
\end{small}the generating functional corresponding to \eqref{FreeMajoranaLagrangian} is defined by
\begin{align}
&W_{0}[\mathcal{J},\mathcal{J}^{\dagger}]=\mathcal{N}\!\!\int\!\mathcal{D}\psi\mathcal{D}\psi^{\dagger}\text{e}^{\text{i}\int d^{4}x\big(\frac{1}{2}(\text{i}\psi^{\dagger}\bar{\sigma}^{\mu}\partial_{\mu}\psi+\text{i}\psi\sigma^{\mu}\partial_{\mu}\psi^{\dagger}-m\psi\psi-m\psi^{\dagger}\psi^{\dagger})+\mathcal{J}\cdot\psi+\psi^{\dagger}\cdot\mathcal{J}^{\dagger}\big)}
\end{align}For later convenience, and in order to obtain the result $W_{0}[\mathcal{J},\mathcal{J}^{\dagger}]$ directly in momentum space, we choose to work within the framework of the Fourier transformation
\begin{align}
\label{FTforField}
&\psi_{\mathrm{I}}(x)=\int\frac{d^{4}p}{(2\pi)^{4}}\text{e}^{-\text{i}p\cdot x}\tilde{\psi}_{\mathrm{I}}(p)~,~\psi_{\dot{\mathrm{I}}}^{\dagger}(x)=\int\frac{d^{4}p}{(2\pi)^{4}}\text{e}^{\text{i}p\cdot x}\tilde{\psi}_{\dot{\mathrm{I}}}^{\dagger}(p)\\
\label{FTforCurrent}
&\mathcal{J}_{\mathrm{I}}(x)=\int\frac{d^{4}p}{(2\pi)^{4}}\text{e}^{-\text{i}p\cdot x}\tilde{\mathcal{J}}_{\mathrm{I}}(p)~,~\mathcal{J}_{\dot{\mathrm{I}}}^{\dagger}(x)=\int\frac{d^{4}p}{(2\pi)^{4}}\text{e}^{\text{i}p\cdot x}\tilde{\mathcal{J}}_{\dot{\mathrm{I}}}^{\dagger}(p)
\end{align}Note that the delta function can be generated via the following integral expression in momentum space
\begin{align}
\label{DeltaFunctionConven}
&\delta^{4}(x-x^{\prime})=\int\frac{d^{4}p}{(2\pi)^{4}}\text{e}^{-\text{i}p\cdot(x-x^{\prime})}\,,\,\delta^{4}(p-p^{\prime})=\int\frac{d^{4}x}{(2\pi)^{4}}\text{e}^{-\text{i}x\cdot(p-p^{\prime})}
\end{align}Therefore, by making use of the inverse Fourier transformation, we also obtain
\begin{align}
\label{InverseFTCurrentPositiveMomen}
&\tilde{\mathcal{J}}_{\mathrm{I}}(q)=\int d^{4}x\text{e}^{\text{i}q\cdot x}\mathcal{J}_{\mathrm{I}}(x)~,~\tilde{\mathcal{J}}_{\dot{\mathrm{I}}}^{\dagger}(q)=\int d^{4}x\text{e}^{-\text{i}q\cdot x}\mathcal{J}_{\dot{\mathrm{I}}}^{\dagger}(x)\\
\label{InverseFTCurrentNegativeMomen}
&\tilde{\mathcal{J}}_{\mathrm{I}}(-q)=\int d^{4}x\text{e}^{-\text{i}q\cdot x}\mathcal{J}_{\mathrm{I}}(x)~,~\tilde{\mathcal{J}}_{\dot{\mathrm{I}}}^{\dagger}(-q)=\int d^{4}x\text{e}^{\text{i}q\cdot x}\mathcal{J}_{\dot{\mathrm{I}}}^{\dagger}(x)
\end{align}Besides, the functional calculus for $\mathcal{J}_{\mathrm{I}},\mathcal{J}_{\dot{\mathrm{I}}}^{\dagger},\tilde{\mathcal{J}}_{\mathrm{I}},\tilde{\mathcal{J}}_{\dot{\mathrm{I}}}^{\dagger}$ follows the rules
\begin{small}
\begin{align}
\nonumber
&\frac{\overrightarrow{\delta}}{\delta\mathcal{J}^{\mathrm{I}}(x)}\mathcal{J}^{\mathrm{J}}(y)\!=\!\mathcal{J}^{\mathrm{J}}(y)\frac{\overleftarrow{\delta}}{\delta\mathcal{J}^{\mathrm{I}}(x)}\!=\!\delta_{\mathrm{I}}^{~\mathrm{J}}\delta^{4}(y-x)\,,\,\frac{\overrightarrow{\delta}}{\delta\mathcal{J}_{\mathrm{I}}(x)}\mathcal{J}_{\mathrm{J}}(y)\!=\!\mathcal{J}_{\mathrm{J}}(y)\frac{\overleftarrow{\delta}}{\delta\mathcal{J}_{\mathrm{I}}(x)}\!=\!\delta_{\mathrm{J}}^{~\mathrm{I}}\delta^{4}(y-x)\\
\nonumber
&\frac{\overrightarrow{\delta}}{\delta\mathcal{J}^{\mathrm{I}}(x)}\mathcal{J}_{\mathrm{J}}(y)\!=\!\frac{\overrightarrow{\delta}}{\delta\mathcal{J}^{\mathrm{I}}(x)}\varepsilon_{\mathrm{J}\mathrm{J}_{1}}\!\mathcal{J}^{\mathrm{J}_{1}}(y)\!=\!\mathcal{J}_{\mathrm{J}}(y)\frac{\overleftarrow{\delta}}{\delta\mathcal{J}^{\mathrm{I}}(x)}=\!\varepsilon_{\mathrm{J}\mathrm{J}_{1}}\!\mathcal{J}^{\mathrm{J}_{1}}(y)\frac{\overleftarrow{\delta}}{\delta\mathcal{J}^{\mathrm{I}}(x)}=\!\varepsilon_{\mathrm{J}\mathrm{I}}\delta^{4}(y-x)\\
\nonumber
&\frac{\overrightarrow{\delta}}{\delta\mathcal{J}_{\mathrm{I}}(x)}\mathcal{J}^{\mathrm{J}}(y)=\frac{\overrightarrow{\delta}}{\delta\mathcal{J}_{\mathrm{I}}(x)}\varepsilon^{\mathrm{J}\mathrm{J}_{1}}\mathcal{J}_{\mathrm{J}_{1}}(y)=\mathcal{J}^{\mathrm{J}}(y)\frac{\overleftarrow{\delta}}{\delta\mathcal{J}_{\mathrm{I}}(x)}=\varepsilon^{\mathrm{J}\mathrm{J}_{1}}\mathcal{J}_{\mathrm{J}_{1}}(y)\frac{\overleftarrow{\delta}}{\delta\mathcal{J}_{\mathrm{I}}(x)}=\varepsilon^{\mathrm{J}\mathrm{I}}\delta^{4}(y-x)\\
\nonumber
&\frac{\overrightarrow{\delta}}{\delta\mathcal{J}^{\dagger\dot{\mathrm{I}}}(x)}\!\mathcal{J}^{\dagger\dot{\mathrm{J}}}(y)\!=\!\mathcal{J}^{\dagger\dot{\mathrm{J}}}(y)\!\frac{\overleftarrow{\delta}}{\delta\mathcal{J}^{\dagger\dot{\mathrm{I}}}(x)}\!=\!\delta_{~\,\dot{\mathrm{I}}}^{\dot{\mathrm{J}}}\delta^{4}(y-x),\frac{\overrightarrow{\delta}}{\delta\mathcal{J}_{\dot{\mathrm{I}}}^{\dagger}(x)}\!\mathcal{J}_{\dot{\mathrm{J}}}^{\dagger}(y)\!=\!\mathcal{J}_{\dot{\mathrm{J}}}^{\dagger}(y)\!\frac{\overleftarrow{\delta}}{\delta\mathcal{J}_{\dot{\mathrm{I}}}^{\dagger}(x)}\!=\!\delta_{\,~\dot{\mathrm{J}}}^{\dot{\mathrm{I}}}\delta^{4}(y-x)\\
\nonumber
&\frac{\overrightarrow{\delta}}{\delta\mathcal{J}^{\dagger\dot{\mathrm{I}}}(x)}\!\mathcal{J}_{\dot{\mathrm{J}}}^{\dagger}(y)\!=\!\frac{\overrightarrow{\delta}}{\delta\mathcal{J}^{\dagger\dot{\mathrm{I}}}(x)}\!\varepsilon_{\dot{\mathrm{J}}\dot{\mathrm{J}}_{1}}\!\mathcal{J}^{\dagger\dot{\mathrm{J}}_{1}}(y)\!=\!\mathcal{J}_{\dot{\mathrm{J}}}^{\dagger}(y)\frac{\overleftarrow{\delta}}{\delta\mathcal{J}^{\dagger\dot{\mathrm{I}}}(x)}\!=\!\varepsilon_{\dot{\mathrm{J}}\dot{\mathrm{J}}_{1}}\mathcal{J}^{\dagger\dot{\mathrm{J}}_{1}}(y)\frac{\overleftarrow{\delta}}{\delta\mathcal{J}^{\dagger\dot{\mathrm{I}}}(x)}\!=\!\varepsilon_{\dot{\mathrm{J}}\dot{\mathrm{I}}}\delta^{4}(y-x)\\
\label{CurrentFuncDerivaMajoraCoordinate}
&\frac{\overrightarrow{\delta}}{\delta\mathcal{J}_{\dot{\mathrm{I}}}^{\dagger}(x)}\!\mathcal{J}^{\dagger\dot{\mathrm{J}}}(y)\!=\!\frac{\overrightarrow{\delta}}{\delta\mathcal{J}_{\dot{\mathrm{I}}}^{\dagger}(x)}\!\varepsilon^{\dot{\mathrm{J}}\dot{\mathrm{J}}_{1}}\mathcal{J}_{\dot{\mathrm{J}}_{1}}^{\dagger}(y)\!=\!\mathcal{J}^{\dagger\dot{\mathrm{J}}}(y)\frac{\overleftarrow{\delta}}{\delta\mathcal{J}_{\dot{\mathrm{I}}}^{\dagger}(x)}\!=\!\varepsilon^{\dot{\mathrm{J}}\dot{\mathrm{J}}_{1}}\mathcal{J}_{\dot{\mathrm{J}}_{1}}^{\dagger}(y)\frac{\overleftarrow{\delta}}{\delta\mathcal{J}_{\dot{\mathrm{I}}}^{\dagger}(x)}\!=\!\varepsilon^{\dot{\mathrm{J}}\dot{\mathrm{I}}}\delta^{4}(y-x)
\end{align}
\end{small}Similarly, by transforming the above identities into momentum space, we get
\begin{small}
\begin{align}
\nonumber
&\frac{\overrightarrow{\delta}}{\delta\tilde{\mathcal{J}}^{\mathrm{I}}(p)}\tilde{\mathcal{J}}^{\mathrm{J}}(q)=\tilde{\mathcal{J}}^{\mathrm{J}}(q)\frac{\overleftarrow{\delta}}{\delta\tilde{\mathcal{J}}^{\mathrm{I}}(p)}=(2\pi)^{4}\delta_{\mathrm{I}}^{~\mathrm{J}}\delta^{4}(q-p) \\
\nonumber
&\frac{\overrightarrow{\delta}}{\delta\tilde{\mathcal{J}}_{\mathrm{I}}(p)}\tilde{\mathcal{J}}_{\mathrm{J}}(q)=\tilde{\mathcal{J}}_{\mathrm{J}}(q)\frac{\overleftarrow{\delta}}{\delta\tilde{\mathcal{J}}_{\mathrm{I}}(p)}=(2\pi)^{4}\delta_{\mathrm{J}}^{~\mathrm{I}}\delta^{4}(q-p)\\
\nonumber
&\frac{\overrightarrow{\delta}}{\delta\tilde{\mathcal{J}}^{\mathrm{I}}(p)}\tilde{\mathcal{J}}_{\mathrm{J}}(q)=\tilde{\mathcal{J}}_{\mathrm{J}}(q)\frac{\overleftarrow{\delta}}{\delta\tilde{\mathcal{J}}^{\mathrm{I}}(p)}=(2\pi)^{4}\varepsilon_{\mathrm{J}\mathrm{I}}\delta^{4}(q-p)\\
\nonumber
&\frac{\overrightarrow{\delta}}{\delta\tilde{\mathcal{J}}_{\mathrm{I}}(p)}\tilde{\mathcal{J}}^{\mathrm{J}}(q)=\tilde{\mathcal{J}}^{\mathrm{J}}(q)\frac{\overleftarrow{\delta}}{\delta\tilde{\mathcal{J}}_{\mathrm{I}}(p)}=(2\pi)^{4}\varepsilon^{\mathrm{J}\mathrm{I}}\delta^{4}(q-p)\\
\nonumber
&\frac{\overrightarrow{\delta}}{\delta\mathcal{J}^{\dagger\dot{\mathrm{I}}}(p)}\mathcal{J}^{\dagger\dot{\mathrm{J}}}(q)=\mathcal{J}^{\dagger\dot{\mathrm{J}}}(q)\frac{\overleftarrow{\delta}}{\delta\mathcal{J}^{\dagger\dot{\mathrm{I}}}(p)}=(2\pi)^{4}\delta_{~\,\dot{\mathrm{I}}}^{\dot{\mathrm{J}}}\delta^{4}(q-p)\\
\nonumber
&\frac{\overrightarrow{\delta}}{\delta\mathcal{J}_{\dot{\mathrm{I}}}^{\dagger}(p)}\mathcal{J}_{\dot{\mathrm{J}}}^{\dagger}(q)=\mathcal{J}_{\dot{\mathrm{J}}}^{\dagger}(q)\frac{\overleftarrow{\delta}}{\delta\mathcal{J}_{\dot{\mathrm{I}}}^{\dagger}(p)}=(2\pi)^{4}\delta_{\,~\dot{\mathrm{J}}}^{\dot{\mathrm{I}}}\delta^{4}(q-p)\\
\nonumber
&\frac{\overrightarrow{\delta}}{\delta\mathcal{J}^{\dagger\dot{\mathrm{I}}}(p)}\mathcal{J}_{\dot{\mathrm{J}}}^{\dagger}(q)=\mathcal{J}_{\dot{\mathrm{J}}}^{\dagger}(q)\frac{\overleftarrow{\delta}}{\delta\mathcal{J}^{\dagger\dot{\mathrm{I}}}(p)}=(2\pi)^{4}\varepsilon_{\dot{\mathrm{J}}\dot{\mathrm{I}}}\delta^{4}(q-p)\\
\label{CurrentFuncDerivaMajoraMomentum}
&\frac{\overrightarrow{\delta}}{\delta\mathcal{J}_{\dot{\mathrm{I}}}^{\dagger}(p)}\mathcal{J}^{\dagger\dot{\mathrm{J}}}(q)=\mathcal{J}^{\dagger\dot{\mathrm{J}}}(q)\frac{\overleftarrow{\delta}}{\delta\mathcal{J}_{\dot{\mathrm{I}}}^{\dagger}(p)}=(2\pi)^{4}\varepsilon^{\dot{\mathrm{J}}\dot{\mathrm{I}}}\delta^{4}(q-p)
\end{align}
\end{small}With the aid of the derivative rules above, the Fourier transformation of the functional derivative from coordinate space to momentum space is given by
\begin{small}
\begin{align}
\nonumber
\frac{\delta}{\delta\mathcal{J}^{\mathrm{I}}(x_{1})}&=\int d^{4}q\frac{\delta\tilde{\mathcal{J}}^{\mathrm{J}}(q)}{\delta\mathcal{J}^{\mathrm{I}}(x_{1})}\frac{\delta}{\delta\tilde{\mathcal{J}}^{\mathrm{J}}(q)}=\int d^{4}q\frac{\delta\big(\int d^{4}y\text{e}^{\text{i}q\cdot y}\mathcal{J}^{\mathrm{J}}(y)\big)}{\delta\mathcal{J}^{\mathrm{I}}(x_{1})}\frac{\delta}{\delta\tilde{\mathcal{J}}^{\mathrm{J}}(q)}\\
&=\int d^{4}q\,\text{e}^{\text{i}q\cdot x_{1}}\frac{\delta}{\delta\tilde{\mathcal{J}}^{\mathrm{I}}(q)}=\int d^{4}q\,\text{e}^{-\text{i}q\cdot x_{1}}\frac{\delta}{\delta\tilde{\mathcal{J}}^{\mathrm{I}}(-q)}\\
\nonumber
\frac{\delta}{\delta\mathcal{J}^{\dagger\dot{\mathrm{I}}}(x_{1})}&=\int d^{4}q\frac{\delta\mathcal{J}^{\dagger\dot{\mathrm{J}}}(q)}{\delta\mathcal{J}^{\dagger\dot{\mathrm{I}}}(x_{1})}\frac{\delta}{\delta\mathcal{J}^{\dagger\dot{\mathrm{J}}}(q)}=\int d^{4}q\frac{\delta\big(\int d^{4}y\text{e}^{-\text{i}q\cdot y}\mathcal{J}^{\dagger\dot{\mathrm{J}}}(y)\big)}{\delta\mathcal{J}^{\dagger\dot{\mathrm{I}}}(x_{1})}\frac{\delta}{\delta\mathcal{J}^{\dagger\dot{\mathrm{J}}}(q)}\\
&=\int d^{4}q\text{e}^{-\text{i}q\cdot x_{1}}\frac{\delta}{\delta\mathcal{J}^{\dagger\dot{\mathrm{I}}}(q)}=\int d^{4}q\text{e}^{\text{i}q\cdot x_{1}}\frac{\delta}{\delta\mathcal{J}^{\dagger\dot{\mathrm{I}}}(-q)}
\end{align}
\end{small}Note that the above results also apply to $\frac{\delta}{\delta\mathcal{J}_{\mathrm{I}}}$ and $\frac{\delta}{\delta\mathcal{J}_{\dot{\mathrm{I}}}^{\dagger}}$. For notational simplicity, we have omitted the directional arrows above the variational symbols $\delta$; the precise expressions should be written as $\frac{\overrightarrow{\delta}}{\delta\mathcal{J}^{\mathrm{I}}}$, $\frac{\overleftarrow{\delta}}{\delta\mathcal{J}^{\mathrm{I}}}$, $\frac{\overrightarrow{\delta}}{\delta\mathcal{J}^{\dagger\dot{\mathrm{I}}}}$, $\frac{\overleftarrow{\delta}}{\delta\mathcal{J}^{\dagger\dot{\mathrm{I}}}}$. To facilitate the use of the Gaussian integral in functional form, it is convenient to rewrite $S_{0}^{\text{(free)}}[J, J^{\dagger}]$ in a matrix representation
\begin{align}
\nonumber
&\quad\quad\quad\quad S_{0}^{\text{(free)}}[\mathcal{J},\mathcal{J}^{\dagger}]=\int d^{4}x\big(\mathcal{L}_{\text{Majorana}}^{\text{(free)}}+\mathcal{J}\cdot\psi+\psi^{\dagger}\cdot\mathcal{J}^{\dagger}\big)\\
\nonumber
&=\frac{1}{2}\int\frac{d^{4}p}{(2\pi)^{4}}\bigg\{\underbrace{\left(\!\!\begin{array}{cc}
	\tilde{\psi}^{\mathrm{I}_{1}}(-p) & \tilde{\psi}_{\dot{\mathrm{I}}_{2}}^{\dagger}(p)\end{array}\!\!\right)}_{\boldsymbol{\Omega}_{p}^{\dagger}(\tilde{\psi})}\underbrace{\left(\!\!\begin{array}{cc}
	(p_{\mu}\sigma^{\mu})_{\mathrm{I}_{1}\dot{\mathrm{I}}_{1}} & -m\delta_{\mathrm{I}_{1}}^{~~\mathrm{I}_{2}}\\
	-m\delta_{~~\dot{\mathrm{I}}_{1}}^{\dot{\mathrm{I}}_{2}} & (p_{\mu}\bar{\sigma}^{\mu})^{\dot{\mathrm{I}}_{2}\mathrm{I}_{2}}
	\end{array}\!\!\right)}_{\boldsymbol{\mathcal{M}}_{p}}\underbrace{\left(\!\!\begin{array}{c}
	\tilde{\psi}^{\dagger\dot{\mathrm{I}}_{1}}(-p)\\
	\tilde{\psi}_{\mathrm{I}_{2}}(p)
	\end{array}\!\!\right)}_{\boldsymbol{\Omega}_{p}(\tilde{\psi})}\\
\label{FreeLagranMajoranaMatrix}
&\quad~~+\underbrace{\left(\!\!\begin{array}{cc}
	\tilde{\psi}^{\mathrm{I}_{1}}(-p) & \tilde{\psi}_{\dot{\mathrm{I}}_{2}}^{\dagger}(p)\end{array}\!\!\right)}_{\boldsymbol{\Omega}_{p}^{\dagger}(\tilde{\psi})}\underbrace{\left(\!\!\begin{array}{c}
	\tilde{\mathcal{J}}_{\mathrm{I}_{1}}(p)\\
	\tilde{\mathcal{J}}^{\dagger\dot{\mathrm{I}}_{2}}(-p)
	\end{array}\!\!\right)}_{\boldsymbol{X}_{p}(\tilde{\mathcal{J}})}\!+\!\underbrace{\left(\!\!\begin{array}{cc}
	\tilde{\mathcal{J}}_{\dot{\mathrm{I}}_{1}}(p) & \tilde{\mathcal{J}}^{\mathrm{I}_{2}}(-p)\end{array}\!\!\right)}_{\boldsymbol{X}^{\dagger}(\tilde{\mathcal{J}})}\underbrace{\left(\!\!\begin{array}{c}
	\tilde{\psi}^{\dot{\mathrm{I}}_{1}}(-p)\\
	\tilde{\psi}_{\mathrm{I}_{2}}(p)
	\end{array}\!\!\right)}_{\boldsymbol{\Omega}(\tilde{\psi})}\bigg\}
\end{align}in which we have used \eqref{FTforField}-\eqref{InverseFTCurrentNegativeMomen}. In order to eliminate the linear terms involving $\boldsymbol{\Omega}_{p}^{\dagger}(\tilde{\psi})$ and $\boldsymbol{\Omega}_{p}(\tilde{\psi})$ in \eqref{FreeLagranMajoranaMatrix}, it is convenient to perform the following redefinition
\begin{align}
\label{RedeFieldVecMajorana}
&\tilde{\boldsymbol{\Omega}}_{p}(\tilde{\psi})=\boldsymbol{\Omega}_{p}(\tilde{\psi})+\boldsymbol{\mathcal{M}}_{p}^{-1}\cdot\boldsymbol{X}_{p}(\tilde{\mathcal{J}})~,~\tilde{\boldsymbol{\Omega}}_{p}^{\dagger}(\tilde{\psi})=\boldsymbol{\Omega}_{p}^{\dagger}(\tilde{\psi})+\boldsymbol{X}_{p}^{\dagger}(\tilde{\mathcal{J}})\cdot(\mathcal{\boldsymbol{M}}_{p}^{-1})^{\dagger}
\end{align}Due to the Hermiticity condition $\boldsymbol{\mathcal{M}}_{p}^{\dagger}=\boldsymbol{\mathcal{M}}_{p}\Rightarrow(\boldsymbol{\mathcal{M}}_{p}^{-1})^{\dagger}=\boldsymbol{\mathcal{M}}_{p}^{-1}$, it is reasonable to assume that the inverse matrix $\boldsymbol{\mathcal{M}}_{p}^{-1}$ takes the following form\begin{align}
&\boldsymbol{\mathcal{M}}_{p}^{-1}=\left(\begin{array}{cc}
c_{1}(p_{\mu}\bar{\sigma}^{\mu})^{\dot{\mathrm{I}}_{1}\mathrm{I}_{1}} & -c_{3}(-m\delta_{~~\dot{\mathrm{I}}_{2}}^{\dot{\mathrm{I}}_{1}})\\
-c_{3}(-m\delta_{\mathrm{I}_{2}}^{~~\mathrm{I}_{1}}) & c_{2}(p_{\mu}\sigma^{\mu})_{\mathrm{I}_{2}\dot{\mathrm{I}}_{2}}
\end{array}\right)
\end{align}By making use of the relation $\boldsymbol{\mathcal{M}}_{p}\boldsymbol{\mathcal{M}}_{p}^{-1}=I$, we could find
\begin{small}
\begin{align}
\nonumber
&\left(\!\begin{array}{cc}
\delta_{\mathrm{I}_{1}}^{~~\mathrm{J}_{1}} & 0\\
0 & \delta_{~~\dot{\mathrm{J}}_{2}}^{\dot{\mathrm{I}}_{2}}
\end{array}\!\right)\!=\!\left(\!\begin{array}{cc}
(p_{\mu}\sigma^{\mu})_{\mathrm{I}_{1}\dot{\mathrm{I}}_{1}} & -m\delta_{\mathrm{I}_{1}}^{~~\mathrm{I}_{2}}\\
-m\delta_{~~\dot{\mathrm{I}}_{1}}^{\dot{\mathrm{I}}_{2}} & (p_{\mu}\bar{\sigma}^{\mu})^{\dot{\mathrm{I}}_{2}\mathrm{I}_{2}}
\end{array}\!\right)\left(\!\begin{array}{cc}
c_{1}(p_{\nu}\bar{\sigma}^{\nu})^{\dot{\mathrm{I}}_{1}\mathrm{J}_{1}} & -c_{3}(-m\delta_{~~\dot{\mathrm{J}}_{2}}^{\dot{\mathrm{I}}_{1}})\\
-c_{3}(-m\delta_{\mathrm{I}_{2}}^{~~\mathrm{J}_{1}}) & c_{2}(p_{\nu}\sigma^{\nu})_{\mathrm{I}_{2}\dot{\mathrm{J}}_{2}}
\end{array}\!\right)\\
&\quad\quad=\!\left(\!\!\begin{array}{cc}
c_{1}p_{\mu}p_{\nu}\eta^{\mu\nu}\delta_{\mathrm{I}_{1}}^{~~\mathrm{J}_{1}}\!-\!c_{3}m^{2}\delta_{\mathrm{I}_{1}}^{~~\mathrm{J}_{1}} & c_{3}m(p_{\mu}\sigma^{\mu})_{\mathrm{I}_{1}\dot{\mathrm{J}}_{2}}\!-\!c_{2}m(p_{\nu}\sigma^{\nu})_{\mathrm{I}_{1}\dot{\mathrm{J}}_{2}}\\
c_{3}m(p_{\mu}\bar{\sigma}^{\mu})^{\dot{\mathrm{I}}_{2}\mathrm{J}_{1}}\!-\!c_{1}m(p_{\nu}\bar{\sigma}^{\nu})^{\dot{\mathrm{I}}_{2}\mathrm{J}_{1}} & c_{2}p_{\mu}p_{\nu}\eta^{\mu\nu}\delta_{~~\dot{\mathrm{J}}_{2}}^{\dot{\mathrm{I}}_{2}}\!-\!c_{3}m^{2}\delta_{~~\dot{\mathrm{J}}_{2}}^{\dot{\mathrm{I}}_{2}}
\end{array}\!\!\right)\\
\nonumber 
&\quad\quad\quad\quad\quad\quad\quad\quad\quad\quad\quad\quad\quad\quad \Longrightarrow \Longrightarrow \\
&\quad\quad\quad\quad\quad\quad\quad\quad\quad\quad c_{1}=c_{2}=c_{3}=1/(p^{2}-m^{2})
\end{align}
\end{small}In summary, the distinct expression for $\boldsymbol{\mathcal{M}}_{p}^{-1}$ is
\begin{align}
\label{InverseHessianMajorana}
&\boldsymbol{\mathcal{M}}_{p}^{-1}=\frac{1}{p^{2}-m^{2}}\left(\begin{array}{cc}
(p_{\mu}\bar{\sigma}^{\mu})^{\dot{\mathrm{I}}_{1}\mathrm{I}_{1}} & m\delta_{~~\dot{\mathrm{I}}_{2}}^{\dot{\mathrm{I}}_{1}}\\
m\delta_{\mathrm{I}_{2}}^{~~\mathrm{I}_{1}} & (p_{\mu}\sigma^{\mu})_{\mathrm{I}_{2}\dot{\mathrm{I}}_{2}}
\end{array}\right)
\end{align}By combining \eqref{FreeLagranMajoranaMatrix}, \eqref{RedeFieldVecMajorana}, and \eqref{InverseHessianMajorana}, the free Majorana action can be reformulated as
\begin{align}
\label{RedefineFreeMajoranaAction}
&S_{0}^{\text{(free)}}[\mathcal{J},\mathcal{J}^{\dagger}]=\frac{1}{2}\int\frac{d^{4}p}{(2\pi)^{4}}\big(\tilde{\boldsymbol{\Omega}}_{p}^{\dagger}(\tilde{\psi})\cdot\boldsymbol{\mathcal{M}}_{p}\cdot\tilde{\boldsymbol{\Omega}}_{p}(\tilde{\psi})-\boldsymbol{X}_{p}^{\dagger}(\tilde{\mathcal{J}})\cdot\boldsymbol{\mathcal{M}}_{p}^{-1}\cdot\boldsymbol{X}_{p}(\tilde{\mathcal{J}})\big)
\end{align}Basing on the quadratic action $\eqref{RedefineFreeMajoranaAction}$, the generating functional can be evaluated as
\begin{small}
\begin{align}
\nonumber
&W_{0}[\mathcal{J},\mathcal{J}^{\dagger}]=\!\text{e}^{-\frac{\text{i}}{2}\int\frac{d^{4}p}{(2\pi)^{4}}\boldsymbol{X}_{p}^{\dagger}(\tilde{\mathcal{J}})\cdot\boldsymbol{\mathcal{M}}_{p}^{-1}\cdot\boldsymbol{X}_{p}(\tilde{\mathcal{J}})}\!\times\!\mathcal{N}\!\!\int\mathcal{D}\tilde{\psi}\mathcal{D}\tilde{\psi}^{\dagger}\!\text{e}^{\frac{\text{i}}{2}\int\frac{d^{4}p}{(2\pi)^{4}}\tilde{\boldsymbol{\Omega}}_{p}^{\dagger}(\tilde{\psi})\cdot\boldsymbol{\mathcal{M}}_{p}\cdot\tilde{\boldsymbol{\Omega}}_{p}(\tilde{\psi})}\\
\nonumber
&=\!\text{e}^{-\frac{\text{i}}{2}\int\frac{d^{4}p}{(2\pi)^{4}}\boldsymbol{X}_{p}^{\dagger}(\tilde{\mathcal{J}})\cdot\boldsymbol{\mathcal{M}}_{p}^{-1}\cdot\boldsymbol{X}_{p}(\tilde{\mathcal{J}})}=\exp\bigg\{-\frac{\text{i}}{2}\int\frac{d^{4}p}{(2\pi)^{4}}\frac{1}{(p^{2}-m^{2})}\bigg(\tilde{\mathcal{J}}_{\dot{\mathrm{I}}_{1}}^{\dagger}(p)(p_{\mu}\bar{\sigma}^{\mu})^{\dot{\mathrm{I}}_{1}\mathrm{I}_{1}}\tilde{\mathcal{J}}_{\mathrm{I}_{1}}(p)\\
\label{GeneFunctionalFreeMajoranaV1}
&\quad+\!\tilde{\mathcal{J}}^{\mathrm{I}_{2}}\!(-p)(p_{\mu}\sigma^{\mu})_{\mathrm{I}_{2}\dot{\mathrm{I}}_{2}}\!\tilde{\mathcal{J}}^{\dagger\dot{\mathrm{I}}_{2}}(-p)\!+\!m\!\tilde{\mathcal{J}}_{\dot{\mathrm{I}}_{1}}^{\dagger}(p)\delta_{~~\dot{\mathrm{I}}_{2}}^{\dot{\mathrm{I}}_{1}}\!\tilde{\mathcal{J}}^{\dagger\dot{\mathrm{I}}_{2}}(-p)\!+\!m\tilde{\mathcal{J}}^{\mathrm{I}_{1}}(-p)\delta_{\mathrm{I}_{1}}^{~~\mathrm{I}_{2}}\tilde{\mathcal{J}}_{\mathrm{I}_{2}}(p)\bigg)\bigg\}
\end{align}
\end{small}in which the constant $\ensuremath{\mathcal{N}}$ is introduced to properly normalize the functional Gaussian integral with respect to $\tilde{\psi}$. Note that the first and second terms in \eqref{GeneFunctionalFreeMajoranaV1} are identical due to the identity $\tilde{\mathcal{J}}^{\mathrm{I}_{2}}(\sigma^{\mu})_{\mathrm{I}_{2}\dot{\mathrm{I}}_{2}}\tilde{\mathcal{J}}^{\dagger\dot{\mathrm{I}}_{2}}\!=\!-\tilde{\mathcal{J}}_{\dot{\mathrm{I}}_{2}}^{\dagger}(\bar{\sigma}^{\mu})^{\dot{\mathrm{I}}_{2}\mathrm{I}_{2}}\tilde{\mathcal{J}}_{\mathrm{I}_{2}}$ and a redefinition of the integration variable, i.e., $p\to-p$. As a consequence, the equivalent form of \eqref{GeneFunctionalFreeMajoranaV1} can also be expressed as
\begin{scriptsize}
\begin{align}
\nonumber
&W_{0}[\mathcal{J},\mathcal{J}^{\dagger}]\!=\!\exp\bigg\{\!-\!\int\frac{d^{4}p}{(2\pi)^{4}}\bigg(\!\tilde{\mathcal{J}}_{\dot{\mathrm{I}}_{1}}^{\dagger}(p)\frac{(\text{i}p_{\mu}\bar{\sigma}^{\mu})^{\dot{\mathrm{I}}_{1}\mathrm{I}_{1}}}{(p^{2}-m^{2})}\tilde{\mathcal{J}}_{\mathrm{I}_{1}}(p)\!+\!\tilde{\mathcal{J}}_{\dot{\mathrm{I}}_{1}}^{\dagger}(p)\frac{\text{i}m\delta_{~~\dot{\mathrm{I}}_{2}}^{\dot{\mathrm{I}}_{1}}}{2(p^{2}-m^{2})}\!\tilde{\mathcal{J}}^{\dagger\dot{\mathrm{I}}_{2}}(-p)\!+\!\tilde{\mathcal{J}}^{\mathrm{I}_{1}}\!(-p)\frac{\text{i}m\delta_{\mathrm{I}_{1}}^{~~\mathrm{I}_{2}}}{2(p^{2}-m^{2})}\!\tilde{\mathcal{J}}_{\mathrm{I}_{2}}(p)\bigg)\bigg\}\\
\label{GeneFunctionalFreeMajoranaV2}
&\quad=\!\exp\bigg\{\!-\!\int\frac{d^{4}p}{(2\pi)^{4}}\!\big(\!\tilde{\mathcal{J}}^{\mathrm{I}_{2}}\!(-p)\frac{(\text{i}p_{\mu}\sigma^{\mu})_{\mathrm{I}_{2}\dot{\mathrm{I}}_{2}}}{(p^{2}-m^{2})}\!\tilde{\mathcal{J}}^{\dagger\dot{\mathrm{I}}_{2}}\!(-p)\!+\!\tilde{\mathcal{J}}_{\dot{\mathrm{I}}_{1}}^{\dagger}\!(p)\frac{\text{i}m\delta_{~~\dot{\mathrm{I}}_{2}}^{\dot{\mathrm{I}}_{1}}}{2(p^{2}-m^{2})}\!\tilde{\mathcal{J}}^{\dagger\dot{\mathrm{I}}_{2}}\!(-p)\!+\!\tilde{\mathcal{J}}^{\mathrm{I}_{1}}\!(-p)\frac{\text{i}m\delta_{\mathrm{I}_{1}}^{~~\mathrm{I}_{2}}}{2(p^{2}-m^{2})}\!\tilde{\mathcal{J}}_{\mathrm{I}_{2}}\!(p)\big)\bigg\}
\end{align}
\end{scriptsize}And then, the various two-point correlation functions for the free Majorana theory are evaluated as
\begin{align}
\nonumber
\langle 0\vert\mathcal{T}\{\psi_{\mathrm{I}_{1}}(x_{1})\psi_{\dot{\mathrm{I}}_{2}}^{\dagger}(x_{2})\}\vert0\rangle =&\bigg\{\frac{1}{W_{0}[0,0]}\int\frac{d^{4}q_{1}}{(2\pi)^{4}}\text{e}^{\text{i}q_{1}\cdot x_{1}}\frac{\int d^{4}q_{2}}{(2\pi)^{4}}\text{e}^{-\text{i}q_{2}\cdot x_{2}}\\
\label{CorrelaMajoPsiPsidagger}
&(\frac{\overrightarrow{\delta}}{\text{i}\delta\tilde{\mathcal{J}}^{\mathrm{I}_{1}}(q_{1})})W_{0}[\tilde{\mathcal{J}},\tilde{\mathcal{J}}^{\dagger}](\frac{\overleftarrow{\delta}}{\text{i}\delta\tilde{\mathcal{J}}^{\dagger\dot{\mathrm{I}}_{2}}(q_{2})})\bigg\}\bigg\vert_{\mathcal{J}=\mathcal{J}^{\dagger}=0}^{\text{connected}}\\
\nonumber
\langle 0\vert\mathcal{T}\{\psi^{\dagger\dot{\mathrm{I}}_{1}}(x_{1})\psi^{\mathrm{I}_{2}}(x_{2})\}\vert0\rangle=&\bigg\{\frac{1}{W_{0}[0,0]}\int\frac{d^{4}q_{1}}{(2\pi)^{4}}\text{e}^{-\text{i}q_{1}\cdot x_{1}}\int\frac{d^{4}q_{2}}{(2\pi)^{4}}\text{e}^{\text{i}q_{2}\cdot x_{2}}\\
\label{CorrelaMajoPsidaggerPsi}
&(\frac{\overrightarrow{\delta}}{\text{i}\delta\tilde{\mathcal{J}}_{\dot{\mathrm{I}}_{1}}^{\dagger}(q_{1})})W_{0}[\tilde{\mathcal{J}},\tilde{\mathcal{J}}^{\dagger}](\frac{\overleftarrow{\delta}}{\text{i}\delta\tilde{\mathcal{J}}_{\mathrm{I}_{2}}(q_{2})})\bigg\}\bigg\vert_{\mathcal{J}=\mathcal{J}^{\dagger}=0}^{\text{connected}}\\
\nonumber
\langle 0\vert\mathcal{T}\{\psi^{\dagger\dot{\mathrm{I}}_{1}}(x_{1})\psi_{\dot{\mathrm{I}}_{2}}^{\dagger}(x_{2})\}\vert0\rangle=&\bigg\{\frac{1}{W_{0}[0,0]}\int d^{4}q_{1}\text{e}^{-\text{i}q_{1}\cdot x_{1}}\int d^{4}q_{2}\text{e}^{-\text{i}q_{2}\cdot x_{2}}\\
\label{CorrelaMajoPsidaggerPsidagger}
&(\frac{\overrightarrow{\delta}}{\text{i}\delta\tilde{\mathcal{J}}_{\dot{\mathrm{I}}_{1}}^{\dagger}(q_{1})})W_{0}[\tilde{\mathcal{J}},\tilde{\mathcal{J}}^{\dagger}](\frac{\overleftarrow{\delta}}{\text{i}\delta\tilde{\mathcal{J}}^{\dagger\dot{\mathrm{I}}_{2}}(q_{2})})\bigg\}\bigg\vert_{\mathcal{J}=\mathcal{J}^{\dagger}=0}^{\text{connect}}\\
\nonumber
\langle 0\vert\mathcal{T}\{\psi_{\mathrm{I}_{1}}(x_{1})\psi^{\mathrm{I}_{2}}(x_{2})\}\vert0\rangle=&\bigg\{\frac{1}{W_{0}[0,0]}\int d^{4}q_{1}\text{e}^{\text{i}q_{1}\cdot x_{1}}\int d^{4}q_{2}\text{e}^{\text{i}q_{2}\cdot x_{2}}\\
\label{CorrelaMajoPsiPsi}
&(\frac{\overrightarrow{\delta}}{\text{i}\delta\tilde{\mathcal{J}}^{\mathrm{I}_{1}}(q_{1})})W_{0}[\tilde{\mathcal{J}},\tilde{\mathcal{J}}^{\dagger}](\frac{\overleftarrow{\delta}}{\text{i}\delta\tilde{\mathcal{J}}_{\mathrm{I}_{2}}(q_{2})})\bigg\}\bigg\vert_{\mathcal{J}=\mathcal{J}^{\dagger}=0}^{\text{connected}}
\end{align}After plugging \eqref{GeneFunctionalFreeMajoranaV2} into \eqref{CorrelaMajoPsiPsidagger}–\eqref{CorrelaMajoPsiPsi}, and using the following functional derivatives with respect to the spinor indices (neglecting momentum for brevity; the precise computational rules should follow \eqref{CurrentFuncDerivaMajoraMomentum}),
\begin{small}
\begin{align}
\nonumber
2\delta_{\mathrm{I}_{1}}^{~~\mathrm{I}_{2}}&\!=\!\frac{\overrightarrow{\delta}}{\delta\tilde{\mathcal{J}}^{\mathrm{I}_{1}}}(\tilde{\mathcal{J}}^{\mathrm{J}_{1}}\!\delta_{\mathrm{J}_{1}}^{~~\mathrm{J}_{2}}\!\tilde{\mathcal{J}}_{\mathrm{J}_{2}})\frac{\overleftarrow{\delta}}{\delta\tilde{\mathcal{J}}_{\mathrm{I}_{2}}}\!=\!\big((\frac{\delta}{\delta\tilde{\mathcal{J}}^{\mathrm{I}_{1}}}\tilde{\mathcal{J}}^{\mathrm{J}_{1}})\delta_{\mathrm{J}_{1}}^{~~\mathrm{J}_{2}}\tilde{\mathcal{J}}_{\mathrm{J}_{2}}\!-\!\tilde{\mathcal{J}}^{\mathrm{J}_{1}}\!\delta_{\mathrm{J}_{1}}^{~~\mathrm{J}_{2}}(\frac{\delta}{\delta\tilde{\mathcal{J}}^{\mathrm{I}_{1}}}\varepsilon_{\mathrm{J}_{2}\mathrm{J}_{3}}\!\tilde{\mathcal{J}}^{\mathrm{J}_{3}})\big)\frac{\overleftarrow{\delta}}{\delta\tilde{\mathcal{J}}_{\mathrm{I}_{2}}}\\
&\overset{\text{or}}{=\!=}\frac{\overrightarrow{\delta}}{\delta\tilde{\mathcal{J}}^{\mathrm{I}_{1}}}\big(-(\varepsilon^{\mathrm{J}_{1}\mathrm{J}_{3}}\tilde{\mathcal{J}}_{\mathrm{J}_{3}}\frac{\overleftarrow{\delta}}{\delta\tilde{\mathcal{J}}_{\mathrm{I}_{2}}})\delta_{\mathrm{J}_{1}}^{~~\mathrm{J}_{2}}\tilde{\mathcal{J}}_{\mathrm{J}_{2}}+\tilde{\mathcal{J}}^{\mathrm{J}_{1}}\delta_{\mathrm{J}_{1}}^{~~\mathrm{J}_{2}}(\tilde{\mathcal{J}}_{\mathrm{J}_{2}}\frac{\overleftarrow{\delta}}{\delta\tilde{\mathcal{J}}_{\mathrm{I}_{2}}})\big)\\
\nonumber
2\delta_{~~\dot{\mathrm{I}}_{2}}^{\dot{\mathrm{I}}_{1}}&=\frac{\overrightarrow{\delta}}{\delta\tilde{\mathcal{J}}_{\dot{\mathrm{I}}_{1}}^{\dagger}}\big(\tilde{\mathcal{J}}_{\dot{\mathrm{J}}_{1}}^{\dagger}\delta_{~~\dot{\mathrm{J}}_{2}}^{\dot{\mathrm{J}}_{1}}\tilde{\mathcal{J}}^{\dagger\dot{\mathrm{J}}_{2}}\big)\frac{\overleftarrow{\delta}}{\delta\tilde{\mathcal{J}}^{\dagger\dot{\mathrm{I}}_{2}}}\!=\!\big(\frac{\delta}{\delta\tilde{\mathcal{J}}_{\dot{\mathrm{I}}_{1}}^{\dagger}}\tilde{\mathcal{J}}_{\dot{\mathrm{J}}_{1}}^{\dagger}\delta_{~~\dot{\mathrm{J}}_{2}}^{\dot{\mathrm{J}}_{1}}\!\tilde{\mathcal{J}}^{\dagger\dot{\mathrm{J}}_{2}}\!-\!\tilde{\mathcal{J}}_{\dot{\mathrm{J}}_{1}}^{\dagger}\delta_{~~\dot{\mathrm{J}}_{2}}^{\dot{\mathrm{J}}_{1}}\varepsilon^{\dot{\mathrm{J}}_{2}\dot{\mathrm{J}}_{3}}\frac{\delta}{\delta\tilde{\mathcal{J}}_{\dot{\mathrm{I}}_{1}}^{\dagger}}\tilde{\mathcal{J}}_{\dot{\mathrm{J}}_{3}}^{\dagger}\big)\frac{\overleftarrow{\delta}}{\delta\tilde{\mathcal{J}}^{\dagger\dot{\mathrm{I}}_{2}}}\\
&\overset{\text{or}}{=\!=}\frac{\overrightarrow{\delta}}{\delta\tilde{\mathcal{J}}_{\dot{\mathrm{I}}_{1}}^{\dagger}}\big(-(\varepsilon_{\dot{\mathrm{J}}_{1}\dot{\mathrm{J}}_{3}}\tilde{\mathcal{J}}^{\dagger\dot{\mathrm{J}}_{3}}\frac{\overleftarrow{\delta}}{\delta\tilde{\mathcal{J}}^{\dagger\dot{\mathrm{I}}_{2}}})\delta_{~~\dot{\mathrm{J}}_{2}}^{\dot{\mathrm{J}}_{1}}\tilde{\mathcal{J}}^{\dagger\dot{\mathrm{J}}_{2}}+\tilde{\mathcal{J}}_{\dot{\mathrm{J}}_{1}}^{\dagger}\delta_{~~\dot{\mathrm{J}}_{2}}^{\dot{\mathrm{J}}_{1}}(\tilde{\mathcal{J}}^{\dagger\dot{\mathrm{J}}_{2}}\frac{\overleftarrow{\delta}}{\delta\tilde{\mathcal{J}}^{\dagger\dot{\mathrm{I}}_{2}}})\big)
\end{align}
\end{small}and
\begin{small}
	\begin{align}
\nonumber
(\text{i}p_{\mu}\sigma^{\mu})_{\mathrm{I}_{1}\dot{\mathrm{I}}_{2}}&=\frac{\overrightarrow{\delta}}{\delta\tilde{\mathcal{J}}^{\mathrm{I}_{1}}}\tilde{\mathcal{J}}^{\mathrm{J}_{1}}(\text{i}p_{\mu}\sigma^{\mu})_{\mathrm{J}_{1}\dot{\mathrm{J}}_{2}}\tilde{\mathcal{J}}^{\dagger\dot{\mathrm{J}}_{2}}\frac{\overleftarrow{\delta}}{\delta\tilde{\mathcal{J}}^{\dagger\dot{\mathrm{I}}_{2}}}\\
&\overset{\text{or}}{=\!=}\!-\frac{\overrightarrow{\delta}}{\delta\tilde{\mathcal{J}}^{\mathrm{I}_{1}}}\big(\!\tilde{\mathcal{J}}_{\dot{\mathrm{J}}_{2}}^{\dagger}(\text{i}p_{\mu}\bar{\sigma}^{\mu})^{\dot{\mathrm{J}}_{2}\mathrm{J}_{1}}\tilde{\mathcal{J}}_{\mathrm{J}_{1}}\!\big)\frac{\overleftarrow{\delta}}{\delta\mathcal{J}^{\dagger\dot{\mathrm{I}}_{2}}}=\!-\varepsilon_{\dot{\mathrm{I}}_{2}\dot{\mathrm{J}}_{2}}(\text{i}p_{\mu}\bar{\sigma}^{\mu})^{\dot{\mathrm{J}}_{2}\mathrm{J}_{1}}\!\varepsilon_{\mathrm{J}_{1}\mathrm{I}_{1}}\\
\nonumber
(\text{i}p_{\mu}\bar{\sigma}^{\mu})^{\dot{\mathrm{I}}_{1}\mathrm{I}_{2}}&=\frac{\overrightarrow{\delta}}{\delta\tilde{\mathcal{J}}_{\dot{\mathrm{I}}_{1}}^{\dagger}}\tilde{\mathcal{J}}_{\dot{\mathrm{J}}_{1}}^{\dagger}(\text{i}p_{\mu}\bar{\sigma}^{\mu})^{\dot{\mathrm{J}}_{1}\mathrm{J}_{1}}\tilde{\mathcal{J}}_{\mathrm{J}_{1}}\frac{\overleftarrow{\delta}}{\delta\tilde{\mathcal{J}}_{\mathrm{I}_{2}}}\\
&\overset{\text{or}}{=\!=}-\frac{\overrightarrow{\delta}}{\delta\tilde{\mathcal{J}}_{\dot{\mathrm{I}}_{1}}^{\dagger}}\big(\tilde{\mathcal{J}}^{\mathrm{J}_{1}}(\text{i}p_{\mu}\sigma^{\mu})_{\mathrm{J}_{1}\dot{\mathrm{J}}_{1}}\tilde{\mathcal{J}}^{\dagger\dot{\mathrm{J}}_{1}}\big)\frac{\overleftarrow{\delta}}{\delta\tilde{\mathcal{J}}_{\mathrm{I}_{2}}}=-\varepsilon^{\mathrm{I}_{2}\mathrm{J}_{1}}(\text{i}p_{\mu}\sigma^{\mu})_{\mathrm{J}_{1}\dot{\mathrm{J}}_{1}}\varepsilon^{\dot{\mathrm{J}}_{1}\dot{\mathrm{I}}_{1}}
\end{align}
\end{small}we could arrive at
\begin{align}
\label{TwoPointPsiPsidagger}
&\langle 0\vert\mathcal{T}\{\psi_{\mathrm{I}_{1}}(x_{1})\psi_{\dot{\mathrm{I}}_{2}}^{\dagger}(x_{2})\}\vert0\rangle=\int\frac{d^{4}p}{(2\pi)^{4}}\text{e}^{-\text{i}p\cdot(x_{1}-x_{2})}\frac{(\text{i}p_{\mu}\sigma^{\mu})_{\mathrm{I}_{1}\dot{\mathrm{I}}_{2}}}{p^{2}-m^{2}}\\
\label{TwoPointPsidaggerPsi}
&\langle 0\vert\mathcal{T}\{\psi^{\dagger\dot{\mathrm{I}}_{1}}(x_{1})\psi^{\mathrm{I}_{2}}(x_{2})\}\vert0\rangle=\int\frac{d^{4}p}{(2\pi)^{4}}\text{e}^{-\text{i}p\cdot(x_{1}-x_{2})}\frac{(\text{i}p_{\mu}\bar{\sigma}^{\mu})^{\dot{\mathrm{I}}_{1}\mathrm{I}_{2}}}{p^{2}-m^{2}}\\
\label{TwoPointPsidaggerPsidagger}
&\langle 0\vert\mathcal{T}\{\psi^{\dagger\dot{\mathrm{I}}_{1}}(x_{1})\psi_{\dot{\mathrm{I}}_{2}}^{\dagger}(x_{2})\}\vert0\rangle=\int\frac{d^{4}p}{(2\pi)^{4}}\text{e}^{-\text{i}p\cdot(x_{1}-x_{2})}\frac{\text{i}m\delta_{~~\dot{\mathrm{I}}_{2}}^{\dot{\mathrm{I}}_{1}}}{p^{2}-m^{2}}\\
\label{TwoPointPsiPsi}
&\langle 0\vert\mathcal{T}\{\psi_{\mathrm{I}_{1}}(x_{1})\psi^{\mathrm{I}_{2}}(x_{2})\}\vert0\rangle=\int\frac{d^{4}p}{(2\pi)^{4}}\text{e}^{-\text{i}p\cdot(x_{1}-x_{2})}\frac{\text{i}m\delta_{\mathrm{I}_{1}}^{~~\mathrm{I}_{2}}}{p^{2}-m^{2}}
\end{align}Regarding the momentum-space representations of the Feynman diagrams, the propagators corresponding to the various two-point correlations \eqref{TwoPointPsiPsidagger}–\eqref{TwoPointPsiPsi} are summarized in Fig. \ref{VariousTwoPointCorrelationLabel}.
\subsection{Feynman rules and two-body decay rate corresponding to interaction $S_{\text{int}}^{(\varphi\psi\psi)}$ \label{CubicInflaMajorana}}

Subsequently, we employ the path integral formalism exhibited in the previous section to derive the Feynman rules associated with the three-point interaction \eqref{YukawaInflatonMajorana}, which involves Majorana fermions. With inclusion of the interaction \eqref{YukawaInflatonMajorana}, the generating functional expanded at $\mathcal{O}(\kappa)$ level is given by
\begin{small}
\begin{align}
\nonumber
&W[J_{(\varphi)},\mathcal{J}^{\dagger},\mathcal{J}]=\mathcal{N}\int\mathcal{D}\psi\mathcal{D}\psi^{\dagger}\mathcal{D}\varphi~\text{e}^{\text{i}\int d^{4}y\big(\frac{\text{i}\kappa}{2}\varphi(\psi^{\dagger}\bar{\sigma}^{\mu}\partial_{\mu}\psi-\partial_{\mu}\psi^{\dagger}\bar{\sigma}^{\mu}\psi)\big)}\\
\nonumber
&\quad\times \text{e}^{\text{i}\int d^{4}x\big(\frac{1}{2}(\text{i}\psi^{\dagger}\bar{\sigma}^{\mu}\partial_{\mu}\psi+\text{i}\psi\sigma^{\mu}\partial_{\mu}\psi^{\dagger}-m\psi\psi-m\psi^{\dagger}\psi^{\dagger})-\frac{1}{2}\varphi(\Box+m^{2})\varphi+\mathcal{J}\cdot\psi+\psi^{\dagger}\cdot\mathcal{J}^{\dagger}+\varphi J_{(\varphi)}\big)}\\
\nonumber
&\backsimeq W_{0}[J_{(\varphi)}]W_{0}[\mathcal{J}^{\dagger},\mathcal{J}]\!-\frac{\kappa}{2}(\bar{\sigma}^{\mu})^{\dot{\mathrm{I}}_{1}\mathrm{J}_{1}}\!\int d^{4}y(\frac{1}{\text{i}}\frac{\delta}{\delta J_{(\varphi)}(y)})W_{0}[J_{(\varphi)}]\!\times\!\bigg((\frac{1}{\text{i}}\frac{\overrightarrow{\delta}}{\delta\mathcal{J}^{\dagger\dot{\mathrm{I}}_{1}}(y)})W_{0}[\mathcal{J}^{\dagger},\mathcal{J}]\partial_{\mu}(\frac{1}{\text{i}}\frac{\overleftarrow{\delta}}{\delta\mathcal{J}^{\mathrm{J}_{1}}(y)})\\
\label{YukawaOneScalarTwoMajorana}
&\quad-\partial_{\mu}(\frac{1}{\text{i}}\frac{\overrightarrow{\delta}}{\delta\mathcal{J}^{\dagger\dot{\mathrm{I}}_{1}}(y)})W_{0}[\mathcal{J}^{\dagger},\mathcal{J}](\frac{1}{\text{i}}\frac{\overleftarrow{\delta}}{\delta\mathcal{J}^{\mathrm{J}_{1}}(y)})\bigg)+\mathcal{O}(\kappa^{2})
\end{align}
\end{small}After substituting \eqref{YukawaOneScalarTwoMajorana} into the relevant three-point correlation functions involving one scalar and two Majorana fermions, we obtain
\begin{small}
\begin{align}
\nonumber
&\langle\Omega\vert\mathcal{T}\{\psi_{\mathrm{I}_{1}}(x_{1})\psi^{\mathrm{I}_{2}}(x_{2})\varphi(x_{3})\}\vert\Omega\rangle\big\vert_{\mathcal{O}(\kappa)}\!=\!\big\{\frac{1}{W[0,0,0]}\frac{\overrightarrow{\delta}}{\text{i}\delta\mathcal{J}^{\mathrm{I}_{1}}(x_{1})}\frac{\delta W[J_{(\varphi)},\mathcal{J}^{\dagger},\mathcal{J}]\big\vert_{\mathcal{O}(\kappa)}}{\text{i}\delta J_{(\varphi)}(x_{3})}\frac{\overleftarrow{\delta}}{\text{i}\delta\mathcal{J}_{\mathrm{I}_{2}}(x_{2})}\big\}\bigg\vert_{J_{(\varphi)}\!=\!\mathcal{J}^{\dagger}\!=\!\mathcal{J}\!=0}^{\text{connected}}\\
\nonumber
&=\bigg\{\frac{\text{i}\kappa}{2}(\bar{\sigma}^{\mu})^{\dot{\tilde{\mathrm{I}}}_{1}\tilde{\mathrm{J}}_{1}}\int d^{4}y\int\frac{d^{4}q_{3}}{(2\pi)^{4}}\int\frac{d^{4}k_{3}}{(2\pi)^{4}}\text{e}^{\text{i}q_{3}\cdot x_{3}}\text{e}^{\text{i}k_{3}\cdot y}\big(\frac{\delta}{\delta\tilde{J}_{(\varphi)}(q_{3})}\frac{\delta}{\delta\tilde{J}_{(\varphi)}(k_{3})}W_{0}[\tilde{J}_{(\varphi)}]\big)\\
\nonumber
&\times\int\frac{d^{4}q_{1}}{(2\pi)^{4}}\int\frac{d^{4}q_{2}}{(2\pi)^{4}}\int\frac{d^{4}k_{1}}{(2\pi)^{4}}\int\frac{d^{4}k_{2}}{(2\pi)^{4}}\text{e}^{\text{i}q_{2}\cdot x_{2}}\text{e}^{\text{i}q_{1}\cdot x_{1}}\text{e}^{\text{i}(k_{1}-k_{2})\cdot y}(k_{1}+k_{2})_{\mu}\\
\label{ThreePointCorrelationI1I2}
&\times\big(\frac{\overrightarrow{\delta}}{\delta\tilde{\mathcal{J}}^{\mathrm{I}_{1}}(q_{1})}\frac{\overrightarrow{\delta}}{\delta\tilde{\mathcal{J}}^{\dagger\dot{\tilde{\mathrm{I}}}_{1}}(k_{2})}W_{0}[\tilde{\mathcal{J}}^{\dagger},\tilde{\mathcal{J}}]\frac{\overleftarrow{\delta}}{\delta\tilde{\mathcal{J}}^{\tilde{\mathrm{J}}_{1}}(k_{1})}\frac{\overleftarrow{\delta}}{\delta\tilde{\mathcal{J}}_{\mathrm{I}_{2}}(q_{2})}\big)\bigg\}\bigg\vert_{\tilde{J}_{(\varphi)}=\tilde{\mathcal{J}}^{\dagger}=\tilde{\mathcal{J}}=0}^{\text{connected}}
\end{align}
\end{small}
\begin{small}
\begin{align}
\nonumber
&\langle\Omega\vert\mathcal{T}\{\psi^{\dagger\dot{\mathrm{I}}_{1}}(x_{1})\psi^{\mathrm{I}_{2}}(x_{2})\varphi(x_{3})\}\vert\Omega\rangle\!\big\vert_{\mathcal{O}(\kappa)}\!=\!\big\{\!\frac{1}{W[0,0,0]}\frac{\overrightarrow{\delta}}{\text{i}\delta\mathcal{J}_{\dot{\mathrm{I}}_{1}}^{\dagger}(x_{1})}\frac{\delta W[J_{(\varphi)},\mathcal{J}^{\dagger},\mathcal{J}]\big\vert_{\mathcal{O}(\kappa)}}{\text{i}\delta J_{(\varphi)}(x_{3})}\frac{\overleftarrow{\delta}}{\text{i}\delta\mathcal{J}_{\mathrm{I}_{2}}(x_{2})}\!\big\}\!\bigg\vert_{J_{(\varphi)}\!=\!\mathcal{J}^{\dagger}\!=\!\mathcal{J}\!=0}^{\text{connected}}\\
\nonumber
&=\bigg\{\frac{\text{i}\kappa}{2}(\bar{\sigma}^{\mu})^{\dot{\tilde{\mathrm{I}}}_{1}\tilde{\mathrm{J}}_{1}}\int d^{4}y\int\frac{d^{4}q_{3}}{(2\pi)^{4}}\int\frac{d^{4}k_{3}}{(2\pi)^{4}}\text{e}^{\text{i}q_{3}\cdot x_{3}}\text{e}^{\text{i}k_{3}\cdot y}\frac{\delta}{\delta\tilde{J}_{(\varphi)}(q_{3})}\frac{\delta}{\delta\tilde{J}_{(\varphi)}(k_{3})}W_{0}[\tilde{J}_{(\varphi)}]\\
\nonumber
&\times\int\frac{d^{4}q_{1}}{(2\pi)^{4}}\int\frac{d^{4}q_{2}}{(2\pi)^{4}}\int\frac{d^{4}k_{1}}{(2\pi)^{4}}\int\frac{d^{4}k_{2}}{(2\pi)^{4}}\text{e}^{-\text{i}q_{1}\cdot x_{1}}\text{e}^{\text{i}q_{2}\cdot x_{2}}\text{e}^{\text{i}(k_{1}-k_{2})\cdot y}(k_{1}+k_{2})_{\mu}\\
\label{ThreePointCorrelationDaggerI1I2}
&\times\big(\frac{\overrightarrow{\delta}}{\delta\tilde{\mathcal{J}}_{\dot{\mathrm{I}}_{1}}^{\dagger}(q_{1})}\frac{\overrightarrow{\delta}}{\delta\tilde{\mathcal{J}}^{\dagger\dot{\tilde{\mathrm{I}}}_{1}}(k_{2})}W_{0}[\tilde{\mathcal{J}}^{\dagger},\tilde{\mathcal{J}}]\frac{\overleftarrow{\delta}}{\delta\tilde{\mathcal{J}}^{\tilde{\mathrm{J}}_{1}}(k_{1})}\frac{\overleftarrow{\delta}}{\delta\tilde{\mathcal{J}}_{\mathrm{I}_{2}}(q_{2})}\big)\bigg\}\bigg\vert_{\tilde{J}_{(\varphi)}=\tilde{\mathcal{J}}^{\dagger}=\tilde{\mathcal{J}}=0}^{\text{connected}}
\end{align}	
\end{small}

\begin{small}
\begin{align}
\nonumber
&\langle\Omega\vert\mathcal{T}\{\psi^{\dagger\dot{\mathrm{I}}_{1}}(x_{1})\psi_{\dot{\mathrm{I}}_{2}}^{\dagger}(x_{2})\varphi(x_{3})\}\vert\Omega\rangle\!\big\vert_{\mathcal{O}(\kappa)}\!=\!\big\{\!\frac{1}{W[0,0,0]}\frac{\overrightarrow{\delta}}{\text{i}\delta\mathcal{J}_{\dot{\mathrm{I}}_{1}}^{\dagger}\!(x_{1})}\frac{\delta W[J_{(\varphi)},\mathcal{J}^{\dagger},\mathcal{J}]\!\big\vert_{\mathcal{O}(\kappa)}}{\text{i}\delta J_{(\varphi)}(x_{3})}\frac{\overleftarrow{\delta}}{\text{i}\delta\mathcal{J}^{\dagger\dot{\mathrm{I}}_{2}}\!(x_{2})}\big\}\bigg\vert_{J_{(\varphi)}\!=\!\mathcal{J}^{\dagger}\!=\!\mathcal{J}\!=\!0}^{\text{connected}}\\
\nonumber
&=\bigg\{\frac{\text{i}\kappa}{2}(\bar{\sigma}^{\mu})^{\dot{\tilde{\mathrm{I}}}_{1}\tilde{\mathrm{J}}_{1}}\int d^{4}y\int\frac{d^{4}q_{3}}{(2\pi)^{4}}\int\frac{d^{4}k_{3}}{(2\pi)^{4}}\text{e}^{\text{i}q_{3}\cdot x_{3}}\text{e}^{\text{i}k_{3}\cdot y}\frac{\delta}{\delta\tilde{J}_{(\varphi)}(q_{3})}\frac{\delta}{\delta\tilde{J}_{(\varphi)}(k_{3})}W_{0}[\tilde{J}_{(\varphi)}]\\
\nonumber
&\times\int\frac{d^{4}q_{1}}{(2\pi)^{4}}\int\frac{d^{4}q_{2}}{(2\pi)^{4}}\int\frac{d^{4}k_{1}}{(2\pi)^{4}}\int\frac{d^{4}k_{2}}{(2\pi)^{4}}\text{e}^{-\text{i}q_{1}\cdot x_{1}}\text{e}^{-\text{i}q_{2}\cdot x_{2}}\text{e}^{\text{i}(k_{1}-k_{2})\cdot y}(k_{1}+k_{2})_{\mu}\\
\label{ThreePointCorrelationDaggerI1DaggerI2}
&\times\big(\frac{\overrightarrow{\delta}}{\delta\tilde{\mathcal{J}}_{\dot{\mathrm{I}}_{1}}^{\dagger}(q_{1})}\frac{\overrightarrow{\delta}}{\delta\tilde{\mathcal{J}}^{\dagger\dot{\tilde{\mathrm{I}}}_{1}}(k_{2})}W_{0}[\tilde{\mathcal{J}}^{\dagger},\tilde{\mathcal{J}}]\frac{\overleftarrow{\delta}}{\delta\tilde{\mathcal{J}}^{\tilde{\mathrm{J}}_{1}}(k_{1})}\frac{\overleftarrow{\delta}}{\delta\tilde{\mathcal{J}}^{\dagger\dot{\mathrm{I}}_{2}}(q_{2})}\big)\bigg\}\bigg\vert_{\tilde{J}_{(\varphi)}\!=\!\tilde{\mathcal{J}}^{\dagger}\!=\!\tilde{\mathcal{J}}\!=\!0}^{\text{connected}}
\end{align}	
\end{small}

\begin{figure}[ht]
	\begin{center}
		\includegraphics[scale=0.4]{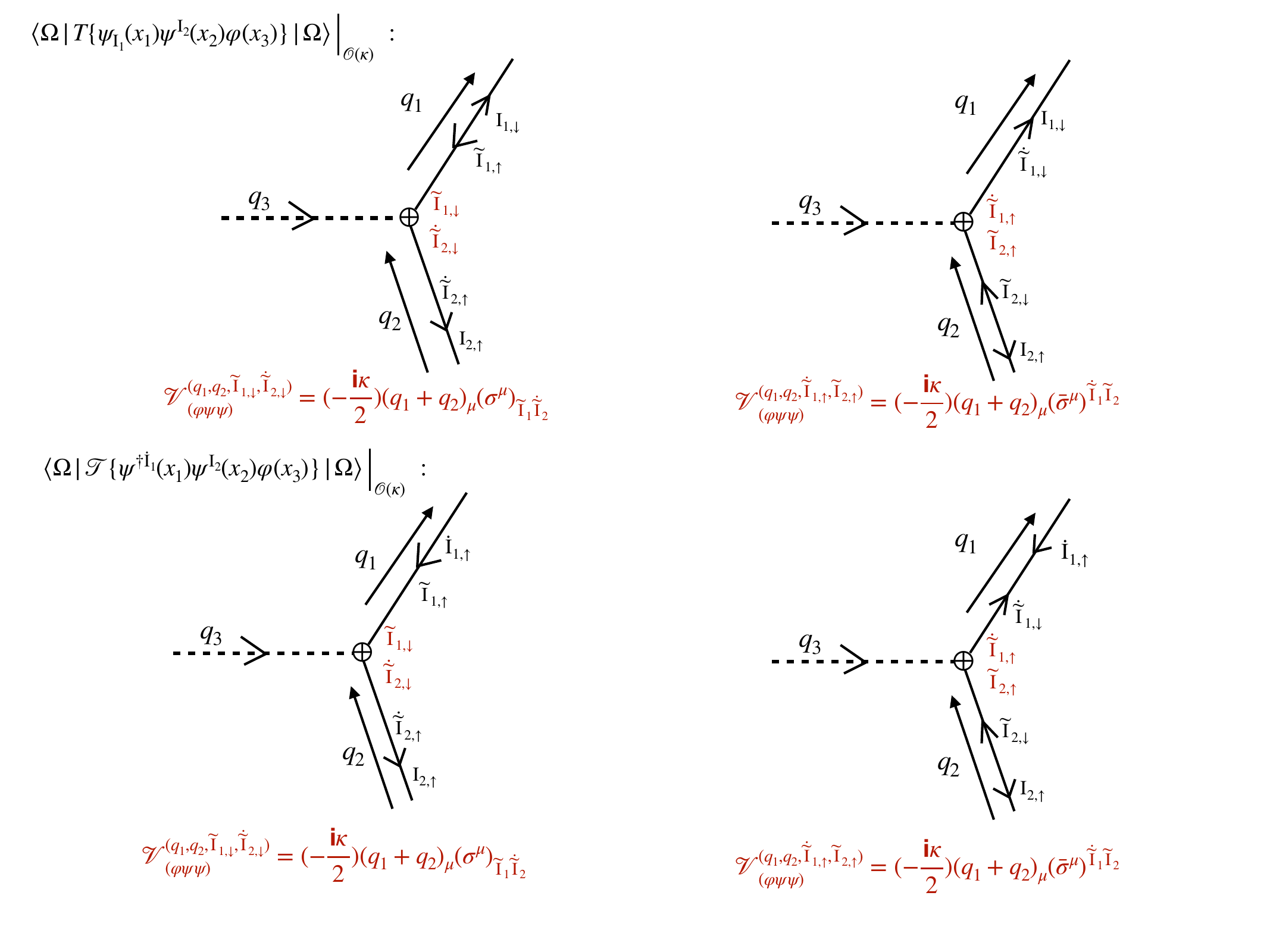}\\
		\includegraphics[scale=0.4]{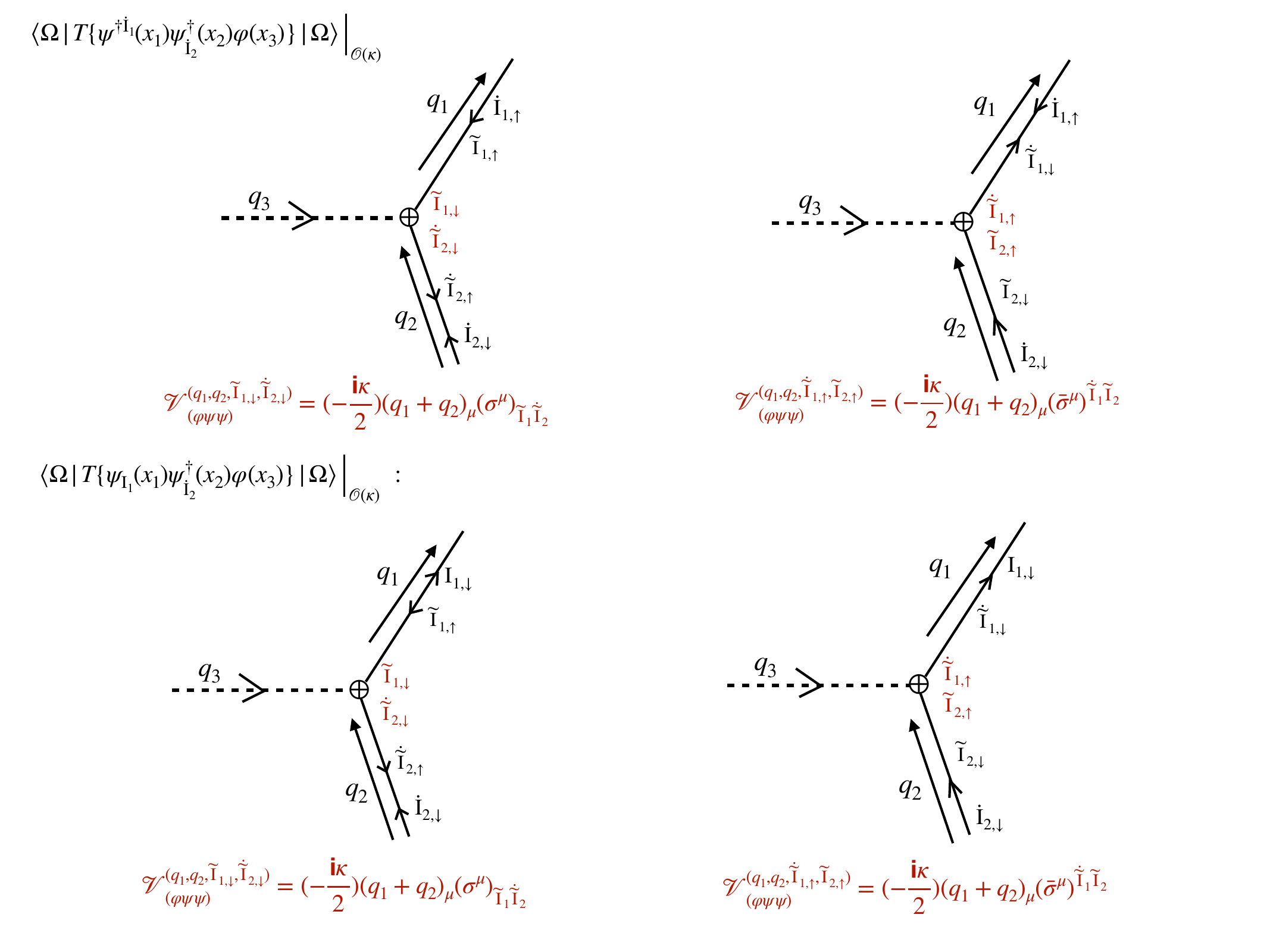}
		\caption{Feynman rules describing the cubic interaction between one inflaton and two Majorana fermions, as encoded in Eq. \eqref{YukawaInflatonMajoranaMain}. These rules are extracted from the three-point correlation functions given in \eqref{ThreePointCorrelationI1I2Result}–\eqref{ThreePointCorrelationI1DaggerI2Result}. Note that each of these three-point functions gives rise to two distinct types of interaction vertices in momentum space, as indicated by the red labels.}
		\label{FigPsiI1PsiI2Inflatonpsipsi}
	\end{center}
\end{figure}

\begin{small}
\begin{align}
\nonumber
&\langle\Omega\vert\!\mathcal{T}\{\!\psi_{\mathrm{I}_{1}}(x_{1})\psi_{\dot{\mathrm{I}}_{2}}^{\dagger}(x_{2})\varphi(x_{3})\!\}\!\vert\Omega\rangle\!\big\vert_{\mathcal{O}(\kappa)}\!=\!\big\{\frac{1}{W[0,0,0]}\frac{\overrightarrow{\delta}}{\text{i}\delta\!\mathcal{J}^{\!\mathrm{I}_{1}}(x_{1})}\frac{\delta W[J_{(\varphi)},\mathcal{J}^{\dagger},\mathcal{J}]\!\big\vert_{\mathcal{O}(\kappa)}}{\text{i}\delta J_{(\varphi)}(x_{3})}\frac{\overleftarrow{\delta}}{\text{i}\delta\!\mathcal{J}^{\dagger\dot{\mathrm{I}}_{2}}(x_{2})}\big\}\bigg\vert_{J_{(\varphi)}\!=\!\mathcal{J}^{\dagger}\!=\!\mathcal{J}\!=\!0}^{\text{connected}}\\
\nonumber
&=\bigg\{\frac{\text{i}\kappa}{2}(\bar{\sigma}^{\mu})^{\dot{\tilde{\mathrm{I}}}_{1}\tilde{\mathrm{J}}_{1}}\int d^{4}y\,\int\frac{d^{4}q_{3}}{(2\pi)^{4}}\int\frac{d^{4}k_{3}}{(2\pi)^{4}}\text{e}^{\text{i}q_{3}\cdot x_{3}}\text{e}^{\text{i}k_{3}\cdot y}\frac{\delta}{\delta\tilde{J}_{(\varphi)}(q_{3})}\frac{\delta}{\delta\tilde{J}_{(\varphi)}(k_{3})}W_{0}[\tilde{J}_{(\varphi)}]\\
\nonumber
&\times\int\frac{d^{4}q_{1}}{(2\pi)^{4}}\int\frac{d^{4}q_{2}}{(2\pi)^{4}}\int\frac{d^{4}k_{1}}{(2\pi)^{4}}\int\frac{d^{4}k_{2}}{(2\pi)^{4}}\text{e}^{\text{i}q_{1}\cdot x_{1}}\text{e}^{-\text{i}q_{2}\cdot x_{2}}\text{e}^{\text{i}(k_{1}-k_{2})\cdot y}(k_{1}+k_{2})_{\mu}\\
\label{ThreePointCorrelationI1DaggerI2}
&\times\big(\frac{\overrightarrow{\delta}}{\delta\tilde{\mathcal{J}}^{\mathrm{I}_{1}}(q_{1})}\frac{\overrightarrow{\delta}}{\delta\tilde{\mathcal{J}}^{\dagger\dot{\tilde{\mathrm{I}}}_{1}}(k_{2})}W_{0}[\tilde{\mathcal{J}}^{\dagger},\tilde{\mathcal{J}}]\frac{\overleftarrow{\delta}}{\delta\tilde{\mathcal{J}}^{\tilde{\mathrm{J}}_{1}}(k_{1})}\frac{\overleftarrow{\delta}}{\delta\tilde{\mathcal{J}}^{\dagger\dot{\mathrm{I}}_{2}}(q_{2})}\big)\bigg\}\bigg\vert_{\tilde{J}_{(\varphi)}=\tilde{\mathcal{J}}^{\dagger}=\tilde{\mathcal{J}}=0}^{\text{connected}}
\end{align}	
\end{small}Note that the Majorana-type functional derivatives corresponding to the connected Feynmann diagrams in momentum space  in \eqref{ThreePointCorrelationI1I2}–\eqref{ThreePointCorrelationI1DaggerI2} can be explicitly expanded as follows
\begin{align}
	\nonumber
	&\quad\frac{\overrightarrow{\delta}}{\delta\tilde{\mathcal{J}}^{\mathrm{I}_{1}}(q_{1})}\frac{\overrightarrow{\delta}}{\delta\tilde{\mathcal{J}}^{\dagger\dot{\tilde{\mathrm{I}}}_{1}}(k_{2})}W_{0}[\tilde{\mathcal{J}}_{(\psi^{\dagger})}^{\dagger},\tilde{\mathcal{J}}_{(\psi)}]\frac{\overleftarrow{\delta}}{\delta\tilde{\mathcal{J}}^{\tilde{\mathrm{J}}_{1}}(k_{1})}\frac{\overleftarrow{\delta}}{\delta\tilde{\mathcal{J}}_{\mathrm{I}_{2}}(q_{2})}\bigg\vert_{\tilde{\mathcal{J}}^{\dagger}=\tilde{\mathcal{J}}=0}^{\text{connected}}\\
	\nonumber
	&=(2\pi)^{4\times2}\big\{\frac{\varepsilon_{\dot{\tilde{\mathrm{I}}}_{1}\dot{\mathrm{I}}_{3}}(\text{i}k_{2,\alpha}\bar{\sigma}^{\alpha})^{\dot{\mathrm{I}}_{3}\mathrm{I}_{3}}\varepsilon_{\mathrm{I}_{3}\mathrm{I}_{1}}}{k_{2}^{2}-m^{2}}\delta^{4}(q_{1}-k_{2})\times\frac{\text{i}m\delta_{\tilde{\mathrm{J}}_{1}}^{~~\mathrm{I}_{2}}}{k_{1}^{2}-m^{2}}\delta^{4}(-q_{2}-k_{1})\\
	\label{FuncDerivaI1DaggI2I3I4}
&-\frac{\varepsilon_{\dot{\tilde{\mathrm{I}}}_{1}\dot{\mathrm{I}}_{3}}(\text{i}k_{2,\beta}\bar{\sigma}^{\beta})^{\dot{\mathrm{I}}_{3}\mathrm{I}_{2}}}{k_{2}^{2}-m^{2}}\delta^{4}(q_{2}-k_{2})\times\frac{\text{i}m\varepsilon_{\tilde{\mathrm{J}}_{1}\mathrm{I}_{1}}}{k_{1}^{2}-m^{2}}\delta^{4}(-q_{1}-k_{1})\big\}
\end{align}

\begin{align}
\nonumber
	&\quad\frac{\overrightarrow{\delta}}{\delta\tilde{\mathcal{J}}_{\dot{\mathrm{I}}_{1}}^{\dagger}(q_{1})}\frac{\overrightarrow{\delta}}{\delta\tilde{\mathcal{J}}^{\dagger\dot{\tilde{\mathrm{I}}}_{1}}(k_{2})}W_{0}[\tilde{\mathcal{J}}^{\dagger},\tilde{\mathcal{J}}]\frac{\overleftarrow{\delta}}{\delta\tilde{\mathcal{J}}^{\tilde{\mathrm{J}}_{1}}(k_{1})}\frac{\overleftarrow{\delta}}{\delta\tilde{\mathcal{J}}_{\mathrm{I}_{2}}(q_{2})}\bigg\vert_{\tilde{\mathcal{J}}^{\dagger}=\tilde{\mathcal{J}}=0}^{\text{connected}}\\
	\nonumber
	&=(2\pi)^{4\times2}\big\{\frac{\text{i}m\delta_{~~\dot{\tilde{\mathrm{I}}}_{1}}^{\dot{\mathrm{I}}_{1}}}{(k_{2}^{2}-m^{2})}\delta^{4}(-k_{2}-q_{1})\times\frac{\text{i}m\delta_{\tilde{\mathrm{J}}_{1}}^{~~\mathrm{I}_{2}}}{(k_{1}^{2}-m^{2})}\delta^{4}(-k_{1}-q_{2})\\
	\label{FuncDerivaDaggI1DaggI2I3I4}
	&+\delta^{4}(k_{2}-q_{2})\frac{\varepsilon^{\mathrm{I}_{2}\mathrm{I}_{3}}(\text{i}k_{2,\alpha}\sigma^{\alpha})_{\mathrm{I}_{3}\dot{\tilde{\mathrm{I}}}_{1}}}{(k_{2}^{2}-m^{2})}\times\frac{(\text{i}k_{1,\beta}\sigma^{\beta})_{\tilde{\mathrm{J}}_{1}\dot{\mathrm{I}}_{4}}\varepsilon^{\dot{\mathrm{I}}_{4}\dot{\mathrm{I}}_{1}}}{(k_{1}^{2}-m^{2})}\delta^{4}(k_{1}-q_{1})\big\}
\end{align}

\begin{align}
	\nonumber
	\\
	\nonumber
	&\quad\frac{\overrightarrow{\delta}}{\delta\tilde{\mathcal{J}}_{\dot{\mathrm{I}}_{1}}^{\dagger}(q_{1})}\frac{\overrightarrow{\delta}}{\delta\tilde{\mathcal{J}}^{\dagger\dot{\tilde{\mathrm{I}}}_{1}}(k_{2})}W_{0}[\tilde{\mathcal{J}}^{\dagger},\tilde{\mathcal{J}}]\frac{\overleftarrow{\delta}}{\delta\tilde{\mathcal{J}}^{\tilde{\mathrm{J}}_{1}}(k_{1})}\frac{\overleftarrow{\delta}}{\delta\tilde{\mathcal{J}}^{\dagger\dot{\mathrm{I}}_{2}}(q_{2})}\bigg\vert_{\tilde{\mathcal{J}}^{\dagger}=\tilde{\mathcal{J}}=0}^{\text{connected}}\\
	\nonumber
	&=(2\pi)^{4\times2}\big\{\frac{\text{i}m\delta_{~~\,\dot{\tilde{\mathrm{I}}}_{1}}^{\dot{\mathrm{I}}_{1}}}{k_{2}^{2}-m^{2}}\delta^{4}(-k_{2}-q_{1})\times\frac{\varepsilon_{\dot{\mathrm{I}}_{2}\dot{\mathrm{I}}_{4}}(\text{i}k_{1,\beta}\bar{\sigma}^{\beta})^{\dot{\mathrm{I}}_{4}\mathrm{I}_{4}}\varepsilon_{\mathrm{I}_{4}\tilde{\mathrm{J}}_{1}}}{k_{1}^{2}-m^{2}}\delta^{4}(k_{1}-q_{2})\\
	\label{FuncDerivaDaggI1DaggI2I3DaggerI4}
	&+\frac{\text{i}m\varepsilon_{\dot{\tilde{\mathrm{I}}}_{1}\dot{\mathrm{I}}_{2}}}{k_{2}^{2}-m^{2}}\delta^{4}(-k_{2}-q_{2})\times\frac{(\text{i}k_{1,\beta}\bar{\sigma}^{\beta})^{\dot{\mathrm{I}}_{1}\mathrm{I}_{4}}\varepsilon_{\mathrm{I}_{4}\tilde{\mathrm{J}}_{1}}}{k_{1}^{2}-m^{2}}\delta^{4}(k_{1}-q_{1})\big\}
\end{align}
\begin{align}
\nonumber
&\quad\frac{\overrightarrow{\delta}}{\delta\tilde{\mathcal{J}}^{\mathrm{I}_{1}}(q_{1})}\frac{\overrightarrow{\delta}}{\delta\tilde{\mathcal{J}}^{\dagger\dot{\tilde{\mathrm{I}}}_{1}}(k_{2})}W_{0}[\tilde{\mathcal{J}}^{\dagger},\tilde{\mathcal{J}}]\frac{\overleftarrow{\delta}}{\delta\tilde{\mathcal{J}}^{\tilde{\mathrm{J}}_{1}}(k_{1})}\frac{\overleftarrow{\delta}}{\delta\tilde{\mathcal{J}}^{\dagger\dot{\mathrm{I}}_{2}}(q_{2})}\bigg\vert_{\tilde{\mathcal{J}}^{\dagger}=\tilde{\mathcal{J}}=0}^{\text{connected}}\\
\nonumber
&=(2\pi)^{4\times2}\big\{\frac{(\text{i}k_{2,\alpha}\sigma^{\alpha})_{\mathrm{I}_{1}\dot{\tilde{\mathrm{I}}}_{1}}}{k_{2}^{2}-m^{2}}\delta^{4}(k_{2}-q_{1})\times\frac{(\text{i}k_{1,\beta}\sigma^{\beta})_{\tilde{\mathrm{J}}_{1}\dot{\mathrm{I}}_{2}}}{k_{1}^{2}-m^{2}}\delta^{4}(k_{1}-q_{2})\\
\label{FuncDerivaI1DaggI2I3DaggerI4}
&-\frac{\text{i}m\varepsilon_{\dot{\tilde{\mathrm{I}}}_{1}\dot{\mathrm{I}}_{2}}}{k_{2}^{2}-m^{2}}\delta^{4}(-k_{2}-q_{2})\times\frac{\text{i}m\varepsilon_{\tilde{\mathrm{J}}_{1}\mathrm{I}_{1}}}{k_{1}^{2}-m^{2}}\delta^{4}(-k_{1}-q_{1})\big\}
\end{align}	
After substituting \eqref{FuncDerivaI1DaggI2I3I4}-\eqref{FuncDerivaI1DaggI2I3DaggerI4} into \eqref{ThreePointCorrelationI1I2}–\eqref{ThreePointCorrelationI1DaggerI2}, and performing a series of tedious calculations, we ultimately arrive at the following results
\begin{small}
\begin{align}
\nonumber
&\langle\Omega\vert\mathcal{T}\{\psi_{\mathrm{I}_{1}}(x_{1})\psi^{\mathrm{I}_{2}}(x_{2})\varphi(x_{3})\}\vert\Omega\rangle\big\vert_{\mathcal{O}(\kappa)}=\int\frac{d^{4}q_{1}}{(2\pi)^{4}}\int\frac{d^{4}q_{2}}{(2\pi)^{4}}\int\frac{d^{4}q_{3}}{(2\pi)^{4}}\int d^{4}y \text{e}^{-\text{i}q_{1}\cdot(x_{1}-y)}\text{e}^{\text{i}q_{2}\cdot(x_{2}-y)}\text{e}^{\text{i}q_{3}\cdot(x_{3}-y)}\\
\nonumber
&\quad\quad\quad\quad\quad\quad\quad\times\frac{\text{i}}{q_{3}^{2}-m^{2}} \big(\frac{\text{i}m\delta_{\mathrm{I}_{1}}^{~~\tilde{\mathrm{I}}_{1}}}{q_{1}^{2}-m^{2}}\times\underbrace{(-\frac{\text{i}\kappa}{2})(q_{1}+q_{2})_{\mu}(\sigma^{\mu})_{\tilde{\mathrm{I}}_{1}\dot{\tilde{\mathrm{I}}}_{2}}}_{\mathcal{V}_{(\varphi\psi\psi)}^{(q_{1},q_{2},\widetilde{\mathrm{I}}_{1,\downarrow},\dot{\widetilde{\mathrm{I}}}_{2,\downarrow})}}\times\frac{(\text{i}q_{2,\beta}\bar{\sigma}^{\beta})^{\dot{\tilde{\mathrm{I}}}_{2}\mathrm{I}_{2}}}{q_{2}^{2}-m^{2}}\\
\label{ThreePointCorrelationI1I2Result}
&\quad\quad\quad\quad\quad\quad\quad+\frac{(\text{i}q_{1,\alpha}\sigma^{\alpha})_{\mathrm{I}_{1}\dot{\tilde{\mathrm{I}}}_{1}}}{q_{1}^{2}-m^{2}}\times\underbrace{(-\frac{\text{i}\kappa}{2})(q_{1}+q_{2})_{\mu}(\bar{\sigma}^{\mu})^{\dot{\tilde{\mathrm{I}}}_{1}\tilde{\mathrm{I}}_{2}}}_{\mathcal{V}_{(\varphi\psi\psi)}^{(q_{1},q_{2},\dot{\widetilde{\mathrm{I}}}_{1,\uparrow},\widetilde{\mathrm{I}}_{2,\uparrow})}}\times\frac{\text{i}m\delta_{\tilde{\mathrm{I}}_{2}}^{~~\mathrm{I}_{2}}}{q_{2}^{2}-m^{2}}\big)\\
\nonumber
&\langle\Omega\vert\mathcal{T}\{\psi^{\dagger\dot{\mathrm{I}}_{1}}(x_{1})\psi^{\mathrm{I}_{2}}(x_{2})\varphi(x_{3})\}\vert\Omega\rangle\big\vert_{\mathcal{O}(\kappa)}\!=\!\int\frac{d^{4}q_{1}}{(2\pi)^{4}}\int\frac{d^{4}q_{2}}{(2\pi)^{4}}\int\frac{d^{4}q_{3}}{(2\pi)^{4}}\int d^{4}y\text{e}^{-\text{i}q_{1}\cdot(x_{1}-y)}\text{e}^{\text{i}q_{2}\cdot(x_{2}-y)}\text{e}^{\text{i}q_{3}\cdot(x_{3}-y)}\\
\nonumber
&\quad\quad\quad\quad\quad\quad\quad\times\frac{\text{i}}{q_{3}^{2}-m^{2}}\bigg(\frac{(\text{i}q_{1,\beta}\bar{\sigma}^{\beta})^{\dot{\mathrm{I}}_{1}\tilde{\mathrm{I}}_{1}}}{q_{1}^{2}-m^{2}}\times\underbrace{(-\frac{\text{i}\kappa}{2})(q_{2}+q_{1})_{\mu}(\sigma^{\mu})_{\tilde{\mathrm{I}}_{1}\dot{\tilde{\mathrm{I}}}_{2}}}_{\mathcal{V}_{(\varphi\psi\psi)}^{(q_{1},q_{2},\widetilde{\mathrm{I}}_{1,\downarrow},\dot{\widetilde{\mathrm{I}}}_{2,\downarrow})}}\times\frac{(\text{i}q_{2,\alpha}\bar{\sigma}^{\alpha})^{\dot{\tilde{\mathrm{I}}}_{2}\mathrm{I}_{2}}}{q_{2}^{2}-m^{2}}\\
\label{ThreePointCorrelationDaggerI1I2Result}
&\quad\quad\quad\quad\quad\quad\quad+\frac{\text{i}m\delta_{~~\dot{\tilde{\mathrm{I}}}_{1}}^{\dot{\mathrm{I}}_{1}}}{q_{1}^{2}-m^{2}}\times\underbrace{(-\frac{\text{i}\kappa}{2})(q_{2}+q_{1})_{\mu}(\bar{\sigma}^{\mu})^{\dot{\tilde{\mathrm{I}}}_{1}\tilde{\mathrm{I}}_{2}}}_{\mathcal{V}_{(\varphi\psi\psi)}^{(q_{1},q_{2},\dot{\widetilde{\mathrm{I}}}_{1,\uparrow},\widetilde{\mathrm{I}}_{2,\uparrow})}}\times\frac{\text{i}m\delta_{\tilde{\mathrm{I}}_{2}}^{~~\mathrm{I}_{2}}}{q_{2}^{2}-m^{2}}\bigg)
\end{align}	
\end{small}and
\begin{small}
\begin{align}
\nonumber
&\langle\Omega\vert\mathcal{T}\{\psi^{\dagger\dot{\mathrm{I}}_{1}}(x_{1})\psi_{\!\dot{\mathrm{I}}_{2}}^{\dagger}(x_{2})\varphi(x_{3})\}\vert\Omega\rangle\big\vert_{\mathcal{O}(\kappa)}\!=\!\int\frac{d^{4}q_{1}}{(2\pi)^{4}}\!\int\frac{d^{4}q_{2}}{(2\pi)^{4}}\!\int\frac{d^{4}q_{3}}{(2\pi)^{4}}\!\int d^{4}y\text{e}^{-\text{i}q_{1}\cdot(x_{1}-y)}\!\text{e}^{\text{i}q_{2}\cdot(x_{2}-y)}\!\text{e}^{\text{i}q_{3}\cdot(x_{3}-y)}\\
\nonumber
&\quad\quad\quad\quad\quad\quad\quad\times\frac{\text{i}}{q_{3}^{2}-m^{2}}\bigg(\frac{(\text{i}q_{1,\beta}\bar{\sigma}^{\beta})^{\dot{\mathrm{I}}_{1}\tilde{\mathrm{I}}_{1}}}{q_{1}^{2}-m^{2}}\times\underbrace{(-\frac{\text{i}\kappa}{2})(q_{2}+q_{1})_{\mu}(\sigma^{\mu})_{\tilde{\mathrm{I}}_{1}\dot{\tilde{\mathrm{I}}}_{2}}}_{\mathcal{V}_{(\varphi\psi\psi)}^{(q_{1},q_{2},\widetilde{\mathrm{I}}_{1,\downarrow},\dot{\widetilde{\mathrm{I}}}_{2,\downarrow})}}\times\frac{\text{i}m\delta_{~~\dot{\mathrm{I}}_{2}}^{\dot{\tilde{\mathrm{I}}}_{2}}}{q_{2}^{2}-m^{2}}\\
\label{ThreePointCorrelationDaggerI1DaggerI2Result}
&\quad\quad\quad\quad\quad\quad\quad+\frac{\text{i}m\delta_{~~\,\dot{\tilde{\mathrm{I}}}_{1}}^{\dot{\mathrm{I}}_{1}}}{q_{1}^{2}-m^{2}}\times\underbrace{(-\frac{\text{i}\kappa}{2})(q_{2}+q_{1})_{\mu}(\bar{\sigma}^{\mu})^{\dot{\tilde{\mathrm{I}}}_{1}\tilde{\mathrm{I}}_{2}}}_{\mathcal{V}_{(\varphi\psi\psi)}^{(q_{1},q_{2},\dot{\widetilde{\mathrm{I}}}_{1,\uparrow},\widetilde{\mathrm{I}}_{2,\uparrow})}}\times\frac{(\text{i}q_{2,\beta}\sigma^{\beta})_{\tilde{\mathrm{I}}_{2}\dot{\mathrm{I}}_{2}}}{q_{2}^{2}-m^{2}}\bigg)\\
\nonumber
&\langle\Omega\vert\mathcal{T}\{\psi_{\mathrm{I}_{1}}\!(x_{1})\psi_{\dot{\mathrm{I}}_{2}}^{\dagger}\!(x_{2})\varphi(x_{3})\}\vert\Omega\rangle\big\vert_{\mathcal{O}(\kappa)}\!=\!\int\frac{d^{4}q_{1}}{(2\pi)^{4}}\int\frac{d^{4}q_{2}}{(2\pi)^{4}}\int\frac{d^{4}q_{3}}{(2\pi)^{4}}\int d^{4}y\text{e}^{-\text{i}q_{1}\cdot(x_{1}-y)}\text{e}^{\text{i}q_{2}\cdot(x_{2}-y)}\text{e}^{\text{i}q_{3}\cdot(x_{3}-y)}\\
\nonumber
&\quad\quad\quad\quad\quad\quad\quad\times\frac{\text{i}}{q_{3}^{2}-m^{2}}\bigg(\frac{\text{i}m\delta_{\mathrm{I}_{1}}^{~~\tilde{\mathrm{I}}_{1}}}{q_{1}^{2}-m^{2}}\times\underbrace{(-\frac{\text{i}\kappa}{2})(q_{2}+q_{1})_{\mu}(\sigma^{\mu})_{\tilde{\mathrm{I}}_{1}\dot{\tilde{\mathrm{I}}}_{2}}}_{\mathcal{V}_{(\varphi\psi\psi)}^{(q_{1},q_{2},\widetilde{\mathrm{I}}_{1,\downarrow},\dot{\widetilde{\mathrm{I}}}_{2,\downarrow})}}\times\frac{\text{i}m\delta_{~~\,\dot{\mathrm{I}}_{2}}^{\dot{\tilde{\mathrm{I}}}_{2}}}{q_{2}^{2}-m^{2}}\\
\label{ThreePointCorrelationI1DaggerI2Result}
&\quad\quad\quad\quad\quad\quad\quad+\frac{(\text{i}q_{1,\alpha}\sigma^{\alpha})_{\mathrm{I}_{1}\dot{\tilde{\mathrm{I}}}_{1}}}{q_{1}^{2}-m^{2}}\times\underbrace{(-\frac{\text{i}\kappa}{2})(q_{2}+q_{1})_{\mu}(\bar{\sigma}^{\mu})^{\dot{\tilde{\mathrm{I}}}_{1}\tilde{\mathrm{I}}_{2}}}_{\mathcal{V}_{(\varphi\psi\psi)}^{(q_{1},q_{2},\dot{\widetilde{\mathrm{I}}}_{1,\uparrow},\widetilde{\mathrm{I}}_{2,\uparrow})}}\times\frac{(\text{i}q_{2,\beta}\sigma^{\beta})_{\tilde{\mathrm{I}}_{2}\dot{\mathrm{I}}_{2}}}{q_{2}^{2}-m^{2}}\bigg)
\end{align}
\end{small}From the three-point correlation functions \eqref{ThreePointCorrelationI1I2Result}–\eqref{ThreePointCorrelationI1DaggerI2Result}, by discarding the propagator contributions, two distinct types of interaction vertices in momentum space can be extracted, as illustrated in Fig. \ref{FigPsiI1PsiI2Inflatonpsipsi}. In summary, for the interaction \eqref{YukawaInflatonMajorana} that describes the coupling between one scalar and two Majorana fermions, it is straightforward to observe that there exist only two types of vertices in momentum space, namely $\mathcal{V}_{(\varphi\psi\psi)}^{(q_{1},q_{2},\widetilde{\mathrm{I}}_{1,\downarrow},\dot{\widetilde{\mathrm{I}}}_{2,\downarrow})}$ and $\mathcal{V}_{(\varphi\psi\psi)}^{(q_{1},q_{2},\dot{\widetilde{\mathrm{I}}}_{1,\uparrow},\widetilde{\mathrm{I}}_{2,\uparrow})}$. Here, the momenta $q_1$ and $q_2$ represent the incoming and outgoing momenta of the Majorana fermions, respectively, while the spinor indices $\tilde{\mathrm{I}},\tilde{\mathrm{J}}$, marked with upward and downward arrows, is used to distinguish between the upper and lower spinor components, respectively.

Using the Feynman rules provided in Fig. \ref{FigPsiI1PsiI2Inflatonpsipsi}, the amplitude for the two-body decay process $\varphi(k)\to\psi^{\mathrm{I}_{1}}(s,p)\psi_{\mathrm{I}_{2}}(s^{\prime},q)$—depicted in the upper panel of Fig. \ref{RefTwoThreeBodyMixMajorana}—can be expressed as
\begin{align}
\label{TwoBodyAmplitude}
&\mathcal{M}_{\varphi(k)\to\psi^{\mathrm{I}_{1}}(s,p)\psi_{\mathrm{I}_{2}}(s^{\prime},q)}\!=-\frac{\text{i}\kappa\mathsf{y}_{\text{D}}}{2}(p\!-\!q)_{\mu}\big(y(\boldsymbol{p},s)\cdot\sigma^{\mu}\cdot x^{\dagger}(\boldsymbol{q},s^{\prime})\!+\! x^{\dagger}(\boldsymbol{p},s)\cdot\bar{\sigma}^{\mu}\cdot y(\boldsymbol{q},s^{\prime})\big)
\end{align}And the corresponding squared amplitude derived from \eqref{TwoBodyAmplitude} is obtained as
\begin{align}
\nonumber
&\sum_{s}\sum_{s^{\prime}}\vert\mathcal{M}_{\varphi(k)\to\psi^{\mathrm{I}_{1}}(s,p)\psi_{\mathrm{I}_{2}}(s^{\prime},q)}\vert^{2}\!=\!\frac{\kappa^{2}\mathsf{y}_{\text{D}}^{2}(p-q)_{\mu_{1}}(p-q)_{\mu_{2}}}{4}\\
\nonumber
&\quad\quad\quad\quad\times\big\{\sum_{s}y_{\boldsymbol{p},s}^{\dagger\dot{\mathrm{J}}_{2}}y_{\boldsymbol{p},s}^{\mathrm{I}_{1}}\sum_{s^{\prime}}x_{\boldsymbol{q},s^{\prime}}^{\dagger\dot{\mathrm{J}}_{1}}x_{\boldsymbol{q},s^{\prime}}^{\mathrm{I}_{2}}(\sigma^{\mu_{2}})_{\mathrm{I}_{2}\dot{\mathrm{J}}_{2}}(\sigma^{\mu_{1}})_{\mathrm{I}_{1}\dot{\mathrm{J}}_{1}}\\
\nonumber
&\quad\quad\quad\quad+\sum_{s}y_{\boldsymbol{p},s}^{\dagger\dot{\mathrm{J}}_{2}}x_{\boldsymbol{p},s,\dot{\mathrm{I}}_{1}}^{\dagger}\sum_{s^{\prime}}y_{\boldsymbol{q},s^{\prime},\mathrm{J}_{1}}x_{\boldsymbol{q},s^{\prime}}^{\mathrm{I}_{2}}(\sigma^{\mu_{2}})_{\mathrm{I}_{2}\dot{\mathrm{J}}_{2}}(\bar{\sigma}^{\mu_{1}})^{\dot{\mathrm{I}}_{1}\mathrm{J}_{1}}\\
\nonumber
&\quad\quad\quad\quad+\sum_{s}x_{\boldsymbol{p},s,\mathrm{J}_{2}}y_{\boldsymbol{p},s}^{\mathrm{I}_{1}}\sum_{s^{\prime}}x_{\boldsymbol{q},s^{\prime}}^{\dagger\dot{\mathrm{J}}_{1}}y_{\boldsymbol{q},s^{\prime},\dot{\mathrm{I}}_{2}}^{\dagger}(\bar{\sigma}^{\mu_{2}})^{\dot{\mathrm{I}}_{2}\mathrm{J}_{2}}(\sigma^{\mu_{1}})_{\mathrm{I}_{1}\dot{\mathrm{J}}_{1}}\\
&\quad\quad\quad\quad+\sum_{s}x_{\boldsymbol{p},s,\mathrm{J}_{2}}x_{\boldsymbol{p},s,\dot{\mathrm{I}}_{1}}^{\dagger}\sum_{s^{\prime}}y_{\boldsymbol{q},s^{\prime},\mathrm{J}_{1}}y_{\boldsymbol{q},s^{\prime},\dot{\mathrm{I}}_{2}}^{\dagger}(\bar{\sigma}^{\mu_{2}})^{\dot{\mathrm{I}}_{2}\mathrm{J}_{2}}(\bar{\sigma}^{\mu_{1}})^{\dot{\mathrm{I}}_{1}\mathrm{J}_{1}}\big\}
\end{align}After summing over the spin states of the wave functions $x_{\boldsymbol{p},s}$ and $y_{\boldsymbol{p},s}$, and making use of the decomposition rules and trace identities for multiple products of $\sigma_\mu$ and $\bar{\sigma}_\nu$ matrices summarized in \cite{Dreiner:2008tw}, namely 
\begin{align}
\label{ThreeSigmaProdv1}
&\sigma_{\mu}\bar{\sigma}_{\nu}\sigma_{\alpha}=\eta_{\mu\nu}\sigma_{\alpha}-\eta_{\mu\alpha}\sigma_{\nu}+\eta_{\nu\alpha}\sigma_{\mu}+\text{i}\epsilon_{\mu\nu\alpha\beta}\sigma^{\beta}\\
\label{ThreeSigmaProdv2}
&\bar{\sigma}_{\mu}\sigma_{\nu}\bar{\sigma}_{\alpha}=\eta_{\mu\nu}\bar{\sigma}_{\alpha}-\eta_{\mu\alpha}\bar{\sigma}_{\nu}+\eta_{\nu\alpha}\bar{\sigma}_{\mu}-\text{i}\epsilon_{\mu\nu\alpha\beta}\bar{\sigma}^{\beta}\\
\label{TraceFourSigmaProd}
&\text{Tr}(\sigma_{\mu}\bar{\sigma}_{\nu}\sigma_{\alpha}\bar{\sigma}_{\beta})=2(\eta_{\mu\nu}\eta_{\alpha\beta}-\eta_{\mu\alpha}\eta_{\nu\beta}+\eta_{\mu\beta}\eta_{\nu\alpha}+\text{i}\epsilon_{\mu\nu\alpha\beta})
\end{align}
we arrive at 
\begin{align}
\nonumber
&\sum_{s}\sum_{s^{\prime}}\vert\mathcal{M}_{\varphi(k)\to\psi^{\mathrm{I}_{1}}(s,p)\psi_{\mathrm{I}_{2}}(s^{\prime},q)}\vert^{2}=\frac{\kappa^{2}\mathsf{y}_{\text{D}}^{2}(p-q)^{\mu_{1}}(p-q)^{\mu_{2}}}{4}\big\{ q^{\beta_{1}}p^{\beta_{2}}\text{Tr}\big(\sigma_{(\mu_{1}}\bar{\sigma}_{\beta_{1}}\sigma_{\mu_{2})}\bar{\sigma}_{\beta_{2}}\big)\\
\nonumber
&\quad\quad\quad\quad\quad\quad -m^{2}\text{Tr}(\sigma_{\mu_{2}}\bar{\sigma}_{\mu_{1}})+p^{\beta_{2}}q^{\beta_{1}}\text{Tr}\big(\sigma_{\beta_{2}}\bar{\sigma}_{(\mu_{1}}\sigma_{\beta_{1}}\bar{\sigma}_{\mu_{2})}\big)-m^{2}\text{Tr}(\sigma_{\mu_{1}}\bar{\sigma}_{\mu_{2}})\big\}\\
\nonumber
&\quad\quad\quad\quad\quad\quad=\kappa^{2}\mathsf{y}_{\text{D}}^{2}(p-q)^{\mu_{1}}(p-q)^{\mu_{2}}\big\{ p_{\mu_{2}}q_{\mu_{1}}+p_{\mu_{1}}q_{\mu_{2}}-p\cdot q\eta_{\mu_{1}\mu_{2}}-m^{2}\eta_{\mu_{1}\mu_{2}}\big\}\\
&\quad\quad\quad\quad\quad\quad=4\kappa^{2}\mathsf{y}_{\text{D}}^{2}m^{2}\times(p\cdot q-m^{2})
\end{align}In the rest frame of the inflaton, i.e., $k^\mu = (M, 0, 0, 0)$, in addition to the on-shell conditions $p^2 = q^2 = m^2$, the conservation of energy and momentum further gives
\begin{align}
&\tilde{p}^\mu=(\frac{M}{2},\boldsymbol{p}^\star)\,,\,\tilde{q}^{\mu}=(\frac{M}{2},-\boldsymbol{p}^\star)\,,\,\vert \boldsymbol{p}^\star \vert =\sqrt{\frac{M^{2}}{4}-m^{2}}\,,\,(\tilde{p}\cdot\tilde{q})=\frac{M^{2}}{2}-m^{2}
\end{align}Based on the above kinematic considerations in the rest frame of the inflaton, the corresponding two-body decay can be computed as
\begin{align}
\nonumber
\Gamma_{\varphi\to\psi\psi}^{(0)}&=\int d\Omega\,\frac{\vert\boldsymbol{p}^{\star}\vert}{32\pi^{2}M^{2}}\big(\sum_{s}\sum_{s^{\prime}}\vert\mathcal{M}_{\varphi(k)\to\psi^{\mathrm{I}_{1}}(s,p)\psi_{\mathrm{I}_{2}}(s^{\prime},q)}\vert^{2}\big)_{\tilde{p},\tilde{q}}\\
&=\frac{\kappa^{2}\mathsf{y}_{\text{D}}^{2}M^{3}}{8\pi}y^{2}(1-4y^{2})^{3/2}
\end{align}in which $y$ is used to denote $y=m/M$.

\subsection{Two-body and three-body decay rates (including graviton emission) associated with the interaction $S_{\text{int}}^{(\varphi\phi\phi)}$ \label{CubicInflaComplexScalar}}

Let us focus on the scalar interaction \eqref{YukawaInflatonComScalar}. The generating functional associated with the interaction term $S^{\varphi\phi\phi}_{\text{int}}$ can be expressed as
\begin{small}
\begin{align}
\nonumber
W[J_{(\varphi)},\mathsf{J}_{(\phi)}^{\star},\mathsf{J}_{(\phi)}]=\exp&\bigg(\frac{\text{i}\kappa\mathsf{y}_{\text{D}}}{4}\int d^{4}x\,\frac{1}{\text{i}}\frac{\delta}{\delta J_{(\varphi)}(x)}\big\{2\eta^{\mu\nu}\partial_{\mu}^{(x)}\big(\frac{1}{\text{i}}\frac{\delta}{\delta\mathsf{J}_{(\phi)}(x)}\big)\partial_{\nu}^{(x)}\big(\frac{1}{\text{i}}\frac{\delta}{\delta\mathsf{J}_{(\phi)}^{\star}(x)}\big)\\
\nonumber
&+4m^{2}\big(\frac{1}{\text{i}}\frac{\delta}{\delta\mathsf{J}_{(\phi)}(x)}\big)\big(\frac{1}{\text{i}}\frac{\delta}{\delta\mathsf{J}_{(\phi)}^{\star}(x)}\big)-\big(\frac{1}{\text{i}}\frac{\delta}{\delta\mathsf{J}_{(\phi)}(x)}\big)\partial_{(x)}^{2}\big(\frac{1}{\text{i}}\frac{\delta}{\delta\mathsf{J}_{(\phi)}^{\star}(x)}\big)\\
&-\big(\frac{1}{\text{i}}\frac{\delta}{\delta\mathsf{J}_{(\phi)}^{\star}(x)}\big)\partial_{(x)}^{2}\big(\frac{1}{\text{i}}\frac{\delta}{\delta\mathsf{J}_{(\phi)}(x)}\big)\big\}\bigg)W_{0}[\mathsf{J}_{(\phi)}^{\star},\mathsf{J}_{(\phi)}]W_{0}[J_{(\varphi)}]
\end{align}
\end{small}At the $\mathcal{O}(\kappa)$ order, this generating functional admits the following expansion
\begin{small}
\begin{align}
\nonumber
W[J_{(\varphi)},\mathsf{J}_{(\phi)}^{\star},\mathsf{J}_{(\phi)}]\big\vert_{\mathcal{O}(\kappa)}\!=&-\frac{\kappa\mathsf{y}_{\text{D}}}{4}\!\int d^{4}x\!\int\frac{d^{4}k_{1}}{(2\pi)^{4}}\text{e}^{\text{i}k_{1}\cdot x}\!\int\frac{d^{4}k_{2}}{(2\pi)^{4}}\text{e}^{-\text{i}k_{2}\cdot x}\!\int\frac{d^{4}k_{3}}{(2\pi)^{4}}\text{e}^{\text{i}k_{3}\cdot x}\!\big((k_{1}\!+\!k_{2})^{2}\!+\!4m^{2}\big)\\
&\times \frac{\delta}{\delta\tilde{\mathsf{J}}_{(\phi)}(k_{1})}\frac{\delta}{\delta\tilde{\mathsf{J}}_{(\phi)}^{\star}(k_{2})}\frac{\delta}{\delta\tilde{J}_{(\varphi)}(k_{3})}W_{0}[\mathsf{J}_{(\phi)}^{\star},\mathsf{J}_{(\phi)}]W_{0}[J_{(\varphi)}]
\end{align}
\end{small}With the preparatory steps completed, we proceed to compute the three-point correlation functions
\begin{small}
\begin{align}
\nonumber
&\langle\Omega\vert\mathcal{T}\{\phi^{\star}(x_{1})\phi(x_{2})\varphi(x_{3})\}\vert\Omega\rangle\big\vert_{\mathcal{O}(\kappa)}\!=\!\big\{\frac{1}{\text{i}^{3}}\frac{1}{W_{0}[0,0]W_{0}[0]}\!\int\frac{d^{4}q_{1}}{(2\pi)^{4}}\!\text{e}^{\text{i}q_{1}\cdot x_{1}}\!\!\int\frac{d^{4}q_{2}}{(2\pi)^{4}}\!\text{e}^{-\text{i}q_{2}\cdot x_{2}}\!\!\int\frac{d^{4}q_{3}}{(2\pi)^{4}}\!\text{e}^{\text{i}q_{3}\cdot x_{3}}\\
&\quad\quad\quad\quad\quad\quad\quad\quad\quad\times \frac{\delta}{\delta\tilde{\mathsf{J}}_{(\phi)}(q_{1})}\frac{\delta}{\delta\tilde{\mathsf{J}}_{(\phi)}^{\star}(q_{2})}\frac{\delta}{\delta\tilde{J}_{(\varphi)}(q_{3})}W[J_{(\varphi)},\mathsf{J}_{(\phi)}^{\star},\mathsf{J}_{(\phi)}]\big\vert_{\mathcal{O}(\kappa)}\big\}\bigg\vert_{J_{(\varphi)}=\mathsf{J}_{(\phi)}^{\star}=\mathsf{J}_{(\phi)}=0}^{\text{connected}}
\end{align}
\end{small}Following a series of detailed calculations, we ultimately obtain
\begin{small}
\begin{align}
\nonumber
&\langle\Omega\vert\mathcal{T}\{\phi^{\star}(x_{1})\phi(x_{2})\varphi(x_{3})\}\vert\Omega\rangle\big\vert_{\mathcal{O}(\kappa)}\!=\! \int d^{4}x\!\int\!\frac{d^{4}q_{1}}{(2\pi)^{4}}\!\text{e}^{\text{i}q_{1}\cdot(x_{1}-x)}\int\!\frac{d^{4}q_{2}}{(2\pi)^{4}}\!\text{e}^{-\text{i}q_{2}\cdot(x_{2}-x)}\int\!\frac{d^{4}q_{3}}{(2\pi)^{4}}\!\text{e}^{\text{i}q_{3}\cdot(x_{3}-x)}\\
\label{ThreePointScalarPairFinal}
&\quad\quad\quad\quad\times\underbrace{\big(-\frac{\text{i}\kappa\mathsf{y}_{\text{D}}}{4}(2q_{2}\cdot q_{1}+4m^{2}+q_{1}^{2}+q_{2}^{2})\big)}_{\mathcal{V}_{(\varphi\phi\phi)}^{(q_{1},q_{2})}}\times\frac{\text{i}}{q_{2}^{2}-m^{2}+\text{i}\epsilon}\frac{\text{i}}{q_{1}^{2}-m^{2}+\text{i}\epsilon}\frac{\text{i}}{q_{3}^{2}-M^{2}+\text{i}\epsilon}
\end{align}
\end{small}From the resulting three-point correlation function \eqref{ThreePointScalarPairFinal}, the Feynman rule for the interaction vertex in momentum space, denoted as $\mathcal{V}_{(\varphi\phi\phi)}^{(q_{1},q_{2})}$, can be identified. Here, $q_{1}$ and $q_{2}$ represent the ingoing and outgoing momenta, respectively, of the particle–antiparticle pair produced by the complex scalar field $\phi$. Based on the vertex $\mathcal{V}_{(\varphi\phi\phi)}^{(q_{1},q_{2})}$, the amplitude for the two-body decay process $\varphi(k)\to\phi(p)\phi^{\star}(q)$ is given by
\begin{align}
\label{AmplitudeTwoBodyScalarPair}
&\mathcal{M}_{\varphi(k)\to\phi(p)\phi^{\star}(q)}=-\frac{\text{i}\kappa\mathsf{y_D}}{4}(-2p\cdot q+4m^{2}+q^{2}+p^{2})=-\frac{\text{i}\kappa\mathsf{y_D}}{2}(3m^{2}-p\cdot q)
\end{align}In the rest frame of the inflaton, the squared amplitude corresponding to \eqref{AmplitudeTwoBodyScalarPair} is readily obtained as
\begin{align}
&\big(\vert\mathcal{M}_{\varphi(k)\to\phi(p)\phi^{\star}(q)}\vert^{2}\big)_{\tilde{p},\tilde{q}}=\frac{\kappa^{2}\mathsf{y_D}^{2}M^{4}}{4}(4y^{2}-\frac{1}{2})^{2}
\end{align}where $y = m/M$. Based on this result, the two-body decay rate $\Gamma_{\varphi\to\phi\phi^{\star}}^{(0)}$, depicted in the upper panel of Fig. \ref{RefTwoThreeBodyMixScalar}, is given by
\begin{align}
&\Gamma_{\varphi\to\phi\phi^{\star}}^{(0)}=\frac{\kappa^{2}\mathsf{y_D}^{2}M^{3}}{64\pi}(4y^{2}-\frac{1}{2})^{2}(1-4y^{2})^{1/2}
\end{align}

Moreover, let us proceed to consider the three-body decay process $\varphi(k)\to\phi(p)\phi^\star(q)e_{\alpha\beta}(\chi,l)$, which corresponds to the lower panel of Fig. \ref{RefTwoThreeBodyMixScalar}. It should be noted that the relevant Feynman rules for the interaction between one graviton and two scalars can be found in \cite{Gleisberg:2003ue,Barman:2023ymn}, and in the present analysis we adopt the same form as that employed in our previous work \cite{Lee:2025lyk}.
\begin{small}
\begin{align}
\nonumber
&\mathcal{M}_{\varphi(k)\to\phi(p)\phi^{\star}(q)e_{\alpha\beta}(\chi,l)}\!=\!\frac{-\text{i}\mathsf{y_D}\kappa^{2}}{8}\frac{(-2p\cdot q\!+\!4m^{2}\!+\!q^{2}\!+\!p^{2})}{(k-l)^{2}-M^{2}}e_{\alpha\beta}^{(\chi)}(\boldsymbol{l})\big\{2k^{\alpha}(k\!-\!l)^{\beta}\!+\!\eta^{\alpha\beta}\big(M^{2}\!-\!k\cdot(k\!-\!l)\big)\big\}\\
\nonumber
&\quad\quad~-\frac{\text{i}\mathsf{y_D}\kappa^{2}}{8}\frac{\big(\!-\!2p\cdot(q\!+\!l)\!+\!4m^{2}\!+\!(q\!+\!l)^{2}\!+\!p^{2}\big)}{(q+l)^{2}-m^{2}}e_{\alpha\beta}^{(\chi)}(\boldsymbol{l})\big\{2(-q)^{\alpha}(-\!q\!-\!l)^{\beta}\!+\!\eta^{\alpha\beta}\big(m^{2}\!-\!(-q)\cdot(-q\!-\!l)\big)\big\}\\
&\quad\quad~-\frac{\text{i}\mathsf{y_D}\kappa^{2}}{8}\frac{\big(\!-\!2(p\!+\!l)\cdot q\!+\!4m^{2}\!+\!q^{2}\!+\!(p\!+\!l)^{2}\big)}{(p+l)^{2}-m^{2}}e_{\alpha\beta}^{(\chi)}(\boldsymbol{l})\big\{2p^{\alpha}(p\!+\!l)^{\beta}\!+\!\eta^{\alpha\beta}\big(m^{2}\!-\!p\cdot(p\!+\!l)\big)\big\}
\end{align}
\end{small}After imposing the gauge-fixing conditions for the transverse on-shell polarization tensors, namely $\eta^{\alpha\beta}e_{\alpha\beta}^{(\chi)}(\boldsymbol{l})=l^{\alpha}e_{\alpha\beta}^{(\chi)}(\boldsymbol{l})=0$ (see \cite{Gleisberg:2003ue,Barman:2023ymn,Lee:2025lyk} for a detailed derivation of these conditions), the above amplitude can be simplified to
\begin{align}
\nonumber
\mathcal{M}_{\varphi(k)\to\phi(p)\phi^\star(q)e_{\alpha\beta}(\chi,l)}&=\frac{\text{i}\mathsf{y_D}\kappa^{2}}{8}e_{\alpha\beta}^{(\chi)}(\boldsymbol{l})\big\{\frac{2(3m^{2}-p\cdot q)}{k\cdot l}k^{\alpha}k^{\beta}-\frac{(8m^{2}+M^{2}-4k\cdot q)}{p\cdot l}p^{\alpha}p^{\beta}\\
\label{ThreeBodyScalarAmplitude}
&-\frac{(8m^{2}+M^{2}-4k\cdot p)}{q\cdot l}q^{\alpha}q^{\beta}\big\}
\end{align}In the rest frame of the inflaton, the decay width associated with the above three-body decay process can be expressed in the general form
\begin{align}
\nonumber
\Gamma_{\varphi(k)\to\phi(p)\phi^{\star}(q)e_{\alpha\beta}(\chi,l)}^{(1)}&=\!\frac{1}{2M}\int\frac{d^{3}\boldsymbol{l}}{(2\pi)^{3}2E_{l}}\int\frac{d^{3}\boldsymbol{p}}{(2\pi)^{3}2E_{p}}\int\frac{d^{3}\boldsymbol{q}}{(2\pi)^{3}2E_{q}}\\
\label{DefinitionThreeBodyScalar}
&\times(2\pi)^{4}\delta^{4}(k\!-\!p\!-\!q\!-\!l)\big(\sum_{\chi}\big\vert\mathcal{M}_{\varphi(k)\to\phi(p)\phi^{\star}(q)e_{\alpha\beta}(\chi,l)}\big\vert^{2}\big)
\end{align}Alternatively, the same three-body decay process can be reformulated as a sequential two-body decay chain, namely
\begin{align}
\nonumber
&\Gamma_{\varphi(k)\to\phi(p)\phi^{\star}(q)e_{\alpha\beta}(\chi,l)}^{(1)}=\frac{1}{(4\pi)^{5}M}\int dM_{B}^{2}\int\frac{d^{3}\boldsymbol{P}_{B}}{E_{B}}\int\frac{d^{3}\boldsymbol{l}}{E_{l}}\delta^{4}(P_{B}+l-k)\\
\label{SequenThreeBodyScalar}
&\quad\quad\quad\times\int\frac{d^{3}\boldsymbol{p}}{E_{p}}\int\frac{d^{3}\boldsymbol{q}}{E_{q}}\delta^{4}(p+q-P_{B})\times\big(\sum_{\chi}\big\vert\mathcal{M}_{\varphi(k)\to\phi(p)\phi^\star(q)e_{\alpha\beta}(\chi,l)}\big\vert^{2}\big)
\end{align}By sequentially eliminating the two $\delta$-functions in the above expression with respect to $\boldsymbol{P}_{\!\!B}$, $\boldsymbol{q}$, $M_{B}^{2}$, and $z=\cos\theta_{\langle\hat{\boldsymbol{l}},\hat{\boldsymbol{p}}\rangle}$, we finally arrive at
\begin{align}
\nonumber
\Gamma_{\varphi(k)\to\phi(p)\phi^{\star}(q)e_{\alpha\beta}(\chi,l)}^{(1)}&=\frac{1}{16M}\frac{1}{(2\pi)^{5}}\int dE_{l}\int d\theta_{\langle\hat{\boldsymbol{l}},\hat{\boldsymbol{z}}\rangle}\int d\phi_{\langle\hat{\boldsymbol{l}},\hat{\boldsymbol{z}}\rangle}\sin\theta_{\langle\hat{\boldsymbol{l}},\hat{\boldsymbol{z}}\rangle}\\
\label{ScalarThreeBodyDecayFinal}
&\times\int dE_{p}\int d\phi_{\langle\hat{\boldsymbol{p}},\hat{\boldsymbol{l}}\rangle}\big(\sum_{\chi}\big\vert\mathcal{M}_{\varphi(k)\to\phi(p)\phi^{\star}(q)e_{\alpha\beta}(\chi,l)}\big\vert^{2}\big)
\end{align}together with the restrictions
\begin{align}
\label{ScalarThreeBodyRestricFromPB}
&\quad\quad\quad\quad\quad~ M_{B}^{\star2}=M(M-2E_{l})>0\,,\,M-E_{\ell}>M_{B}^{\star}>2m\\
\label{ScalarThreeBodyRestricFromThetalp}
&-1\leq z^{\star}(x)=-\frac{(M-E_{l})}{E_{l}}x+\frac{(M-2E_{l})M}{2E_{l}m}\sqrt{x^{2}-1}\leq+1~,~x=\frac{E_{p}}{\sqrt{E_{p}^{2}-m^{2}}}
\end{align}in which $\theta_{\langle\hat{\boldsymbol{l}},\hat{\boldsymbol{z}}\rangle}$ and $\phi_{\langle\hat{\boldsymbol{l}},\hat{\boldsymbol{z}}\rangle}$ denote the angular coordinates specifying the orientation of the vector $\boldsymbol{l}$ with respect to the $z$-axis in the rest frame of the inflaton, while $\theta_{\langle\hat{\boldsymbol{p}},\hat{\boldsymbol{l}}\rangle}$ denotes the angle between the vectors $\boldsymbol{l}$ and $\boldsymbol{p}$. Actually, under the constraints given in \eqref{ScalarThreeBodyRestricFromPB}, the expression $z^{\star}(x)$ in \eqref{ScalarThreeBodyRestricFromThetalp} is a monotonically increasing function of $x$. Consequently, the minimum and maximum values of $x$ can be determined by imposing the boundary conditions $z^{\star}(x_{\text{min}})=-1$ and $z^{\star}(x_{\text{max}})=1$, respectively
\begin{align}
&x_{min}=\frac{M(M-2E_{l})\sqrt{M(M-2E_{l})(M^{2}-2ME_{l}-4m^{2})}-4m^{2}E_{l}(M-E_{l})}{\big((M^{2}-2ME_{l})^{2}-4m^{2}(M-E_{l})^{2}\big)}\\
&x_{max}=\frac{M(M-2E_{l})\sqrt{M(M-2E_{l})(M^{2}-2ME_{l}-4m^{2})}+4m^{2}E_{l}(M-E_{l})}{\big((M^{2}-2ME_{l})^{2}-4m^{2}(M-E_{l})^{2}\big)}
\end{align}Since $x$ is inversely related to $E_p$, the extremal values $E_{p,\text{max}}$ and $E_{p,\text{min}}$ can be obtained directly from $x_{\text{min}}$ and $x_{\text{max}}$, respectively
\begin{align}
\label{ThreeBodyEpmax}
&E_{p,max}=\frac{1}{2}\big(M-E_{l}+E_{l}\sqrt{\frac{M^{2}-2ME_{l}-4m^{2}}{M(M-2E_{l})}}\big) \\
\label{ThreeBodyEpmin}
&E_{p,min}=\frac{1}{2}\big(M-E_{l}-E_{l}\sqrt{\frac{M^{2}-2ME_{l}-4m^{2}}{M(M-2E_{l})}}\big)
\end{align}Besides, it is worth noting that although \eqref{SequenThreeBodyScalar} appears more complicated than \eqref{DefinitionThreeBodyScalar}, it is convenient to derive the constraint \eqref{ScalarThreeBodyRestricFromPB}. In the rest frame of the inflaton, and making use of the on-shell conditions together with energy–momentum conservation, the fundamental building blocks relevant to the kinematics of the three-body decay take the form
\begin{align}
\nonumber
&k\cdot l=ME_{l}\,,\,k\cdot p=ME_{p}\, , \, k\cdot q=M(M-E_{p}-E_{l})\\
\label{ThreeBodyBuildingBlockV1}
&p\cdot q\!=\!\frac{M^{2}-2m^{2}-2ME_{l}}{2}\, ,\, p\cdot l\!=\!\frac{M(2E_{p}+2E_{l}-M)}{2} \, , \, q\cdot l\!=\! \frac{M(M-2E_{p})}{2}
\end{align}Furthermore, due to the appearance of the vector $\bar{l}^\mu=(E_l , -\boldsymbol{l})$ in the polarization sum of the tensor $e^{\chi}_{\alpha\beta}(\boldsymbol{l})$, it is necessary to introduce
\begin{align}
\nonumber
&q\cdot\bar{l}=2E_{q}E_{l}-q\cdot l=2(M-E_{p}-E_{l})E_{l}-\frac{M(M-2E_{p})}{2}\\
\label{ThreeBodyBuildingBlockV2}
&p\cdot\bar{l}=2E_{p}E_{l}-p\cdot l=2E_{p}E_{l}-\frac{M(2E_{p}+2E_{l}-M)}{2}
\end{align}Therefore, in the rest frame of the inflaton, and after employing the relations \eqref{ThreeBodyBuildingBlockV1}–\eqref{ThreeBodyBuildingBlockV2}, the squared amplitude corresponding to \eqref{ThreeBodyScalarAmplitude} can be expressed as
\begin{align}
\label{ThreeBodyAmplitudeScalar}
&\sum_{\chi}\big\vert\mathcal{M}_{\varphi(k)\to\phi(p)\phi^{\star}(q)e_{\alpha\beta}(\chi,l)}\big\vert^{2}=\left(E_lM(3M-8E_p)-2M(M-2E_p)^2+8E_lm^2\right)^2\\ \nonumber
& \times\frac{\mathsf{y_D}^2\kappa^4\left(4M^2(E_p^2+3E_pE_l+E_l^2)-4M^3(E_p+E_l)-8E_pE_lM(E_p+E_l)+4E_l^2m^2+M^4\right)^2}{128E_{l}^4M^2(M-2E_{l})^2\left(M-2(E_p+E_{l})\right)^2}
\end{align}It is straightforward to see that the squared amplitude \eqref{ThreeBodyAmplitudeScalar} is independent of the angular variables. When evaluating the gravitational-wave spectrum, it is sufficient to consider the differential three-body decay width. Therefore, by substituting \eqref{ThreeBodyAmplitudeScalar} into the general expression for the three-body decay \eqref{ScalarThreeBodyDecayFinal} and performing the angular integrations, we obtain the differential three-body decay width
\begin{align}
\nonumber
\frac{d\Gamma_{\varphi(k)\to\phi(p)\phi^{\star}(q)e_{\alpha\beta}(\chi,l)}^{(1)}}{dE_l}&=\frac{\kappa^4M^4\mathsf{y_D}^2}{122880\pi^3x}\times\bigg\{\\ \nonumber
&\alpha\bigg(1920(1-2x)y^6+32\left(2x(168x-95)+15\right)y^4\\ \nonumber
&+2(2x-1)\left(8x(56x-75)+105\right)y^2
+(1-2x)^2\left(8x(4x-5)+15\right)\bigg)\\
&+60y^2(8y^2-1)\left(3(8x-3)y^2-2x+8y^4+1\right)\ln\left(\frac{1+\alpha}{1-\alpha}\right)\bigg\}
\end{align}

\section{Derivation of the Feynman Rules for interaction between one graviton and two Majorana fermions \label{FeyRuleGravitonMajorana}}

\begin{figure}[ht]
	\begin{center}
		\includegraphics[scale=0.45]{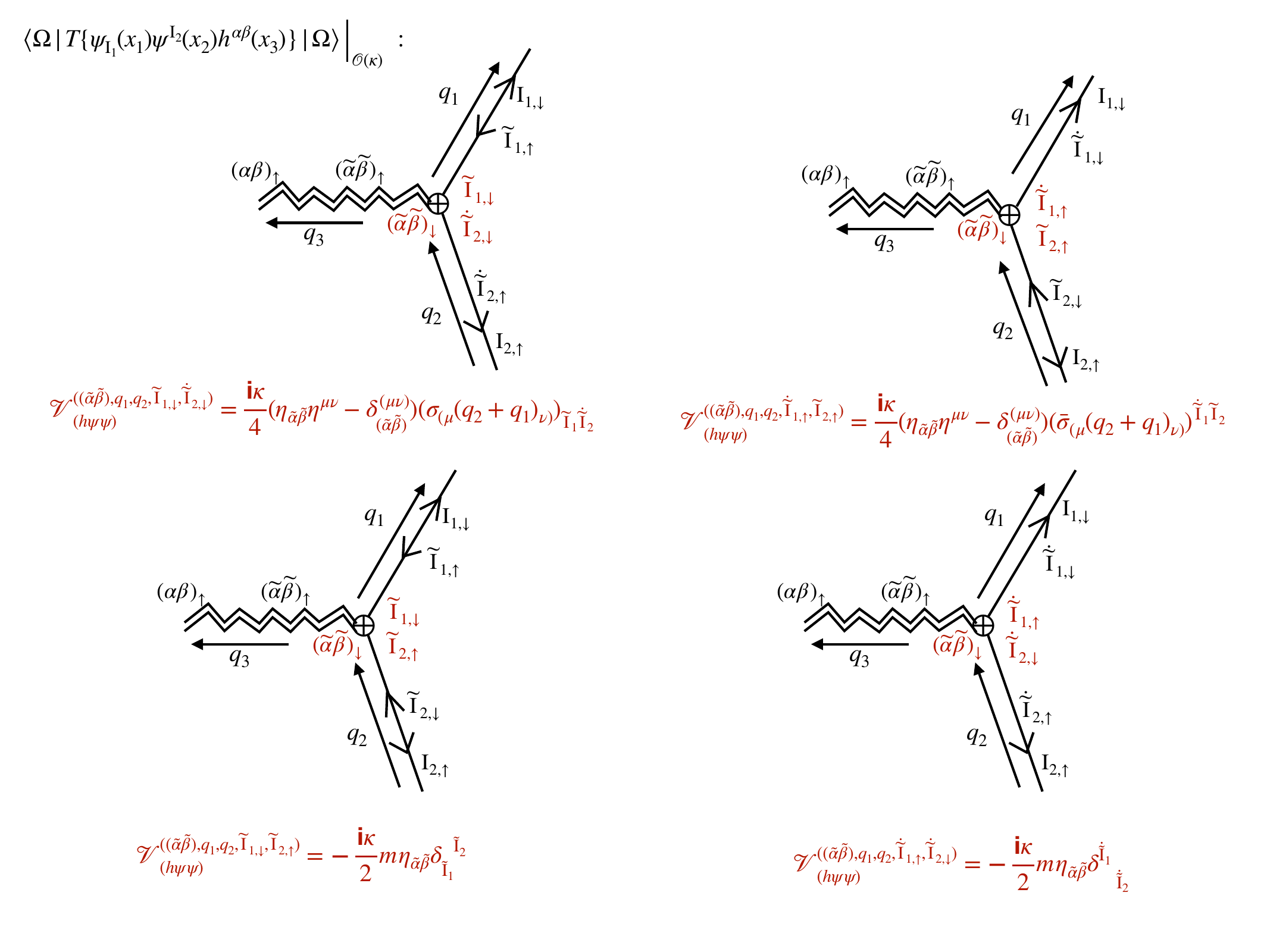}
		\caption{Feynman rules describing the minimal coupling between one graviton and two Majorana fermions, as given in \eqref{YukawaGravitonMajorana}. These rules are extracted from the three-point correlation functions \eqref{GravitonPsiI1PsiI2}–\eqref{GravitonPsiI1PsiDaggerI2}, all of which give rise to four distinct types of interaction vertices in momentum space. As a representative example, we illustrate the vertices obtained from the correlator
$\langle\Omega\vert\mathcal{T}\{\psi_{\mathrm{I}_{1}}(x_{1})\psi^{\mathrm{I}_{2}}(x_{2})h^{\alpha\beta}(x_{3})\}\vert\Omega\rangle\big\vert_{\mathcal{O}(\kappa)}$,
with the corresponding Feynman rules highlighted in red.}
		\label{PsiI1PsiI2GRMajorana}
	\end{center}
\end{figure}

We begin with the interaction between a single graviton and two four-component fermions at the $\mathcal{O}(\kappa)$ level
\begin{align}
\nonumber
S_{\text{int}}&=\frac{\kappa}{2}\int d^{4}x\,\big\{\frac{\text{i}}{4}\bar{\Psi}_{\text{M}}\big(\partial_{\mu}h\gamma^{\mu}-\partial_{(\mu}h^{\mu\nu}\gamma_{\nu)}\big)\Psi_{\text{M}}\\
\label{InteractionGravitionDiracFermion}
&-\frac{\text{i}}{2}\bar{\Psi}_{\text{M}}h^{\mu\nu}\gamma_{(\mu}\partial_{\nu)}\Psi_{\text{M}}+\frac{h}{2}(\text{i}\bar{\Psi}_{\text{M}}\gamma^{\mu}\partial_{\mu}\Psi_{\text{M}}-m_{\psi}\bar{\Psi}_{\text{M}}\Psi_{\text{M}})\big\}
\end{align}Further details of this interaction can be found in the research works \cite{Barman:2023ymn,Lee:2025lyk}. Corresponding to the chiral representation of the Gamma matrices, the four-component wavefunction $\Psi_{\text{M}}$ can be decomposed into two two-component Majorana wavefunctions
\begin{align}
\label{MajoranaWavefunctionToDirac}
&\Psi_{\text{M}}(x)=\left(\begin{array}{c}
\psi_{\mathrm{I}}(x)\\
\psi^{\dagger\dot{\mathrm{I}}}(x)
\end{array}\right)~,~\bar{\Psi}_{\text{M}}(x)=\left(\begin{array}{cc}
\psi^{\mathrm{J}}(x) & \psi_{\dot{\mathrm{J}}}^{\dagger}(x)\end{array}\right)
\end{align}After substituting Eq. \eqref{MajoranaWavefunctionToDirac} into the interaction Lagrangian \eqref{InteractionGravitionDiracFermion}, the interaction between a single graviton and two Majorana fermions is generated by\begin{align}
\label{YukawaGravitonMajorana}
&S_{\text{int}}^{(h\psi\psi)}=\frac{\kappa}{4}\int d^{4}x\,\big\{\text{i}(h\eta^{\mu\nu}-h^{\mu\nu})(\psi\sigma_{\mu}\partial_{\nu}\psi^{\dagger}+\psi^{\dagger}\bar{\sigma}_{\mu}\partial_{\nu}\psi)-mh(\psi\psi+\psi^{\dagger}\psi^{\dagger})\big\}
\end{align}From \eqref{YukawaGravitonMajorana}, the corresponding generating functional up to $\mathcal{O}(\kappa)$ is constructed as
\begin{small}
\begin{align}
\nonumber
&W[J_{\mu\nu},\mathcal{J}^{\dagger},\mathcal{J}]=\mathcal{N}\int\mathcal{D}\psi\mathcal{D}\psi^{\dagger}\mathcal{D}h_{\mu\nu}\,\text{e}^{\frac{\text{i}\kappa}{4}\int d^{4}x\big\{\text{i}(h\eta^{\mu\nu}-h^{\mu\nu})(\psi\sigma_{\mu}\partial_{\nu}\psi^{\dagger}+\psi^{\dagger}\bar{\sigma}_{\mu}\partial_{\nu}\psi)-mh(\psi\psi+\psi^{\dagger}\psi^{\dagger})\big\}}\\
\nonumber
&\quad\quad\quad\quad\quad\quad\times\text{e}^{\text{i}\int d^{4}x\big\{\frac{1}{2}(\text{i}\psi^{\dagger}\bar{\sigma}^{\mu}\partial_{\mu}\psi+\text{i}\psi\sigma^{\mu}\partial_{\mu}\psi^{\dagger}-m\psi\psi-m\psi^{\dagger}\psi^{\dagger})-\frac{1}{2}h^{\alpha\beta}\hat{\Box}_{(\alpha\beta)(\rho\lambda)}h^{\rho\lambda}+\mathcal{J}\cdot\psi+\psi^{\dagger}\cdot\mathcal{J}^{\dagger}+J_{\alpha\beta}h^{\alpha\beta}\big\}}\\
&\quad\quad\quad\quad\quad\quad=W_{0}[J_{\mu\nu},\mathcal{J}^{\dagger},\mathcal{J}]+W[J_{\mu\nu},\mathcal{J}^{\dagger},\mathcal{J}]\big\vert_{\mathcal{O}(\kappa)}+\mathcal{O}(\kappa^{2})
\end{align}
\end{small}where the generating functional at $\mathcal{O}(\kappa)$ is explicitly given by
\begin{scriptsize}
\begin{align}
\nonumber
&W[J_{\alpha\beta},\mathcal{J}^{\dagger},\mathcal{J}]\big\vert_{\mathcal{O}(\kappa)}=-\frac{\text{i}\kappa}{4}\int d^{4}x\big\{(\eta_{\alpha\beta}\eta^{\mu\nu}\frac{\delta}{\delta J_{\alpha\beta}}-\frac{\delta}{\delta J_{\mu\nu}})_{x}\\
\nonumber
&\times\big(\frac{\overrightarrow{\delta}}{\delta\mathcal{J}_{\mathrm{I}}}(\sigma_{\mu})_{\mathrm{I}\dot{\mathrm{I}}}W_{0}[J_{\alpha\beta},\mathcal{J}^{\dagger},\mathcal{J}]\partial_{\nu}(\frac{\overleftarrow{\delta}}{\delta\mathcal{J}_{\dot{\mathrm{I}}}^{\dagger}})+\frac{\overrightarrow{\delta}}{\delta\mathcal{J}^{\dagger\dot{\mathrm{I}}}}(\bar{\sigma}_{\mu})^{\dot{\mathrm{I}}\mathrm{I}}W_{0}[J_{\alpha\beta},\mathcal{J}^{\dagger},\mathcal{J}]\partial_{\nu}(\frac{\overleftarrow{\delta}}{\delta\mathcal{J}^{\mathrm{I}}})\big)_{x}\\
\nonumber
&+\text{i}m\eta_{\alpha\beta}(\frac{\delta}{\delta J_{\alpha\beta}})_{x}\big(\frac{\overrightarrow{\delta}}{\delta\mathcal{J}_{\mathrm{I}}}W_{0}[J_{\alpha\beta},\mathcal{J}^{\dagger},\mathcal{J}]\frac{\overleftarrow{\delta}}{\delta\mathcal{J}^{\mathrm{I}}}+\frac{\overrightarrow{\delta}}{\delta\mathcal{J}^{\dagger\dot{\mathrm{I}}}}W_{0}[J_{\alpha\beta},\mathcal{J}^{\dagger},\mathcal{J}]\frac{\overleftarrow{\delta}}{\delta\mathcal{J}_{\dot{\mathrm{I}}}^{\dagger}}\big)_{x}\big\}\\
\nonumber
&=\frac{\kappa}{4}\int d^{4}y\prod_{i=1}^{3}\int\frac{d^{4}k_{i}}{(2\pi)^{4}}\text{e}^{\text{i}(k_{3}+k_{2}+k_{1})\cdot y}\bigg\{(k_{2}-k_{1})_{\nu}\big\{(\frac{\delta W_{0}[\tilde{J}_{\alpha\beta}]}{\delta\tilde{J}_{\alpha\beta}(k_{3})}\eta_{\alpha\beta}\eta^{\mu\nu}-\frac{\delta W_{0}[\tilde{J}_{\alpha\beta}]}{\delta\tilde{J}_{\mu\nu}(k_{3})})(\sigma_{\mu})_{\mathrm{I}\dot{\mathrm{I}}}\frac{\overrightarrow{\delta}}{\delta\tilde{\mathcal{J}}_{\mathrm{I}}(k_{1})}W_{0}[\tilde{\mathcal{J}}^{\dagger},\tilde{\mathcal{J}}]\frac{\overleftarrow{\delta}}{\delta\tilde{\mathcal{J}}_{\dot{\mathrm{I}}}^{\dagger}(-k_{2})}\big\}\\
\label{GeneraFunGravitonMajorana}
&+\,m\eta_{\alpha\beta}\,\frac{\delta}{\delta\tilde{J}_{\alpha\beta}(k_{3})}W_{0}[\tilde{J}_{\alpha\beta}]\big\{\frac{\overrightarrow{\delta}}{\delta\tilde{\mathcal{J}}_{\mathrm{I}}(k_{1})}W_{0}[\tilde{\mathcal{J}}^{\dagger},\tilde{\mathcal{J}}]\frac{\overleftarrow{\delta}}{\delta\tilde{\mathcal{J}}^{\mathrm{I}}(k_{2})}+\frac{\overrightarrow{\delta}}{\delta\tilde{\mathcal{J}}^{\dagger\dot{\mathrm{I}}}(-k_{1})}W_{0}[\tilde{\mathcal{J}}^{\dagger},\tilde{\mathcal{J}}]\frac{\overleftarrow{\delta}}{\delta\tilde{\mathcal{J}}_{\dot{\mathrm{I}}}^{\dagger}(-k_{2})}\big\}\bigg\}
\end{align}
\end{scriptsize}Based on \eqref{GeneraFunGravitonMajorana}, the various three-point correlation functions can be evaluated as
\begin{small}
\begin{align}
\nonumber
&\langle\Omega\vert\mathcal{T}\{\psi_{\mathrm{I}_{1}}(x_{1})\psi^{\mathrm{I}_{2}}(x_{2})h^{\alpha\beta}(x_{3})\}\vert\Omega\rangle\big\vert_{\mathcal{O}(\kappa)}=\frac{1}{\text{i}^{3}}\frac{1}{W_{0}[0,0,0]}\int\frac{d^{4}q_{1}}{(2\pi)^{4}}\text{e}^{-\text{i}q_{1}\cdot x_{1}}\int\frac{d^{4}q_{2}}{(2\pi)^{4}}\text{e}^{\text{i}q_{2}\cdot x_{2}}\int\frac{d^{4}q_{3}}{(2\pi)^{4}}\text{e}^{-\text{i}q_{3}\cdot x_{3}}\\
&\quad \times \big\{ \frac{\overrightarrow{\delta}}{\delta\tilde{\mathcal{J}}^{\mathrm{I}_{1}}(-q_{1})}\frac{\delta W[J_{\mu\nu},\mathcal{J}^{\dagger},\mathcal{J}]\big\vert_{\mathcal{O}(\kappa)}}{\delta\tilde{J}_{\alpha\beta}(q_{3})}\frac{\overleftarrow{\delta}}{\delta\tilde{\mathcal{J}}_{\mathrm{I}_{2}}(q_{2})}\big\}\bigg\vert_{\tilde{J}_{\bar{\alpha}\bar{\beta}}=\tilde{\mathcal{J}}=\tilde{\mathcal{J}}^{\dagger}=0}^{\text{connected}}\\
\nonumber
&=\int d^{4}y\int\frac{d^{4}q_{1}}{(2\pi)^{4}}\text{e}^{-\text{i}q_{1}\cdot(x_{1}-y)}\int\frac{d^{4}q_{2}}{(2\pi)^{4}}\text{e}^{\text{i}q_{2}\cdot(x_{2}-y)}\int\frac{d^{4}q_{3}}{(2\pi)^{4}}\text{e}^{-\text{i}q_{3}\cdot(x_{3}-y)}\big(\frac{-\text{i}\mathcal{C}^{(\alpha\beta)(\tilde{\alpha}\tilde{\beta})}(q_{3})}{q_{3}^{2}+\text{i}\epsilon}\big)\\
\nonumber
&\quad \times\bigg\{\frac{\text{i}m\delta_{\mathrm{I}_{1}}^{~~\tilde{\mathrm{I}}_{1}}}{q_{1}^{2}-m^{2}}\times\underbrace{\frac{\text{i}\kappa}{4}(\eta_{\tilde{\alpha}\tilde{\beta}}\eta^{\mu\nu}-\delta_{(\tilde{\alpha}\tilde{\beta})}^{(\mu\nu)})(\sigma_{(\mu}(q_{2}+q_{1})_{\nu)})_{\tilde{\mathrm{I}}_{1}\dot{\tilde{\mathrm{I}}}_{2}}}_{\mathcal{V}_{(h\psi\psi)}^{((\tilde{\alpha}\tilde{\beta}),q_{1},q_{2},\widetilde{\mathrm{I}}_{1,\downarrow},\dot{\widetilde{\mathrm{I}}}_{2,\downarrow})}}\times\frac{(\text{i}q_{2}\cdot\bar{\sigma})^{\dot{\tilde{\mathrm{I}}}_{2}\mathrm{I}_{2}}}{q_{2}^{2}-m^{2}}\\
\nonumber
&\quad+\frac{(\text{i}q_{1}\cdot\sigma)_{\mathrm{I}_{1}\dot{\tilde{\mathrm{I}}}_{1}}}{q_{1}^{2}-m^{2}}\times\underbrace{\frac{\text{i}\kappa}{4}(\eta_{\tilde{\alpha}\tilde{\beta}}\eta^{\mu\nu}-\delta_{(\tilde{\alpha}\tilde{\beta})}^{(\mu\nu)})(\bar{\sigma}_{(\mu}(q_{2}+q_{1})_{\nu)})^{\dot{\tilde{\mathrm{I}}}_{1}\tilde{\mathrm{I}}_{2}}}_{\mathcal{V}_{(h\psi\psi)}^{((\tilde{\alpha}\tilde{\beta}),q_{1},q_{2},\dot{\widetilde{\mathrm{I}}}_{1,\uparrow},\widetilde{\mathrm{I}}_{2,\uparrow})}}\times\frac{\text{i}m\delta_{\tilde{\mathrm{I}}_{2}}^{~~\mathrm{I}_{2}}}{q_{2}^{2}-m^{2}}\\
\label{GravitonPsiI1PsiI2}
&\quad+\frac{\text{i}m\delta_{\mathrm{I}_{1}}^{~~\tilde{\mathrm{I}}_{1}}}{q_{1}^{2}-m^{2}}\!\times\!\underbrace{(\boldsymbol{-}\frac{\text{i}\kappa}{2}m\eta_{\tilde{\alpha}\tilde{\beta}}\delta_{\tilde{\mathrm{I}}_{1}}^{~~\tilde{\mathrm{I}}_{2}})}_{\mathcal{V}_{(h\psi\psi)}^{((\tilde{\alpha}\tilde{\beta}),q_{1},q_{2},\widetilde{\mathrm{I}}_{1,\downarrow},\widetilde{\mathrm{I}}_{2,\uparrow})}}\!\times\!\frac{\text{i}m\delta_{\tilde{\mathrm{I}}_{2}}^{~~\mathrm{I}_{2}}}{q_{2}^{2}-m^{2}}+\frac{(\text{i}q_{1}\cdot\sigma)_{\mathrm{I}_{1}\dot{\tilde{\mathrm{I}}}_{1}}}{q_{1}^{2}-m^{2}}\!\times\!\underbrace{(\boldsymbol{-}\frac{\text{i}\kappa}{2}m\eta_{\tilde{\alpha}\tilde{\beta}}\delta_{~~~\dot{\tilde{\mathrm{I}}}_{2}}^{\dot{\tilde{\mathrm{I}}}_{1}})}_{\mathcal{V}_{(h\psi\psi)}^{((\tilde{\alpha}\tilde{\beta}),q_{1},q_{2},\dot{\widetilde{\mathrm{I}}}_{1,\uparrow},\dot{\widetilde{\mathrm{I}}}_{2,\downarrow})}}\!\times\!\frac{(\text{i}q_{2}\cdot\bar{\sigma})^{\dot{\tilde{\mathrm{I}}}_{2}\mathrm{I}_{2}}}{q_{2}^{2}-m^{2}}\bigg\}
\end{align}
\end{small}and
\begin{small}
\begin{align}
\nonumber
&\langle\Omega\vert\mathcal{T}\{\psi^{\dagger\dot{\mathrm{I}}_{1}}(x_{1})\psi^{\mathrm{I}_{2}}(x_{2})h^{\alpha\beta}(x_{3})\}\vert\Omega\rangle\big\vert_{\mathcal{O}(\kappa)}=\frac{1}{\text{i}^{3}}\frac{1}{W_{0}[0,0,0]}\int\frac{d^{4}q_{1}}{(2\pi)^{4}}\text{e}^{-\text{i}q_{1}\cdot x_{1}}\int\frac{d^{4}q_{2}}{(2\pi)^{4}}\text{e}^{\text{i}q_{2}\cdot x_{2}}\int\frac{d^{4}q_{3}}{(2\pi)^{4}}\text{e}^{-\text{i}q_{3}\cdot x_{3}}\\
&\quad\quad \times \big\{\frac{\overrightarrow{\delta}}{\delta\tilde{\mathcal{J}}_{\dot{\mathrm{I}}_{1}}^{\dagger}(q_{1})}\frac{\delta W[J_{\mu\nu},\mathcal{J}^{\dagger},\mathcal{J}]\big\vert_{\mathcal{O}(\kappa)}}{\delta\tilde{J}_{\alpha\beta}(q_{3})}\frac{\overleftarrow{\delta}}{\delta\tilde{\mathcal{J}}_{\mathrm{I}_{2}}(q_{2})}\big\}\bigg\vert_{\tilde{J}_{\bar{\alpha}\bar{\beta}}=\tilde{\mathcal{J}}=\tilde{\mathcal{J}}^{\dagger}=0}^{\text{connected}}\\
\nonumber
&=\int d^{4}y\int\frac{d^{4}q_{1}}{(2\pi)^{4}}\text{e}^{-\text{i}q_{1}\cdot(x_{1}-y)}\int\frac{d^{4}q_{2}}{(2\pi)^{4}}\text{e}^{\text{i}q_{2}\cdot(x_{2}-y)}\int\frac{d^{4}q_{3}}{(2\pi)^{4}}\text{e}^{-\text{i}q_{3}\cdot(x_{3}-y)}\big(\frac{-\text{i}\mathcal{C}^{(\alpha\beta)(\tilde{\alpha}\tilde{\beta})}(q_{3})}{q_{3}^{2}+\text{i}\epsilon}\big)\\
\nonumber
&\quad\quad \times\bigg\{\frac{(\text{i}q_{1}\cdot\bar{\sigma})^{\dot{\mathrm{I}}_{1}\tilde{\mathrm{I}}_{1}}}{q_{1}^{2}-m^{2}}\times\underbrace{\frac{\text{i}\kappa}{4}(\eta_{\tilde{\alpha}\tilde{\beta}}\eta^{\mu\nu}-\delta_{(\tilde{\alpha}\tilde{\beta})}^{(\mu\nu)})(\sigma_{(\mu}(q_{2}+q_{1})_{\nu)})_{\tilde{\mathrm{I}}_{1}\dot{\tilde{\mathrm{I}}}_{2}}}_{\mathcal{V}_{(h\psi\psi)}^{((\tilde{\alpha}\tilde{\beta}),q_{1},q_{2},\widetilde{\mathrm{I}}_{1,\downarrow},\dot{\widetilde{\mathrm{I}}}_{2,\downarrow})}}\times\frac{(\text{i}q_{2}\cdot\bar{\sigma})^{\dot{\tilde{\mathrm{I}}}_{2}\mathrm{I}_{2}}}{q_{2}^{2}-m^{2}}\\
\nonumber
&\quad\quad+\frac{\text{i}m\delta_{~~\dot{\tilde{\mathrm{I}}}_{1}}^{\dot{\mathrm{I}}_{1}}}{q_{1}^{2}-m^{2}}\times\underbrace{\frac{\text{i}\kappa}{4}(\eta_{\tilde{\alpha}\tilde{\beta}}\eta^{\mu\nu}-\delta_{(\tilde{\alpha}\tilde{\beta})}^{(\mu\nu)})(\bar{\sigma}_{(\mu}(q_{2}+q_{1})_{\nu)})^{\dot{\tilde{\mathrm{I}}}_{1}\tilde{\mathrm{I}}_{2}}}_{\mathcal{V}_{(h\psi\psi)}^{((\tilde{\alpha}\tilde{\beta}),q_{1},q_{2},\dot{\widetilde{\mathrm{I}}}_{1,\uparrow},\widetilde{\mathrm{I}}_{2,\uparrow})}}\times\frac{\text{i}m\delta_{\tilde{\mathrm{I}}_{2}}^{~~\mathrm{I}_{2}}}{q_{2}^{2}-m^{2}}\\
\label{GravitonPsiDaggerI1PsiI2}
&\quad\quad+\frac{(\text{i}q_{1}\cdot\bar{\sigma})^{\dot{\mathrm{I}}_{1}\tilde{\mathrm{I}}_{1}}}{q_{1}^{2}-m^{2}}\!\times\!\underbrace{(-\frac{\text{i}\kappa}{2}m\eta_{\tilde{\alpha}\tilde{\beta}}\delta_{\tilde{\mathrm{I}}_{1}}^{~~\tilde{\mathrm{I}}_{2}})}_{\mathcal{V}_{(h\psi\psi)}^{((\tilde{\alpha}\tilde{\beta}),q_{1},q_{2},\widetilde{\mathrm{I}}_{1,\downarrow},\widetilde{\mathrm{I}}_{2,\uparrow})}}\!\times\!\frac{\text{i}m\delta_{\tilde{\mathrm{I}}_{2}}^{~~\mathrm{I}_{2}}}{q_{2}^{2}-m^{2}}+\frac{\text{i}m\delta_{~~\dot{\tilde{\mathrm{I}}}_{1}}^{\dot{\mathrm{I}}_{1}}}{q_{1}^{2}-m^{2}}\!\times\!\underbrace{(-\frac{\text{i}\kappa}{2}m\eta_{\tilde{\alpha}\tilde{\beta}}\delta_{~~\dot{\tilde{\mathrm{I}}}_{2}}^{\dot{\tilde{\mathrm{I}}}_{1}})}_{\mathcal{V}_{(h\psi\psi)}^{((\tilde{\alpha}\tilde{\beta}),q_{1},q_{2},\dot{\widetilde{\mathrm{I}}}_{1,\uparrow},\dot{\widetilde{\mathrm{I}}}_{2,\downarrow})}}\!\times\!\frac{(\text{i}q_{2}\cdot\bar{\sigma})^{\dot{\tilde{\mathrm{I}}}_{2}\mathrm{I}_{2}}}{q_{2}^{2}-m^{2}}\bigg\}
\end{align}
\end{small}and
\begin{small}
\begin{align}
\nonumber
&\langle\Omega\vert\mathcal{T}\{\psi^{\dagger\dot{\mathrm{I}}_{1}}(x_{1})\psi_{\dot{\mathrm{I}}_{2}}^{\dagger}(x_{2})h^{\alpha\beta}(x_{3})\}\vert\Omega\rangle\big\vert_{\mathcal{O}(\kappa)}=\frac{1}{\text{i}^{3}}\frac{1}{W_{0}[0,0,0]}\int\frac{d^{4}q_{1}}{(2\pi)^{4}}\text{e}^{-\text{i}q_{1}\cdot x_{1}}\int\frac{d^{4}q_{2}}{(2\pi)^{4}}\text{e}^{\text{i}q_{2}\cdot x_{2}}\int\frac{d^{4}q_{3}}{(2\pi)^{4}}\text{e}^{-\text{i}q_{3}\cdot x_{3}}\\
&\quad\quad \times\big\{\frac{\overrightarrow{\delta}}{\delta\tilde{\mathcal{J}}_{\dot{\mathrm{I}}_{1}}^{\dagger}(q_{1})}\frac{\delta W[J_{\mu\nu},\mathcal{J}^{\dagger},\mathcal{J}]\big\vert_{\mathcal{O}(\kappa)}}{\delta\tilde{J}_{\alpha\beta}(q_{3})}\frac{\overleftarrow{\delta}}{\delta\tilde{\mathcal{J}}^{\dagger\dot{\mathrm{I}}_{2}}(-q_{2})}\big\}\bigg\vert_{\tilde{J}_{\alpha\beta}=\tilde{\mathcal{J}}=\tilde{\mathcal{J}}^{\dagger}=0}^{\text{connected}}\\
\nonumber
&=\int d^{4}y\int\frac{d^{4}q_{1}}{(2\pi)^{4}}\text{e}^{-\text{i}q_{1}\cdot(x_{1}-y)}\int\frac{d^{4}q_{2}}{(2\pi)^{4}}\text{e}^{\text{i}q_{2}\cdot(x_{2}-y)}\int\frac{d^{4}q_{3}}{(2\pi)^{4}}\text{e}^{-\text{i}q_{3}\cdot(x_{3}-y)}\frac{\big(-\text{i}\mathcal{C}^{(\alpha\beta)(\tilde{\alpha}\tilde{\beta})}(q_{3})\big)}{q_{3}^{2}+\text{i}\epsilon}\\
\nonumber
&\quad\quad\times\bigg\{\frac{\text{i}m\delta_{~~\dot{\widetilde{\mathrm{I}}}_{1}}^{\dot{\mathrm{I}}_{1}}}{q_{1}^{2}-m^{2}}\times\underbrace{\frac{\text{i}\kappa}{4}(\eta_{\tilde{\alpha}\tilde{\beta}}\eta^{\mu\nu}-\delta_{(\tilde{\alpha}\tilde{\beta})}^{(\mu\nu)})(\bar{\sigma}_{(\mu}(q_{1}+q_{2})_{\nu)})^{\dot{\widetilde{\mathrm{I}}}_{1}\widetilde{\mathrm{I}}_{2}}}_{\mathcal{V}_{(h\psi\psi)}^{((\tilde{\alpha}\tilde{\beta}),q_{1},q_{2},\dot{\widetilde{\mathrm{I}}}_{1,\uparrow},\widetilde{\mathrm{I}}_{2,\uparrow})}}\times\frac{(\text{i}q_{2}\cdot\sigma)_{\widetilde{\mathrm{I}}_{2}\dot{\mathrm{I}}_{2}}}{q_{2}^{2}-m^{2}}\\
\nonumber
&\quad\quad+\frac{(\text{i}q_{1}\cdot\bar{\sigma})^{\dot{\mathrm{I}}_{1}\tilde{\mathrm{I}}_{1}}}{q_{1}^{2}-m^{2}}\times\underbrace{\frac{\text{i}\kappa}{4}(\eta_{\tilde{\alpha}\tilde{\beta}}\eta^{\mu\nu}-\delta_{(\tilde{\alpha}\tilde{\beta})}^{(\mu\nu)})(\sigma_{(\mu}(q_{2}+q_{1})_{\nu)})_{\tilde{\mathrm{I}}_{1}\dot{\tilde{\mathrm{I}}}_{2}}}_{\mathcal{V}_{(h\psi\psi)}^{((\tilde{\alpha}\tilde{\beta}),q_{1},q_{2},\widetilde{\mathrm{I}}_{1,\downarrow},\dot{\widetilde{\mathrm{I}}}_{2,\downarrow})}}\times\frac{\text{i}m\delta_{~~~\dot{\mathrm{I}}_{2}}^{\dot{\tilde{\mathrm{I}}}_{2}}}{q_{2}^{2}-m^{2}}\\
\label{GravitonPsiDaggerI1PsiDaggerI2}
&\quad\quad+\frac{(\text{i}q_{1}\cdot\bar{\sigma})^{\dot{\mathrm{I}}_{1}\tilde{\mathrm{I}}_{1}}}{q_{1}^{2}-m^{2}}\!\times\!\underbrace{(-\frac{\text{i}\kappa}{2}m\eta_{\tilde{\alpha}\tilde{\beta}}\delta_{\tilde{\mathrm{I}}_{1}}^{~~\tilde{\mathrm{I}}_{2}})}_{\mathcal{V}_{(h\psi\psi)}^{((\tilde{\alpha}\tilde{\beta}),q_{1},q_{2},\widetilde{\mathrm{I}}_{1,\downarrow},\widetilde{\mathrm{I}}_{2,\uparrow})}}\!\times\!\frac{(\text{i}q_{2}\cdot\sigma)_{\tilde{\mathrm{I}}_{2}\dot{\mathrm{I}}_{2}}}{q_{2}^{2}-m^{2}}+\frac{\text{i}m\delta_{~~\dot{\tilde{\mathrm{I}}}_{1}}^{\dot{\mathrm{I}}_{1}}}{q_{1}^{2}-m^{2}}\!\times\!\underbrace{(-\frac{\text{i}\kappa}{2}m\eta_{\tilde{\alpha}\tilde{\beta}}\delta_{~~\dot{\tilde{\mathrm{I}}}_{2}}^{\dot{\tilde{\mathrm{I}}}_{1}})}_{\mathcal{V}_{(h\psi\psi)}^{((\tilde{\alpha}\tilde{\beta}),q_{1},q_{2},\dot{\widetilde{\mathrm{I}}}_{1,\uparrow},\dot{\widetilde{\mathrm{I}}}_{2,\downarrow})}}\!\times\!\frac{\text{i}m\delta_{~~~\dot{\mathrm{I}}_{2}}^{\dot{\tilde{\mathrm{I}}}_{2}}}{q_{2}^{2}-m^{2}}\bigg\}
\end{align}
\end{small}and
\begin{small}
\begin{align}
\nonumber
&\langle\Omega\vert\mathcal{T}\{\psi_{\mathrm{I}_{1}}(x_{1})\psi_{\dot{\mathrm{I}}_{2}}^{\dagger}(x_{2})h^{\alpha\beta}(x_{3})\}\vert\Omega\rangle\big\vert_{\mathcal{O}(\kappa)}=\frac{1}{\text{i}^{3}}\frac{1}{W_{0}[0,0,0]}\int\frac{d^{4}q_{1}}{(2\pi)^{4}}\text{e}^{-\text{i}q_{1}\cdot x_{1}}\int\frac{d^{4}q_{2}}{(2\pi)^{4}}\text{e}^{\text{i}q_{2}\cdot x_{2}}\int\frac{d^{4}q_{3}}{(2\pi)^{4}}\text{e}^{-\text{i}q_{3}\cdot x_{3}}\\
&\quad\quad\times\big\{\frac{\overrightarrow{\delta}}{\delta\tilde{\mathcal{J}}^{\mathrm{I}_{1}}(-q_{1})}\frac{\delta W[J_{\mu\nu},\mathcal{J}^{\dagger},\mathcal{J}]\big\vert_{\mathcal{O}(\kappa)}}{\delta\tilde{J}_{\alpha\beta}(q_{3})}\frac{\overleftarrow{\delta}}{\delta\tilde{\mathcal{J}}^{\dagger\dot{\mathrm{I}}_{2}}(-q_{2})}\big\}\bigg\vert_{\tilde{J}_{\alpha\beta}=\tilde{\mathcal{J}}=\tilde{\mathcal{J}}^{\dagger}=0}^{\text{connected}}\\
\nonumber
&=\int d^{4}y\int\frac{d^{4}q_{1}}{(2\pi)^{4}}\text{e}^{-\text{i}q_{1}\cdot(x_{1}-y)}\int\frac{d^{4}q_{2}}{(2\pi)^{4}}\text{e}^{\text{i}q_{2}\cdot(x_{2}-y)}\int\frac{d^{4}q_{3}}{(2\pi)^{4}}\text{e}^{-\text{i}q_{3}\cdot(x_{3}-y)}\frac{\big(-\text{i}\mathcal{C}^{(\alpha\beta)(\tilde{\alpha}\tilde{\beta})}(q_{3})\big)}{q_{3}^{2}+\text{i}\epsilon}\\
\nonumber
&\quad\quad\times\bigg\{\frac{(\text{i}q_{1}\cdot\sigma)_{\mathrm{I}_{1}\dot{\widetilde{\mathrm{I}}}_{1}}}{q_{1}^{2}-m^{2}}\times\underbrace{\frac{\text{i}\kappa}{4}(\eta_{\tilde{\alpha}\tilde{\beta}}\eta^{\mu\nu}-\delta_{(\tilde{\alpha}\tilde{\beta})}^{(\mu\nu)})(\bar{\sigma}_{(\mu}(q_{2}+q_{1})_{\nu)})^{\dot{\widetilde{\mathrm{I}}}_{1}\tilde{\mathrm{I}}_{2}}}_{\mathcal{V}_{(h\psi\psi)}^{((\tilde{\alpha}\tilde{\beta}),q_{1},q_{2},\dot{\widetilde{\mathrm{I}}}_{1,\uparrow},\widetilde{\mathrm{I}}_{2,\uparrow})}}\times\frac{(\text{i}q_{2}\cdot\sigma)_{\tilde{\mathrm{I}}_{2}\dot{\mathrm{I}}_{2}}}{q_{2}^{2}-m^{2}}\\
\nonumber
&\quad\quad+\frac{\text{i}m\delta_{\mathrm{I}_{1}}^{~~\tilde{\mathrm{I}}_{1}}}{q_{1}^{2}-m^{2}}\times\underbrace{\frac{\text{i}\kappa}{4}(\eta_{\tilde{\alpha}\tilde{\beta}}\eta^{\mu\nu}-\delta_{(\tilde{\alpha}\tilde{\beta})}^{(\mu\nu)})(\sigma_{(\mu}(q_{2}+q_{1})_{\nu)})_{\tilde{\mathrm{I}}_{1}\dot{\tilde{\mathrm{I}}}_{2}}}_{\mathcal{V}_{(h\psi\psi)}^{((\tilde{\alpha}\tilde{\beta}),q_{1},q_{2},\widetilde{\mathrm{I}}_{1,\downarrow},\dot{\widetilde{\mathrm{I}}}_{2,\downarrow})}}\times\frac{\text{i}m\delta_{~~\dot{\mathrm{I}}_{2}}^{\dot{\tilde{\mathrm{I}}}_{2}}}{q_{2}^{2}-m^{2}}\\
\label{GravitonPsiI1PsiDaggerI2}
&\quad\quad+\frac{\text{i}m\delta_{\mathrm{I}_{1}}^{~~\tilde{\mathrm{I}}_{1}}}{q_{1}^{2}-m^{2}}\!\times\!\underbrace{(-\frac{\text{i}\kappa}{2}m\eta_{\tilde{\alpha}\tilde{\beta}}\delta_{\tilde{\mathrm{I}}_{1}}^{~~\tilde{\mathrm{I}}_{2}})}_{\mathcal{V}_{(h\psi\psi)}^{((\tilde{\alpha}\tilde{\beta}),q_{1},q_{2},\widetilde{\mathrm{I}}_{1,\downarrow},\widetilde{\mathrm{I}}_{2,\uparrow})}}\!\times\!\frac{(\text{i}q_{2}\cdot\sigma)_{\tilde{\mathrm{I}}_{2}\dot{\mathrm{I}}_{2}}}{q_{2}^{2}-m^{2}}+\frac{(\text{i}q_{1}\cdot\sigma)_{\mathrm{I}_{1}\dot{\tilde{\mathrm{I}}}_{1}}}{q_{1}^{2}-m^{2}}\!\times\!\underbrace{(-\frac{\text{i}\kappa}{2}m\eta_{\tilde{\alpha}\tilde{\beta}}\delta_{~~\dot{\tilde{\mathrm{I}}}_{2}}^{\dot{\tilde{\mathrm{I}}}_{1}})}_{\mathcal{V}_{(h\psi\psi)}^{((\tilde{\alpha}\tilde{\beta}),q_{1},q_{2},\dot{\widetilde{\mathrm{I}}}_{1,\uparrow},\dot{\widetilde{\mathrm{I}}}_{2,\downarrow})}}\!\times\!\frac{\text{i}m\delta_{~~\dot{\mathrm{I}}_{2}}^{\dot{\tilde{\mathrm{I}}}_{2}}}{q_{2}^{2}-m^{2}}\!\bigg\}
\end{align}
\end{small}
From the above results concerning the three-point correlation functions at the $\mathcal{O}(\kappa)$ level, it can be seen that there are, in fact, only four independent interaction vertices, i.e., $\mathcal{V}_{(h\psi\psi)}^{((\tilde{\alpha}\tilde{\beta}),q_{1},q_{2},\widetilde{\mathrm{I}}_{1,\downarrow},\dot{\widetilde{\mathrm{I}}}_{2,\downarrow})}$, $\mathcal{V}_{(h\psi\psi)}^{((\tilde{\alpha}\tilde{\beta}),q_{1},q_{2},\dot{\widetilde{\mathrm{I}}}_{1,\uparrow},\widetilde{\mathrm{I}}_{2,\uparrow})}$, $\mathcal{V}_{(h\psi\psi)}^{((\tilde{\alpha}\tilde{\beta}),q_{1},q_{2},\widetilde{\mathrm{I}}_{1,\downarrow},\widetilde{\mathrm{I}}_{2,\uparrow})}$, $\mathcal{V}_{(h\psi\psi)}^{((\tilde{\alpha}\tilde{\beta}),q_{1},q_{2},\dot{\widetilde{\mathrm{I}}}_{1,\uparrow},\dot{\widetilde{\mathrm{I}}}_{2,\downarrow})}$. For instance, from the three-point correlator $\langle\Omega\vert\mathcal{T}\{\psi_{\mathrm{I}_{1}}(x_{1})\psi^{\mathrm{I}_{2}}(x_{2})h^{\alpha\beta}(x_{3})\}\vert\Omega\rangle\big\vert_{\mathcal{O}(\kappa)}$, the relevant Feynman diagrams can be extracted, as shown in Fig. \ref{PsiI1PsiI2GRMajorana}. In Eqs. \eqref{GravitonPsiDaggerI1PsiI2}–\eqref{GravitonPsiI1PsiDaggerI2}, we observe that they involve the same interaction vertices as in \eqref{GravitonPsiI1PsiI2}. Therefore, it is unnecessary to explicitly draw the Feynman diagrams corresponding to Eqs. \eqref{GravitonPsiDaggerI1PsiI2}–\eqref{GravitonPsiI1PsiDaggerI2}; the Fig. \ref{PsiI1PsiI2GRMajorana} associated with \eqref{GravitonPsiI1PsiI2} alone is sufficient.

\section{Three-body decay process for Majorana fermionic case with graviton emission \label{EvaluThreeBodyDecay}}

In this section, we provide a detailed calculation of the differential decay rates for inflaton decays into three-body final states, with particular emphasis on processes in which a pair of Majorana fermions or scalar particles is produced via the interactions \eqref{YukawaInflatonMajorana}-\eqref{YukawaInflatonComScalar}, accompanied by the emission of a graviton.

By combining the Feynman rules presented in Fig. \ref{FigPsiI1PsiI2Inflatonpsipsi} and Fig. \ref{PsiI1PsiI2GRMajorana}, all possible Feynman diagrams describing the three-body decay process $\varphi(k)\to\psi^{\mathrm{I}_{1}}(s,p)\psi_{\mathrm{I}_{2}}(s^{\prime},q)e_{\bar{\alpha}\bar{\beta}}(\chi,l)$ are shown in Fig. \ref{RefTwoThreeBodyMixMajorana}–Fig. \ref{ThreeBodyMajoranaCase3Case4}. After performing the simplifications using the conditions
\begin{align}
&e^{\tilde{\alpha}\tilde{\beta}}(\boldsymbol{l},\chi)\eta_{\tilde{\alpha}\tilde{\beta}}=0~,~l^{\tilde{\alpha}}e_{\tilde{\alpha}\tilde{\beta}}(\boldsymbol{l},\chi)=0
\end{align}
for the polarization tensors $e^{\alpha\beta}(\boldsymbol{l},\chi)$, and simultaneously applying the on-shell conditions
\begin{align}
\label{MajoranaOnshellMomentumV1}
&(p\cdot\bar{\sigma})^{\dot{\mathrm{I}}\,\mathrm{I}_{1}}x_{\boldsymbol{p},s,\mathrm{I}_{1}}=my_{\boldsymbol{p},s}^{\dagger\dot{\mathrm{I}}}~,~(p\cdot\sigma)_{\mathrm{I}\,\dot{\mathrm{I}}_{1}}y_{\boldsymbol{p},s}^{\dagger\dot{\mathrm{I}}_{1}}=mx_{\boldsymbol{p},s,\mathrm{I}}\\
\label{MajoranaOnshellMomentumV2}
&(p\cdot\sigma)_{\mathrm{I}\,\dot{\mathrm{I}}_{1}}x_{\boldsymbol{p},s}^{\dagger\dot{\mathrm{I}}_{1}}=-my_{\boldsymbol{p},s,\mathrm{I}}~,~(p\cdot\bar{\sigma})^{\dot{\mathrm{I}}\,\mathrm{I}_{1}}y_{\boldsymbol{p},s,\mathrm{I}_{1}}=-mx_{\boldsymbol{p},s}^{\dagger\dot{\mathrm{I}}}\\
\label{MajoranaOnshellMomentumV3}
&x_{\boldsymbol{p},s}^{\mathrm{I}_{1}}(p\cdot\sigma)_{\mathrm{I}_{1}\,\dot{\mathrm{I}}}=-my_{\boldsymbol{p},s,\dot{\mathrm{I}}}^{\dagger}~,~y_{\boldsymbol{p},s,\dot{\mathrm{I}}_{1}}^{\dagger}(p\cdot\bar{\sigma})^{\dot{\mathrm{I}}_{1}\,\mathrm{I}}=-mx_{\boldsymbol{p},s}^{\mathrm{I}}\\
\label{MajoranaOnshellMomentumV4}
&x_{\boldsymbol{p},s,\dot{\mathrm{I}}_{1}}^{\dagger}(p\cdot\bar{\sigma})^{\dot{\mathrm{I}}_{1}\,\mathrm{I}}=my_{\boldsymbol{p},s}^{\mathrm{I}}~,~y_{\boldsymbol{p},s}^{\mathrm{I}_{1}}(p\cdot\sigma)_{\mathrm{I}_{1}\,\dot{\mathrm{I}}}=mx_{\boldsymbol{p},s,\dot{\mathrm{I}}}^{\dagger}
\end{align}
for Majorana spinors in momentum space, we obtain the following simplified expression for the amplitude that characterizes the three-body decay processes shown in Fig. \ref{RefTwoThreeBodyMixMajorana}–\ref{ThreeBodyMajoranaCase3Case4}. For clarity of notation, we denote by $\mathcal{M}_{\varphi\to\psi^{\mathrm{I}_{1}}\psi_{\mathrm{I}_{2}}e_{\alpha\beta}}^{\text{(part-I)}}$ the amplitudes of all three-body decay processes shown in Fig. \ref{RefTwoThreeBodyMixMajorana}, where the graviton is emitted from the inflaton scalar. Similarly, we denote by $\mathcal{M}_{\varphi\to\psi^{\mathrm{I}_{1}}\psi_{\mathrm{I}_{2}}e_{\alpha\beta}}^{\text{(part-II)}}$ the amplitudes of all three-body decay processes displayed in Fig. \ref{ThreeBodyMajoranaCase1Case2}-\ref{ThreeBodyMajoranaCase3Case4}, where the graviton is emitted from the Majorana fermion. The explicit expressions for $\mathcal{M}_{\varphi\to\psi^{\mathrm{I}_{1}}\psi_{\mathrm{I}_{2}}e_{\alpha\beta}}^{\text{(part-I)}}$ and $\mathcal{M}_{\varphi\to\psi^{\mathrm{I}_{1}}\psi_{\mathrm{I}_{2}}e_{\alpha\beta}}^{\text{(part-II)}}$ are given by
\begin{small}
\begin{align}
&\mathcal{M}_{\varphi(k)\to\psi^{\mathrm{I}_{1}}(s,p)\psi_{\mathrm{I}_{2}}(s^{\prime},q)e_{\alpha\beta}(\chi,l)}^{\text{(part-I)}}\!=\!\frac{\text{i}\mathsf{y}_{\text{D}}\kappa^{2}e_{\alpha_{1}\beta_{1}}^{(\chi)}\!(\boldsymbol{l})}{4k\cdot l}k^{\alpha_{1}}k^{\beta_{1}}\big(y_{\boldsymbol{p},s}\sigma^{\mu_{1}}x_{\boldsymbol{q},s^{\prime}}^{\dagger}\!+\!x_{\boldsymbol{p},s}^{\dagger}\bar{\sigma}^{\mu_{1}}y_{\boldsymbol{q},s^{\prime}}\big)(p\!-\!q)_{\mu_{1}}
\end{align}
\end{small}and
\begin{small}
\begin{align}
\nonumber
&\mathcal{M}_{\varphi(k)\to\psi^{\mathrm{I}_{1}}\!(s,p)\psi_{\mathrm{I}_{2}}\!(s^{\prime},q)e_{\alpha\beta}(\chi,l)}^{\text{(part-II)}}\!=\!-\frac{\text{i}m\mathsf{y}_{\text{D}}\kappa^{2}e^{(\chi),\mu_{1}\nu_{1}}\!(\boldsymbol{l})}{4(M^{2}-2k\cdot q)}\big(x_{\boldsymbol{p},s}^{\dagger}(\bar{\sigma}_{\mu_{1}}p_{\nu_{1}}k\cdot\sigma)x_{\boldsymbol{q},s^{\prime}}^{\dagger}\!+\!2m\,x_{\boldsymbol{p},s}^{\dagger}(\bar{\sigma}_{\mu_{1}}p_{\nu_{1}})y_{\boldsymbol{q},s^{\prime}}\big)\\
\nonumber
&\quad\quad\quad\quad\quad -\frac{\text{i}\mathsf{y}_{\text{D}}\kappa^{2}e^{(\chi),\mu_{1}\nu_{1}}\!(\boldsymbol{l})}{4(M^{2}\!-\!2k\cdot q)}\big((M^{2}\!+\!2m^{2}\!-\!2k\cdot q)\,y_{\boldsymbol{p},s}(\sigma_{\mu_{1}}p_{\nu_{1}})x_{\boldsymbol{q},s^{\prime}}^{\dagger}\!+\!my_{\boldsymbol{p},s}(\sigma_{\mu_{1}}p_{\nu_{1}}k\cdot\bar{\sigma})y_{\boldsymbol{q},s^{\prime}}\big)\\
\nonumber
&\quad\quad\quad\quad\quad-\frac{\text{i}m\mathsf{y}_{\text{D}}\kappa^{2}e^{(\chi),\mu_{1}\nu_{1}}(\boldsymbol{l})}{4(M^{2}-2k\cdot p)}\big(y_{\boldsymbol{p},s}(k\cdot\sigma\bar{\sigma}_{(\mu_{1}}q_{\nu_{1})})y_{\boldsymbol{q},s^{\prime}}-2mx_{\boldsymbol{p},s}^{\dagger}(\bar{\sigma}_{(\mu_{1}}q_{\nu_{1})})y_{\boldsymbol{q},s^{\prime}}\big)\\
\nonumber
&\quad\quad\quad\quad\quad+\frac{\text{i}\mathsf{y}_{\text{D}}\kappa^{2}e^{(\chi),\mu_{1}\nu_{1}}(\boldsymbol{l})}{4(M^{2}-2k\cdot p)}\big((M^{2}\!+\!2m^{2}\!-\!2k\cdot p)y_{\boldsymbol{p},s}(\sigma_{(\mu_{1}}q_{\nu_{1})})x_{\boldsymbol{q},s^{\prime}}^{\dagger}\!-\!mx_{\boldsymbol{p},s}^{\dagger}(k\cdot\bar{\sigma}\sigma_{(\mu_{1}}q_{\nu_{1})})x_{\boldsymbol{q},s^{\prime}}^{\dagger}\big)\\
\nonumber
&\quad\quad\quad\quad\quad-\frac{\text{i}m\mathsf{y}_{\text{D}}\kappa^{2}e^{(\chi),\mu_{1}\nu_{1}}\!(\boldsymbol{l})}{4(M^{2}-2k\cdot q)}\big(y_{\boldsymbol{p},s}(\sigma_{(\mu_{1}}p_{\nu_{1})}k\cdot\bar{\sigma})y_{\boldsymbol{q},s^{\prime}}+2my_{\boldsymbol{p},s}(\sigma_{(\mu_{1}}p_{\nu_{1})})x_{\boldsymbol{q},s^{\prime}}^{\dagger}\big)\\
\nonumber
&\quad\quad\quad\quad\quad-\frac{\text{i}\mathsf{y}_{\text{D}}\kappa^{2}e^{(\chi),\mu_{1}\nu_{1}}\!(\boldsymbol{l})}{4(M^{2}-2k\cdot q)}\big((M^{2}\!-\!2k\cdot q\!+\!2m^{2})x_{\boldsymbol{p},s}^{\dagger}(\bar{\sigma}_{(\mu_{1}}p_{\nu_{1})})y_{\boldsymbol{q},s^{\prime}}\!+\!mx_{\boldsymbol{p},s}^{\dagger}(\bar{\sigma}_{(\mu_{1}}p_{\nu_{1})}k\cdot\sigma)x_{\boldsymbol{q},s^{\prime}}^{\dagger}\big)\\
\nonumber
&\quad\quad\quad\quad\quad-\frac{\text{i}m\mathsf{y}_{\text{D}}\kappa^{2}e^{(\chi),\mu_{1}\nu_{1}}\!(\boldsymbol{l})}{4(M^{2}-2k\cdot p)}\big(x_{\boldsymbol{p},s}^{\dagger}(k\cdot\bar{\sigma}\sigma_{(\mu_{1}}q_{\nu_{1})})x_{\boldsymbol{q},s^{\prime}}^{\dagger}\!-\!2my_{\boldsymbol{p},s}(\sigma_{(\mu_{1}}q_{\nu_{1})})x_{\boldsymbol{q},s^{\prime}}^{\dagger}\big)\\
\nonumber
&\quad\quad\quad\quad\quad+\frac{\text{i}\mathsf{y}_{\text{D}}\kappa^{2}e^{(\chi),\mu_{1}\nu_{1}}\!(\boldsymbol{l})}{4(M^{2}-2k\cdot p)}\big((M^{2}-2k\cdot p+2m^{2})x_{\boldsymbol{p},s}^{\dagger}(\bar{\sigma}_{(\mu_{1}}q_{\nu_{1})})y_{\boldsymbol{q},s^{\prime}}\!-\!my_{\boldsymbol{p},s}(k\cdot\sigma\bar{\sigma}_{(\mu_{1}}q_{\nu_{1})})y_{\boldsymbol{q},s^{\prime}}\big)\\
\nonumber
&\quad\quad\quad\quad\quad=-\frac{\text{i}\mathsf{y}_{\text{D}}\kappa^{2}e^{\mu_{1}\nu_{1}}(\boldsymbol{l},\chi)}{4}\big\{\frac{M^{2}+4m^{2}-2k\cdot q}{M^{2}-2k\cdot q}\big(y_{\boldsymbol{p},s}\sigma_{(\mu_{1}}p_{\nu_{1})}x_{\boldsymbol{q},s^{\prime}}^{\dagger}+x_{\boldsymbol{p},s}^{\dagger}\bar{\sigma}_{(\mu_{1}}p_{\nu_{1})}y_{\boldsymbol{q},s^{\prime}}\big)\\
\nonumber
&\quad\quad\quad\quad\quad+\frac{2m}{M^{2}-2k\cdot q}\big(y_{\boldsymbol{p},s}\sigma_{(\mu_{1}}p_{\nu_{1})}(k\cdot\bar{\sigma})y_{\boldsymbol{q},s^{\prime}}+x_{\boldsymbol{p},s}^{\dagger}\bar{\sigma}_{(\mu_{1}}p_{\nu_{1})}(k\cdot\sigma)x_{\boldsymbol{q},s^{\prime}}^{\dagger}\big)\big\}\\
\nonumber
&\quad\quad\quad\quad\quad+\frac{\text{i}\mathsf{y}_{\text{D}}\kappa^{2}e^{\mu_{1}\nu_{1}}(\boldsymbol{l},\chi)}{4}\big\{\frac{M^{2}+4m^{2}-2k\cdot p}{M^{2}-2k\cdot p}\big(y_{\boldsymbol{p},s}\sigma_{(\mu_{1}}q_{\nu_{1})}x_{\boldsymbol{q},s^{\prime}}^{\dagger}+x_{\boldsymbol{p},s}^{\dagger}\bar{\sigma}_{(\mu_{1}}q_{\nu_{1})}y_{\boldsymbol{q},s^{\prime}}\big)\\
&\quad\quad\quad\quad\quad-\frac{2m}{M^{2}-2k\cdot p}\big(y_{\boldsymbol{p},s}(k\cdot\sigma)\bar{\sigma}_{(\mu_{1}}q_{\nu_{1})}y_{\boldsymbol{q},s^{\prime}}+x_{\boldsymbol{p},s}^{\dagger}(k\cdot\bar{\sigma})(\sigma_{(\mu_{1}}q_{\nu_{1})})x_{\boldsymbol{q},s^{\prime}}^{\dagger}\big)\big\}
\end{align}
\end{small}After adding these two contributions and performing the helicity–polarization sums of the corresponding squared amplitudes,
\begin{align}
&\sum_{s}x_{\boldsymbol{p},s,\mathrm{I}_{1}}x_{\boldsymbol{p},s,\dot{\mathrm{I}}_{2}}^{\dagger}=p\cdot\sigma_{\mathrm{I}_{1}\dot{\mathrm{I}}_{2}}~,~\sum_{s}x_{\boldsymbol{p},s}^{\dagger\dot{\mathrm{I}}_{1}}x_{\boldsymbol{p},s}^{\mathrm{I}_{2}}=p\cdot\bar{\sigma}^{\dot{\mathrm{I}}_{1}\mathrm{I}_{2}}\\
&\sum_{s}y_{\boldsymbol{p},s}^{\dagger\dot{\mathrm{I}}_{1}}y_{\boldsymbol{p},s}^{\mathrm{I}_{2}}=p\cdot\bar{\sigma}^{\dot{\mathrm{I}}_{1}\mathrm{I}_{2}}~,~\sum_{s}y_{\boldsymbol{p},s,\mathrm{I}_{1}}y_{\boldsymbol{p},s,\dot{\mathrm{I}}_{2}}^{\dagger}=p\cdot\sigma_{\mathrm{I}_{1}\dot{\mathrm{I}}_{2}}\\
&\sum_{s}x_{\boldsymbol{p},s,\mathrm{I}_{1}}y_{\boldsymbol{p},s}^{\mathrm{I}_{2}}=m\delta_{\mathrm{I}_{1}}^{~~\mathrm{I}_{2}}~,~\sum_{s}y_{\boldsymbol{p},s,\mathrm{I}_{1}}x_{\boldsymbol{p},s}^{\mathrm{I}_{2}}=-m\delta_{\mathrm{I}_{1}}^{~~\mathrm{I}_{2}}\\
&\sum_{s}y_{\boldsymbol{p},s}^{\dagger\dot{\mathrm{I}}_{1}}x_{\boldsymbol{p},s,\dot{\mathrm{I}}_{2}}^{\dagger}=m\delta_{~~\dot{\mathrm{I}}_{2}}^{\dot{\mathrm{I}}_{1}}~,~\sum_{s}x_{\boldsymbol{p},s}^{\dagger\dot{\mathrm{I}}_{1}}y_{\boldsymbol{p},s,\dot{\mathrm{I}}_{2}}^{\dagger}=-m\delta_{~~\dot{\mathrm{I}}_{2}}^{\dot{\mathrm{I}}_{1}}
\end{align}we arrive at a remarkably simple result of the form
\begin{align}
\nonumber
&\sum_{s}\sum_{s^{\prime}}\sum_{\chi}\vert\mathcal{M}_{\varphi(k)\to\psi^{\mathrm{I}_{1}}(s,p)\psi_{\mathrm{I}_{2}}(s^{\prime},q)e_{\alpha\beta}(\chi,l)}^{\text{(total)}}\vert^{2}=\frac{-\kappa^{4}\mathsf{y}_{\text{D}}^{2}}{16E_{l}^{4}M^{2}(M-2E_{p})^{2}\big(M-2(E_{p}+E_{l})\big)^{2}}\\
\nonumber
&\times\big(4M^{2}(E_{p}^{2}+3E_{p}E_{l}+E_{l}^{2})-4M^{3}(E_{p}+E_{l})-8E_{l}E_{p}M(E_p+E_l)+4E_l^2m^2+M^{4}\big)^{2}\\
\nonumber
&\times\bigg(M^{2}(M-2E_{p})\big((M-2E_{p})^{3}-4E_{l}^{3}\big)-4E_{l}M(M-2E_{p})^{2}\big(M(M-2E_{p})+2m^{2}\big)\\
\label{SquAmpliThreeBodyMajoranaAppend}
&+2E_{l}^{2}\big(2m^{2}M(3M-8E_{p})+3M^{2}(M-2E_{p})^{2}+8m^{4}\big)\bigg)
\end{align}
Note that if one encounters products involving more than four $\sigma$-matrices, it is necessary to repeatedly apply the triple–$\sigma$ identities \eqref{ThreeSigmaProdv1}–\eqref{ThreeSigmaProdv2} to reduce the number of $\sigma$-matrices in the product until it is brought down to a fourfold product. Only then should the trace identity \eqref{TraceFourSigmaProd} be used. Actually, the cross terms between $\mathcal{M}_{\varphi(k)\to\psi^{\mathrm{I}_{1}}\!(s,p)\psi_{\mathrm{I}_{2}}\!(s^{\prime},q)e_{\alpha\beta}(\chi,l)}^{\text{(part-I)}}$ and 
$\mathcal{M}_{\varphi(k)\to\psi^{\mathrm{I}_{1}}\!(s,p)\psi_{\mathrm{I}_{2}}\!(s^{\prime},q)e_{\alpha\beta}(\chi,l)}^{\text{(part-II)}}$ vanishes, and the squared amplitudes can be rewritten as
\begin{align}
\nonumber
&\sum_{\chi}\sum_{s}\sum_{s^{\prime}}\vert\mathcal{M}_{\varphi(k)\to\psi^{\mathrm{I}_{1}}\!(s,p)\psi_{\mathrm{I}_{2}}\!(s^{\prime},q)e_{\alpha\beta}(\chi,l)}^{\text{(total)}}\vert^{2}=\sum_{\chi}\sum_{s}\sum_{s^{\prime}}\vert\mathcal{M}_{\varphi(k)\to\psi^{\mathrm{I}_{1}}\!(s,p)\psi_{\mathrm{I}_{2}}\!(s^{\prime},q)e_{\alpha\beta}(\chi,l)}^{\text{(part-I)}}\vert^{2}\\
&\quad\quad\quad\quad\quad\quad\quad\quad\quad\quad\quad\quad\quad\quad\quad\quad~+\sum_{\chi}\sum_{s}\sum_{s^{\prime}}\vert\mathcal{M}_{\varphi(k)\to\psi^{\mathrm{I}_{1}}\!(s,p)\psi_{\mathrm{I}_{2}}\!(s^{\prime},q)e_{\alpha\beta}(\chi,l)}^{\text{(part-II)}}\vert^{2}
\end{align}Finally, by substituting \eqref{SquAmpliThreeBodyMajoranaAppend} into the general expression for the three-body decay \eqref{ScalarThreeBodyDecayFinal} and performing the angular integrations, we obtain the differential three-body decay width
\begin{align}\nonumber
\frac{d\Gamma^{(1)}_{\varphi(k)\to\psi^{\mathrm{I}_{1}}\!(p)\psi_{\mathrm{I}_{2}}\!(q)e_{\alpha\beta}(l)}}{d E_{l}}=&\frac{\kappa^4M^4\mathsf{y}_{\text{D}}^{2}}{3840\pi^3x}\bigg[60y^2\left(x(x+2)+5y^2-12xy^2-4y^4-1\right)\log\left[\frac{1+\alpha}{1-\alpha}\right]\\ \nonumber
&+\alpha\bigg(3x^2(1-2x)^2+120y^6(2x-1)-2y^4\left(2x(168x-95)+15\right)\\
&-2xy^2(x-8)(6x-5)+15y^2\bigg)\bigg]
\end{align}in which $y=m/M$ and $x=E_l/M$ denote the dimensionless mass ratio and the dimensionless graviton energy, respectively.

\bibliographystyle{unsrt}	
\bibliography{bibliography}

\end{document}